\documentclass[12pt]{article}
\usepackage{epsfig,amssymb}
\usepackage{latexsym}

\newcommand{\bq}{\begin{equation}}
\newcommand{\eq}{\end{equation}}
\newcommand{\bqa}{\begin{eqnarray}}
\newcommand{\eqa}{\end{eqnarray}}
\newcommand{\ben}{\begin{enumerate}}
\newcommand{\een}{\end{enumerate}}
\newcommand{\bc}{\begin{center}}
\newcommand{\ec}{\end{center}}
\newcommand{\bqb}{\begin{eqnarray*}}
\newcommand{\eqb}{\end{eqnarray*}}
\newcommand{\psl}{\rlap / p}

\newcommand{\eesl}{\rlap / \epsilon}

\def\pr#1#2#3{Phys. Rev. ${\bf{#1}}$, #2 (#3)}
\def\prl#1#2#3{Phys. Rev. Lett. ${\bf{#1}}$, #2 (#3)}
\def\pl#1#2#3{Phys. Lett. ${\bf{#1}}$, #2 (#3)}
\def\prep#1#2#3{Phys. Rept. ${\bf{#1}}$, #2 (#3)}

\def\np#1#2#3{Nucl. Phys. ${\bf{#1}}$, #2 (#3)}
\def\npps#1#2#3{Nucl. Phys. Proc. Suppl. ${\bf{#1}}$, #2 (#3)}
\def\zp#1#2#3{Z. f. Phys. ${\bf{#1}}$, #2 (#3)}

\def\ijmp#1#2#3{Int. J. Mod. Phys. ${\bf{#1}}$, #2 (#3)}

\def\fortp#1#2#3{Fortsch. Phys. ${\bf{#1}}$, #2 (#3)}

\def\aop#1#2#3{Annals of Phys. ${\bf{#1}}$, #2 (#3)}

\def\polon#1#2#3{Acta Phys. Polon. ${\bf{#1}}$, #2 (#3)}
\def\lnc#1#2#3{Lett.Nuov.Cim. ${\bf{#1}}$, #2 (#3)}
\begin{document}
\pagenumbering{arabic}
\thispagestyle{empty}
\def\thefootnote{\fnsymbol{footnote}}
\setcounter{footnote}{1}

%\vspace{2cm}
\begin{flushright}
February 16, 2015\\
arXiv:1409.2596 \\
 \end{flushright}

\vspace{2cm}

\begin{center}
{\Large {\bf Supersimple analysis of $e^-e^+\to  ZH$\\
 at high energy}}.\\
 \vspace{1cm}
%-----------------------------------------------------------------
{\large G.J. Gounaris$^a$ and F.M. Renard$^b$}\\
\vspace{0.2cm}
$^a$Department of Theoretical Physics, Aristotle
University of Thessaloniki,\\
Gr-54124, Thessaloniki, Greece.\\
\vspace{0.2cm}
$^b$Laboratoire Univers et Particules de Montpellier,
UMR 5299\\
Universit\'{e} Montpellier II, Place Eug\`{e}ne Bataillon CC072\\
 F-34095 Montpellier Cedex 5.\\
\end{center}

\vspace*{1.cm}
\begin{center}
{\bf Abstract}
\end{center}

We study the process $e^-e^+\to  ZH$ where $H$ represents the standard model (SM) Higgs particle
 $H_{SM}$, or the MSSM   ones  $h^0$ and  $H^0$. In each case, we  compute the one-loop effects and
 establish very simple expressions, called supersimple (sim),  for the helicity conserving (dominant) and the helicity violating (suppressed) amplitudes. Such  expressions, are then used to construct various cross sections and asymmetries, involving  polarized or unpolarized beams and Z-polarization measurements.
We  examine the adequacy  of such expressions  to distinguish SM from  MSSM effects or from other types
of BSM (beyond the standard model) contributions.

\vspace{0.5cm}
\def\thefootnote{\arabic{footnote}}
\setcounter{footnote}{0}
\clearpage

\section{Introduction}

Our basic motivation for this study is that after the Higgs boson \cite{Higgs}
discovery \cite{Higgsdiscov},
precise analyses of Higgs properties are necessary in order to confirm its origin
and nature;  SM, SUSY, other extensions, compositeness, ... etc.
Such searches will be done in several places and with various processes
\cite{Higgsearch}, in particular at the future ILC \cite{ILC}.

Our aim here is to look for simple tests, using experimental measurements at high energies,
which could immediately distinguish SM or MSSM (minimal supersymmetric standard model) contributions,
from possible additional (small) anomalous BSM effects.

For various processes observable at LHC or  ILC, we have already shown
that one loop effects can be described, at sufficiently high energies, by simple  expressions which reflect
in a clear way the nature of the underlying dynamics. We have called these expressions
"supersimple" (sim) and we have already derived them for
$e^-e^+\to  t\bar t$,  $e^-e^+\to  W^-W^+$ and other processes
\cite{heli1,heli2,super,ttbar,WW}. In all cases, these simple expressions can help
distinguishing SM or  MSSM, from   other BSM (beyond the standard model) dynamics.\\

In this paper we concentrate on $e^-e^+\to  ZH$.
This process has been considered as the most important one for studying
the Higgs boson at $e^-e^+$ colliders \cite{Denner}. As we point out below, there are several points
that may be added  to the previous analyses of this process, even  in the SM case  \cite{Denner}.

In the next sections we examine the contents of the Born and
 one loop amplitudes for this process. At the Born level,  we verify that
the helicity conserving (HC) amplitudes (those involving  longitudinal $Z$-states) are
the dominant ones at high energies, behaving  like constants;
while the helicity violating (HV) amplitudes
(those involving transverse $Z$-states) are high-energy-suppressed, but only like $m_Z/\sqrt{s}$.
This is in agreement with the asymptotic helicity conservation rule, which claims (to all orders),
 that the HC amplitudes are the only ones that may be asymptotically non-vanishing
\cite{heli1,heli2}.

At the  one loop level, the aforementioned high energy behaviour of the various amplitudes,
is only modified by $\ln$- and $\ln^2$-terms  accompanied  by  constant terms. The coefficients of these terms possibly involve  ratios of Mandelstam variables. This way, we establish the aforementioned "sim" expressions.   Such expressions clearly emphasize the dynamics behind the high energy values of the various helicity amplitudes.

  We first work in the SM case.
  Particularly for the HC amplitudes, it is instructive to see how the sim expressions are realized through cancelations among various contributions, and how their various logarithmic terms combine  to produce the  Sudakov forms expected by the general rules \cite{MSSMrules}. For achieving this, it is  advantageous to use
the equivalence theorem  relating the longitudinal $Z$ amplitudes to   the Goldstone
 $e^+e^-\to  G^0H$ ones, not only at
tree-level \cite{equivalence1}, but also to all orders in SM or MSSM  \cite{equivalence2}.

For the HV $e^+e^-\to  ZH$ amplitudes, the rather weak  $m_Z/\sqrt{s}$ Born
suppression, is considerably modified by   $\ln$ and $\ln^2$ one-loop
corrections. Therefore, at intermediately high  energies, we cannot completely
 neglect the one-loop corrections for them,
 as we have done for  $e^-e^+\to t\bar t, ~ W^-W^+$  \cite{ttbar,WW}. Therefore,
 sim expressions for them are also needed.
It turns out that these one-loop corrections  are more complicated than
those for the HC ones, reflecting the fact that there is no Sudakov rule for the
HV amplitudes \cite{MSSMrules}.

In a second step we consider the MSSM cases
$e^-e^+\to  Zh^0$ and $e^-e^+\to ZH^0$.
Apart from a simple mixing factor for the Higgs coupling (which strongly
suppresses the $H^0$ amplitudes)
the corresponding Born terms are similar to those  of the SM case. At one loop level though,
there are specific supersymmetric contributions, involving sfermions, charginos, neutralinos and additional Higgses.  We establish the corresponding MSSM
sim expressions for the HC and HV amplitudes and we compare them with those of
the SM case.

Finally, as the HV amplitudes always give only small contributions to the various observables,
it is sufficient to use simple fits for them, which we  present in subsection B.1\\

In a final step we  compare  the SM and MSSM effects, based on either the exact  one-loop
or sim expressions  for the various amplitudes. In addition the SM effects are compared  to those
of a tree-level  BSM contribution created by some anomalous $HZZ$ and $HZ\gamma$ couplings \cite{anom}.

To this aim, we study whether   the  sim expressions may be  sufficient for a
good description of various $e^-e^+\to  ZH$ observables, in either SM or MSSM.
As such, we consider cross sections involving unpolarized $e^\mp$-beams, as well as forward-backward  and polarization asymmetries constructed by using polarized $e^-$-beams and/or measuring the
final $Z$-polarization. It turns out that some of these observables may be  useful, not only for distinguishing  SM from a BSM model like the one mentioned above, but they may also be sensitive
to  SM -  MSSM differences. \\

The contents of the paper is the following. In Sect.2 we give the expressions for
the Born helicity amplitudes  for $H_{SM},h^0,H^0$ production.
In Sect.3 we present the  one-loop effects in the on-shell renormalization scheme.
Their sim expressions are introduced  in Sect.4 for all HC and  HV
amplitudes, while the complete results are given in Appendices A and B.
In Sect.5 we give an example of an effective  BSM contribution created by some  anomalous couplings of the Standard Model Higgs particle.
The possibility of a complete amplitude analysis using various unpolarized and polarized observables is described in Sect.6; while the corresponding
numerical analysis, including illustrations, is given in Sect.7.
Sect.8 contains the conclusions on the possibility of discriminating  SM or MSSM from  some
types of BSM corrections. The  accuracy of the sim expressions in SM and MSSM is also discussed.

\section{Kinematics, Born helicity amplitudes}

We consider the process
\bq
e^-_\lambda (l) ~ e^+_{\lambda'} (l') \to Z_\tau (p)~ H(p')~~, \label{process}
\eq
where $(\lambda, \lambda')$ denote the helicities of the incoming $(e^-, e^+)$ states,
and $\tau$ the helicity of the outgoing $Z$ with its polarization vector being $\epsilon$.
$H$ represents  either $H_{SM}$, $h^0$ or $H^0$. As shown in (\ref{process}),  $(l,l',p,p')$  are
the various particle momenta satisfying $l+l'=p+p'$. We also use
\bq
  s=(l+l')^2 ~~,~~ t=(l-p)^2 ,~~ u=(l-p')^2~~,~~   p_Z=\beta_Z{s\over2}~~, \label{kinematics}
\eq
where $p_Z=\sqrt{E^2_Z-m^2_Z}$  denotes  the $Z$-three-momentum
in the $ZH$-rest frame. Finally, the angle between the incoming $e^-$
momentum $l$ and the outgoing $Z$ momentum $p$, in the  center of mass frame,
is denoted as $\theta$.

In the standard model (SM) and its minimal supersymmetric extension MSSM,  the Born amplitude
\bq
A^{\rm Born}=-~{e^2f_{ZZH}\over(s-m^2_Z)}~I_1~[g^Z_{eL}P_L+g^Z_{eR}P_R] ~~, \label{Born-amp}
\eq
is only due to s-channel $Z$ exchange. Since the  electron mass is neglected, the
only possible invariant form it can contain is
\bq
I_1=\bar v(e^+_{\lambda'})~\eesl~u(e^-_\lambda) ~~. \label{I1-form}
\eq
  The relevant couplings  in (\ref{Born-amp}) are
\bq
g^Z_{eL}={-1+2s^2_W\over2s_Wc_W} ~~,~~  g^Z_{eR}={s_W\over c_W} ~~, \label{Zee-couplings}
\eq
\bq
f_{ZZH_{SM}}={m_Z\over s_Wc_W}   ~~,~~
f_{ZZh^0}={m_Z\over s_Wc_W}\sin(\beta-\alpha) ~~ , ~~
f_{ZZH^0}={m_Z\over s_Wc_W}\cos(\beta-\alpha) ~~, \label{ZZH-couplings}
\eq
where the first term in (\ref{ZZH-couplings})  applies  to SM, while the rest to MSSM. \\

Since, the neglect of the electron mass   implies
\bq
\lambda =-\lambda'=\mp {1\over 2}  ~~, \label{e-massless}
\eq
and
$(\tau=\pm 1,0)$, there exist  six independent helicity amplitudes, denoted as                               $F_{\lambda \tau}(\theta)$,  in the usual Jacob-Wick convention \cite{JW}. In case CP
is conserved though, the exact relation
\bq
F_{\lambda,\tau}(\theta)=F_{\lambda,-\tau}(\pi-\theta) ~~, \label{CP-conservation}
\eq
 reduces them to only four independent ones. \\

 At the Born level, these are the  transverse-$Z$ amplitudes ($\lambda=\pm 1/2 $, $\tau=\pm1$)
\bq
F^{\rm Born}_{\lambda,\tau}(\theta)= -~{e^2f_{ZZH}\sqrt{s}\over \sqrt{2}(s-m^2_Z)}
\Big [g^Z_{eL}(\tau\cos\theta-1)\delta_{\lambda,-}-g^Z_{eR}(\tau\cos\theta+1)\delta_{\lambda,+} \Big ]
~~, \label{FBorn-T}
\eq
and the  longitudinal-$Z$ amplitudes ($\lambda=\pm 1/2, ~,\tau=0$)
\bq
F^{\rm Born}_{\lambda,0}(\theta) =~{e^2f_{ZZH}E_Z\sqrt{s}\sin\theta\over m_Z(s-m^2_Z)}
\Big [g^Z_{eL}\delta_{\lambda,-}-g^Z_{eR}\delta_{\lambda,+} \Big ] ~~. \label{FBorn-Long}
\eq
Note that the  high energy Helicity Conservation (HCns) rule \cite{heli1, heli2} requires
\bq
\lambda +\lambda'=\tau ~~, \label{heli-cons}
\eq
which, combined with (\ref{e-massless}), implies that the  transverse amplitudes in (\ref{FBorn-T})
 violate HCns, and are indeed suppressed like $m_Z/\sqrt{s}$, according to (\ref{FBorn-T}).
We  call them  HV (helicity violating) amplitudes.

Contrary to this, the  longitudinal-$Z$ amplitudes  in (\ref{FBorn-Long}), which satisfy (\ref{heli-cons})  and are called HC (helicity conserving), tend to  constants  at high energies,  given by
\bqa
F^{\rm Born}_{\lambda,0} &\to&~{e^2f_{ZZH}\sin\theta\over2m_Z}
\Big  [g^Z_{eL}\delta_{\lambda,-}-g^Z_{eR}\delta_{\lambda,+} \Big ]  ~~. \label{FBorn-Long-asym}
\eqa
 We have checked that this result agrees
 with the direct computation
of the Goldstone process $e^-e^+\to G^0H$,
and the asymptotic relation $F^{\rm Born}(Z_{\rm long}H)=iF^{\rm Born}(G^0H)$ implied by the equivalence theorem in  \cite{equivalence1, equivalence2}.\\

\section{Electroweak corrections at one loop}

In all explicit amplitude expressions presented in Sect.3,4, we assume for simplicity that CP symmetry is respected in\footnote{The possibility of CP violating effects are
only considered with respect of a BSM model in Sect.5.}  SM and MSSM, implying according to (\ref{CP-conservation}), four independent helicity amplitudes, two HC amplitudes and two HV ones.

The corresponding  one-loop amplitudes arise from  $ZZ$ and $\gamma Z$ self-energies and counter terms generating renormalized initial and final vertices, triangles (initial and final
in $s$-channel, and up and down in $t$ and $u$ channels), direct, crossed and twisted
boxes and specific diagrams involving 4-leg bosonic couplings, see \cite{Denner}.
In the MSSM case  additional diagrams exist involving supersymmetric partners like
sleptons, squarks, charginos, neutralinos and additional Higgses.
We have recomputed all these contributions in terms of Passarino-Veltman
(PV) functions \cite{PV},  which are then expanded, as explained in the next section,
in order to obtain simple high energy expressions.

All  one loop contributions  appear in 4 invariant forms: the $I_1$-form, already appearing at Born level and  given in (\ref{I1-form}),  and the 4 new ones
\bqa
&& I_2=\bar v(e^+_{\lambda'})~\epsilon\cdot p'\psl~u(e^-_\lambda)~~,~~
I_3=\bar v(e^+_{\lambda'})~\epsilon \cdot l'\psl~~u(e^-_\lambda)~~,~~ \nonumber \\
&& I_4=\bar v(e^+_{\lambda'})~\epsilon\cdot l\psl~u(e^-_\lambda) ~~, ~~
J_1=-i\bar v(e^+_{\lambda'})
\epsilon^{\mu\nu\rho\sigma}\gamma_{\mu}\epsilon_{\nu}p'_{\rho}p_{\sigma} u(e^-_\lambda)
~~, \label{I2I3I4J1-forms}
\eqa
the last of which $J_1$ would only appear in the case of CP-violating couplings;
see Sect.5.

Concerning the counter terms, we note that they are calculated
in the so-called on-shell scheme
\cite{OS}, leading  to the usual forms  for  the renormalized initial
$Zee$- and the final $ZZH$-vertices.
We assume that the mass of the final Higgs bosons are well determined,
such that one can apply the corresponding residue and demixing constraints leading
to a contribution $\delta Z_{H}/2$ factorizing the Born term. In the SM case, we thus get
\bq
\delta Z_{H_{SM}}=-\Sigma'_H(M^2_{H_{SM}})~~. \label{deltaZHsm}
\eq
 Correspondingly in  MSSM, where  $(H_1=h^0, ~ H_2=H^0)$, we have
\bq
{1\over2}\delta Z_{H_i}=-{1\over2} \left [\Sigma'_{H_iH_i}(M^2_{H_i})
+ \sum_j {f_{ZZH_j}\over f_{ZZH_i}}
{\Big [\Sigma^*_{H_jH_i}(M^2_{H_i})-\Sigma_{H_iH_j}(M^2_{H_j}) \Big ]\over(M^2_{H_j}-M^2_{H_i})} \right ]
+\delta c_{H_i} ~~, \label{deltaZHmssm}
\eq
where
\bq
\delta c_{H_1}={\delta \sin(\beta-\alpha)\over \sin(\beta-\alpha)}~~,~~
\delta c_{H_2}={\delta \cos(\beta-\alpha)\over \sin(\beta-\alpha)}~~, \label{deltacH}
\eq
with $\alpha$ being the usual neutral Higgs mixing angle.
Using the purely divergent $\beta$ renormalization,  the quantities in (\ref{deltacH})
may be  calculated using  \cite{Freitas}
\bq
{\delta\tan\beta\over\tan\beta}={\alpha\Delta\over16\pi s^2_Wm^2_W}
\Sigma_f N_f \left ({m^2_f\over \cos^2\beta}\delta_{\rm f,up}
-{m^2_f\over \sin^2\beta}\delta_{\rm f,down} \right ) ~~, \label{deltatan}
\eq
while  $\alpha$ is kept fixed at the value given by the benchmark choice.
We have checked the cancelation of the divergences in such a scheme. The
counter terms in (\ref{deltaZHsm}-\ref{deltatan})  are expressed in terms
of the un-renormalized  self-energies $\Sigma_{H_iH_j}$, $\Sigma_{ZZ}$, $\Sigma_{\gamma Z}$
that can be found in \cite{ttbar,WW,HH}.\\

\section{Supersimple (sim) expressions}

For  deriving these simple expressions one starts from the exact one-loop results in terms of Passarino-Veltman (PV) functions,
and  then  uses their high energy expansions given in  \cite{asPV}.
The sim-results thus obtained  for the HC amplitudes are given in Appendix A,
 while the corresponding ones for the HV amplitudes appear in Appendix B.

Before discussing these results, we first note that the infrared divergencies are regularized by introducing a photon mass $m_\gamma$. As usual, these divergencies are canceled by adding to $d\sigma(e^-e^+\to ZH)/d\cos\theta$, the cross section for bremsstrahlung of an unobservable soft photon
contribution   given by \cite{Denner}
\bq
{d\sigma_{\rm brems}\over d\cos\theta}={d\sigma^{\rm Born}\over d\cos\theta}
\delta_{\rm brems}(m_\gamma,\Delta E)
~~,  \label{brems-sigma}
\eq
with
\bq
\delta_{\rm brems}(m_\gamma,\Delta E)=-  {\alpha\over\pi} \Big \{2 \Big (1-\ln{s\over m^2_e} \Big )
\ln \Big ({2\Delta E\over m_{\gamma}} \Big )
+{1\over2}\ln^2{m^2_e\over s}+\ln{m^2_e\over s}
+{\pi^2\over3} \Big \} ~~, \label{delta-br}
\eq
where  $\Delta E$   describes the highest  energy  of the emitted unobservable
soft photon satisfying
\bq
m_\gamma  \leq \Delta E \ll \sqrt{s} ~~. \label{brems-kin}
\eq
When (\ref{brems-sigma}) is added to  $d\sigma(e^-e^+\to ZH)/d\cos\theta$ (for the same polarizations  of the initial $e^\mp$-beams or the final Z boson), it completely cancels the infrared-photon part, irrespective of the actual value of $m_\gamma$, provided that it can be treated as infinitesimal. As in              \cite{WW, super,ttbar}, we always chose $m_\gamma=m_Z$, which satisfies (\ref{brems-kin}) at the high energies we are interested in, and considerably simplifies the results in Appendices A and B.

This way, the infrared divergencies may be handled, not only in the differential cross sections,
but also in  the various asymmetries defined in Sect.6.  \\

We next turn to the HC and HV amplitudes:\\

\noindent
\underline{(a) The  HC amplitudes}:\\
For the HC amplitudes, which are the leading ones at high energy,
we follow the procedure  used in \cite{super,ttbar,WW}.
In the process $e^-e^+\to  ZH$, these HC amplitudes are
the 2 longitudinal $Z$ amplitudes whose Born expressions are given in
(\ref{FBorn-Long}). At the one-loop level though, the direct derivation of the high energy
expressions is  very delicate, because of huge gauge cancelations
among  individual terms growing like $\sqrt{s}/m_Z$.
But the derivation is facilitated by working with the amplitudes
of the Goldstone process $e^-e^+\to  G^0H$, which are equivalent
up to $m^2/s$ corrections \cite{equivalence2}.

This way, one obtains the sim expressions in Appendix A.
Explicitly, for the   SM case, these are given by the results (\ref{SM-FSIMm0}, \ref{SM-FSIMp0}).
Correspondingly, for the  MSSM cases of  $H=h^0$ and $H=H^0$, the results are given in
 (\ref{MSSM-FSIMm0}, \ref{MSSM-FSIMp0}).
These expressions  clearly indicate the important dynamical contents.

As in \cite{super,ttbar,WW}, these high energy expression consist of combinations of the two augmented Sudakov terms exemplified in (\ref{Sud-ln-form}, \ref{Sud-ln2-form}), and the two energy and mass independent forms (\ref{lnr-form}, \ref{rxy-form}). Such expressions provide precise high energy predictions, where only quantities vanishing like powers of energy are neglected.\\

\noindent
\underline{(b) The  HV amplitudes}\\
The   HV amplitudes involve transverse-$Z$ states, which  at Born level  are given by (\ref{FBorn-T}).
Assuming  for simplicity CP invariance for the SM or MSSM contributions, then the constraint (\ref{CP-HV-constraint}) implies the existence of only two independent amplitudes. As such we take  $F_{--}, ~F_{+-}$ given in (\ref{FHVmm}, \ref{FHVpm}) at the one-loop order.
Since these amplitudes are only suppressed like $m/\sqrt{s}$ at high energy,
 they are  not negligible at intermediate energies, in both SM and MSSM. To obtain satisfactory expressions for them at intermediate energies, we start from the direct one loop expressions in terms of the PV functions, which we subsequently expanded. The resulting expressions are less compact than those in the HC cases, because of more involved mass dependent terms. The reason for this is due to the non-existence
of any Sudakov rule for such mass-suppressed amplitudes \cite{MSSMrules}. This way we obtain the expressions
(\ref{NL1SM}-\ref{NR34SM}) in SM, and (\ref{NL1MSSM}-\ref{NR34MSSM}) in MSSM, which must be used in conjunction with (\ref{FHVmm}, \ref{FHVpm}) mentioned above.

These expressions are quite complicated. Nevertheless, since the HV amplitudes are much less important than the HC ones, it is possible to obtain sufficiently accurate fits for them by neglecting mass differences in the various loop contributions (most importantly the logarithms) and only consider a common mass scale for them.      Such fits are given in (\ref{simple-HV-fit}) and Table B.2, expressed in terms  of squared log, linear log and constant contributions.   \\

\section{Other BSM effects}

BSM effects like anomalous Higgs couplings
to vector bosons may be described by effective operators, like e.g. in  \cite{anom}.
Here we just want to see how one can differentiate such effects
from the one loop SM  contributions,   and whether the
accuracy of the  sim expressions are adequate for that purpose.

For a simple illustration we take the 2 CP-conserving operators
$O_{UB}$, $O_{UW}$, and the 2 CP-violating ones
$\bar O_{UB}$, $\bar O_{UW}$ of \cite{anom} with couplings
$d_{UB}$, $d_{UW}$, $\bar d_{UB}$, $\bar d_{UW}$,
inducing the anomalous $\gamma ZH_{SM}$- and $ZZH_{SM}$-couplings
\bqa
d_{\gamma Z}=s_Wc_W(d_{UB}-d_{UW}) &,& d_{ZZ}=d_{UB}c^2_W+d_{UW})s^2_B ~~, \nonumber \\
\bar d_{\gamma Z}=s_Wc_W(\bar d_{UB}-\bar d_{UW}) &,&
\bar d_{ZZ}=\bar d_{UB}c^2_W+\bar d_{UW})s^2_B ~~. \label{dVV'H-couplings}
\eqa
Combining them with the SM $Zee$  couplings given in (\ref{Zee-couplings}) and the $\gamma ee$
couplings   $g^{\gamma}_{eL}=g^{\gamma}_{eR}=q_e=-1$, we obtain the tree level BSM contributions to the invariant amplitudes
\bq
A(e^-e^+ \to Z H_{SM})=-~\sum_{V=Z,\gamma}{e^2\over(s-m^2_V)}~[g^V_1~I_1+g^V_2~I_2+g^V_3~J_1]~
[g^V_{eL}P_L+g^V_{eR}P_R] ~~, \label{BSM-amplitudes}
\eq
where the $I_1, ~I_2, ~J_1$ forms are defined  in (\ref{I1-form}, \ref{I2I3I4J1-forms}), while the corresponding effective couplings are
\bqa
g^{\gamma}_1=-~{2E_Z\sqrt{s}\over m_Zs_Wc_W}d_{\gamma Z} &,&
g^{Z}_1=~{2E_Z\sqrt{s}\over m_Zs_Wc_W}d_{ZZ} ~~, \nonumber \\
g^{\gamma}_2=-~{2\over m_Zs_Wc_W}d_{\gamma Z} &,&
g^{Z}_2=~{2\over m_Zs_Wc_W}d_{ZZ} ~~, \nonumber \\
g^{\gamma}_3=-~{2\over m_Zs_Wc_W}\bar d_{\gamma Z}  &,&
g^{Z}_3=~{2\over m_Zs_Wc_W}\bar d_{ZZ} ~~. \label{gVi-eff}
\eqa
The last line in (\ref{gVi-eff}) indicates that the presence of CP-violating BSM physics would induce
a $J_1$ contribution coming from a $Z$ or  photon exchange.

For  illustrating   BSM contributions somehow comparable to  the  SM  ones
at energies below 5 TeV, we  use values of the  effective couplings in (\ref{gVi-eff}), of the order
of 0.0005. Such BSM corrections, denoted as "eff",  appear  in several figures  below. \\

\section{Observables and amplitude analysis}

In the case that no CP-conservation constraint is available, there exist six independent $e^-e^+\to ZH$ helicity amplitudes, which means that at least six independent observables are needed.
If  CP conservation holds, this number reduces to four.

Such observables may be defined  by using the two different initial $e^{\pm}$ polarizations
and the three possible final $Z$ polarizations. In constructing them,  we first note  the differential  unpolarized cross section
\bq
{d\sigma\over d\cos\theta}={\beta_Z\over 128\pi s}
\sum_{\lambda \tau}|F_{\lambda \tau}(\theta)|^2 ~~, \label{dsigma-unpol}
\eq
and its   integrated over all angles cross section
\bq
\sigma=\int^{1}_{-1} d\cos\theta {d\sigma\over d\cos\theta}~~, \label{sigma-unpol}
\eq
where $\beta_Z$ is defined in (\ref{kinematics}), and one sums  over
$\lambda=\pm{1\over2}$ and $\tau=\pm1,0$.
Cross sections could also be   integrated over the forward (with respect to the $e^-$-beam) or the backward region, leading to   $\sigma_F$ and $\sigma_B$  respectively.

Another possibility of unpolarized beam cross sections, is when the helicity of the final $Z$ is also observed. The integrated cross section in this  case are denoted as
$\sigma^{Z_\tau}$ with  $(\tau=-1,0,+1)$. Similarly $\sigma^{Z_\tau}_F$ and $\sigma^{Z_\tau}_B$ may also be considered.

Finally, the integrated cross section for  initially polarized $e^-$ beams, in the case the $Z$-helicity is not observed, is denoted as $\sigma_{L,R}$ for the cases  $\lambda=-{1\over2}$ and $\lambda=+{1\over2}$ respectively.\\

In principle we can also consider more detail measurements where polarized $e^-$-beams are used and the  $Z$ helicity is simultaneously measured. The integrated cross section may then be denoted as
$\sigma(\lambda, \tau)$ and  related useful definitions are
\bqa
&& \sigma(\lambda, \tau) ~~ \Rightarrow  ~~ \sigma \left (-\frac{1}{2}, \tau \right) \equiv \sigma_L(\tau) ~,~  \sigma \left (\frac{1}{2}, \tau \right) \equiv \sigma_R(\tau) ~~, \nonumber \\
&& \sigma (\lambda, \tau=-1) \equiv  \sigma^{Z_-}(\lambda) ~,~
\sigma(\lambda , \tau=1) \equiv  \sigma^{Z_+}(\lambda) ~,~ \nonumber \\
&& \sigma (\lambda, \tau=0) \equiv  \sigma^{Z_0}(\lambda) ~.  \label{sigma-general}
\eqa

 Apart from the unpolarized  differential   cross section in (\ref{dsigma-unpol}) and the  angularly
integrated ones, we next enumerate
 some in principle observable asymmetries, for  unpolarized or  polarized beams.
These are:

\begin{itemize}

\item

Initial $e^-$-left-right polarization asymmetries, for the case that the Z polarization is not looked at
\bq
A_{LR}={\sigma_L-\sigma_R\over \sigma_L+\sigma_R} ~~, \label{ALR-unpolZ}
\eq
or separately for the three cases where the Z helicity is also measured
\bq
A_{LR}(\tau)={\sigma_L(\tau)-\sigma_R(\tau)\over \sigma_L(\tau)+\sigma_R(\tau)}
\equiv {\sigma_L^{Z_\tau}-\sigma_R^{Z_\tau}\over \sigma_L^{Z_\tau}+\sigma_R^{Z_\tau}} ~~, \label{ALR-tau}
\eq
where $\tau=\pm 1, 0$.

\item

Left-Right asymmetries obtained by restricting angular integrations in the forward  direction,
\bq
A_{LR}^F(\tau)={\sigma_{LF}(\tau)-\sigma_{RF}(\tau) \over \sigma_{LF}(\tau)+\sigma_{RF}(\tau)}
\equiv {\Big [\sigma_L^{Z_\tau}-\sigma_R^{Z_\tau} \Big ]_F\over \Big [\sigma_L^{Z_\tau}+\sigma_R^{Z_\tau} \Big ]_F} ~~, \label{ALRF-tau}
\eq
and correspondingly in the backward direction.

\item

Final $Z$ transverse polarization asymmetry for unpolarized $e^{\mp}$ beams
\bq
A^{{\rm pol}~Z}={\sigma^{Z_-}-\sigma^{Z_+}\over \sigma^{Z_-}+\sigma^{Z_+}}  ~~, \label{A-polZ}
\eq
and  for a definite $e^-$-helicity $\lambda$
\bq
A^{{\rm pol}~Z}(\lambda)={\sigma^{Z_-}(\lambda)-\sigma^{Z_+}(\lambda)\over \sigma^{Z_-}(\lambda)+\sigma^{Z_+}(\lambda)}
\equiv {\sigma(\lambda, \tau=-1)-\sigma(\lambda, \tau=+1)\over
\sigma(\lambda, \tau=-1)+\sigma(\lambda, \tau=+1)}  ~~.  \label{Alambdatau-polZ}
\eq

\item

Forward-backward asymmetries, in the unpolarized beam case
\bq
A_{FB}={\sigma_F-\sigma_B\over \sigma} ~~, \label{AFB-polZ}
\eq
or
\bq
A_{FB}(\lambda, \tau)={\sigma_F(\lambda, \tau)-\sigma_B(\lambda, \tau)
\over \sigma(\lambda, \tau)} ~~, \label{AFB-lambdatau}
\eq
 for any definite $e^-$ and $Z$  polarizations.

\item

Combining (\ref{A-polZ}-\ref{AFB-lambdatau}), one also obtains a peculiar forward-backward asymmetry of the
above $Z$ transverse polarization asymmetry,
\bq
A^{{\rm pol}~Z}_{FB}={(\sigma^{Z_-}-\sigma^{Z_+})_F-(\sigma^{Z_-}-\sigma^{Z_+})_B
\over \sigma^{Z_-}+\sigma^{Z_+}} ~~, \label{AFB-polZ-tau}
\eq
for unpolarized $e^{\pm}$ beams, and
\bq
A^{{\rm pol}~Z}_{FB}(\lambda )
={[\sigma^{Z_-}(\lambda )-\sigma^{Z_+}(\lambda )]_F-[\sigma^{Z_-}(\lambda )-\sigma^{Z_+}(\lambda )]_B
\over \sigma^{Z_-}(\lambda )+\sigma^{Z_+}(\lambda )} ~~, \label{AFB-polZ-lambdatau}
\eq
for any definite electron helicity $\lambda=L,R$. It turns out  that the asymmetries in
(\ref{AFB-polZ-tau}, \ref{AFB-polZ-lambdatau}) are very useful for disentangling SM, MSSM and BSM corrections.

\end{itemize}

\vspace*{1cm}
In  case  CP is  conserved, then (\ref{CP-conservation}) would relate the forward and backward cross sections by
\bq
\sigma_F(\lambda, \tau)=\sigma_B(\lambda, -\tau) ~~, \label{CP-conservation-for-sigma}
\eq
 implying through  (\ref{AFB-lambdatau}) the following conditions:

\begin{itemize}

\item

\bq
A_{FB}(\lambda, -\tau)=-A_{FB}(\lambda, \tau) ~~, \label{CP-conservation-AFBlambdatau}
\eq
which  remains true also for unpolarized $e^\mp$-beams, where one sums over $\lambda=\mp 1/2$,
obtaining
\bq
  A_{FB}(-\tau)=-A_{FB}(\tau) ~ \Rightarrow ~
  A_{FB}(Z_+)=-A_{FB}(Z_-) ~ , ~  A_{FB}(Z_0)=0 ~~. \label{CP-conservation-AFBtau}
\eq
 If the Z polarization is not measured, then (\ref{CP-conservation-AFBtau}) leads to $A_{FB}=0 $.

\item

\bq
A^F_{LR}(\tau)=A^B_{LR}(-\tau)  ~~, \label{ALRFB-tau}
\eq
induced by the Left-Right asymmetries
in the forward and backward directions and  the definition (\ref{ALRF-tau}). This means
\bq
A^F_{LR}(Z_-)=A^B_{LR}(Z_+)~~~,~~~A^F_{LR}(Z_0)=A^B_{LR}(Z_0) ~~. \label{ALRFB-tau1}
\eq

\item
And the condition
\bq
A_F^{{\rm pol} ~Z}(\lambda)=-A_B^{{\rm pol}~Z}(\lambda) ~~, \label{AF-polZ}
\eq
arising from (\ref{Alambdatau-polZ}, \ref{CP-conservation-for-sigma}),
for any  $\lambda=L,R$. Note that the totally integrated asymmetry
$A^{{\rm pol}~Z}(\lambda)$ vanishes.

\end{itemize}

In the next section we illustrate the above properties for
CP conserving one loop corrections to SM or MSSM models; as well for  effective, possibly
CP violating anomalous couplings of the SM  Higgs particle.
It turns out that  some of the above asymmetries are particularly sensitive
to the dynamical details,  and may be very useful for disentangling
SM, BSM and  MSSM.\\

\section{Numerical analysis}

For the  MSSM illustrations , we use benchmarks  S1  and S2 of \cite{bench}. In both of them,
the electroweak (EW)  scale values of  all squark masses are at the 2 TeV level, $A_t=A_b=2.3$ TeV and
\bq
 \mu =0.4~~,~~ M_1=0.25 ~~,~~ M_2=0.5 ~~,~~ M_3=2 ~~, \label{bench-common-param}
\eq
where all masses are in TeV.
These two benchmarks differ only in  the leptonic and Higgs-sector EW scale parameters given by
\bqa
S1 & \Rightarrow & m_{\tilde l}=A_\tau= 0.5 ~, ~  m_{A^0}=0.5 ~,~  \tan\beta=20  ~,~ \nonumber \\
S2 & \Rightarrow & m_{\tilde l}=A_\tau= 0.75 ~, ~  m_{A^0}=1. ~,~  \tan\beta=30  ~.~ \label{bench-dif-param}
\eqa
Such  benchmarks are   consistent with present LHC constraints \cite{bench}.

The computations and  various comparisons have been made for both benchmarks  S1  and S2.
The resulting amplitudes are very similar, the differences being at the order of 0.5\%.
So we only present illustrations for the the S1 case.

\subsection{Comparison of basic amplitudes}

\begin{figure}[h]
%\vspace{-1cm}
\[
\epsfig{file=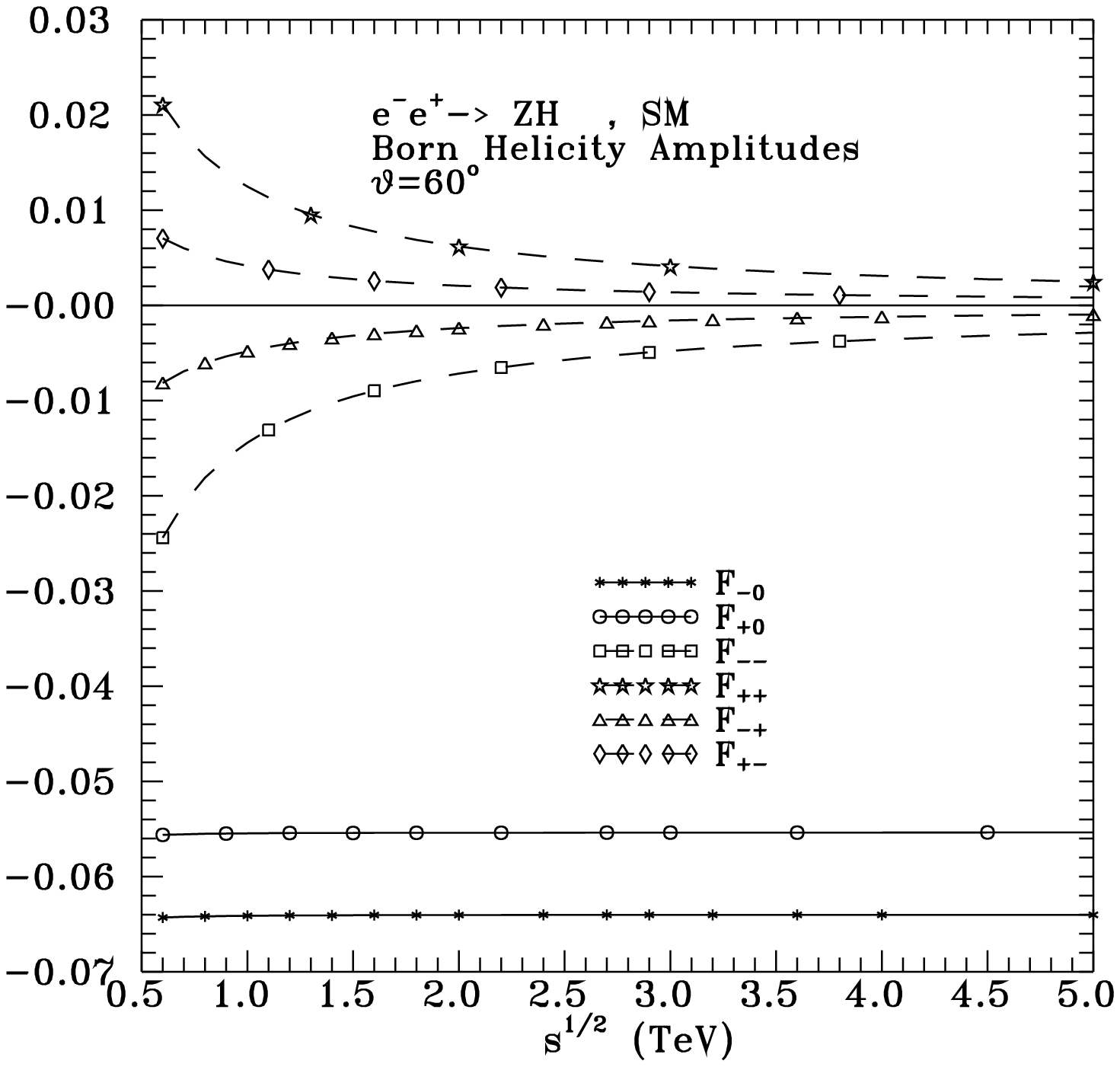, height=6.cm}\hspace{1.cm}
\epsfig{file=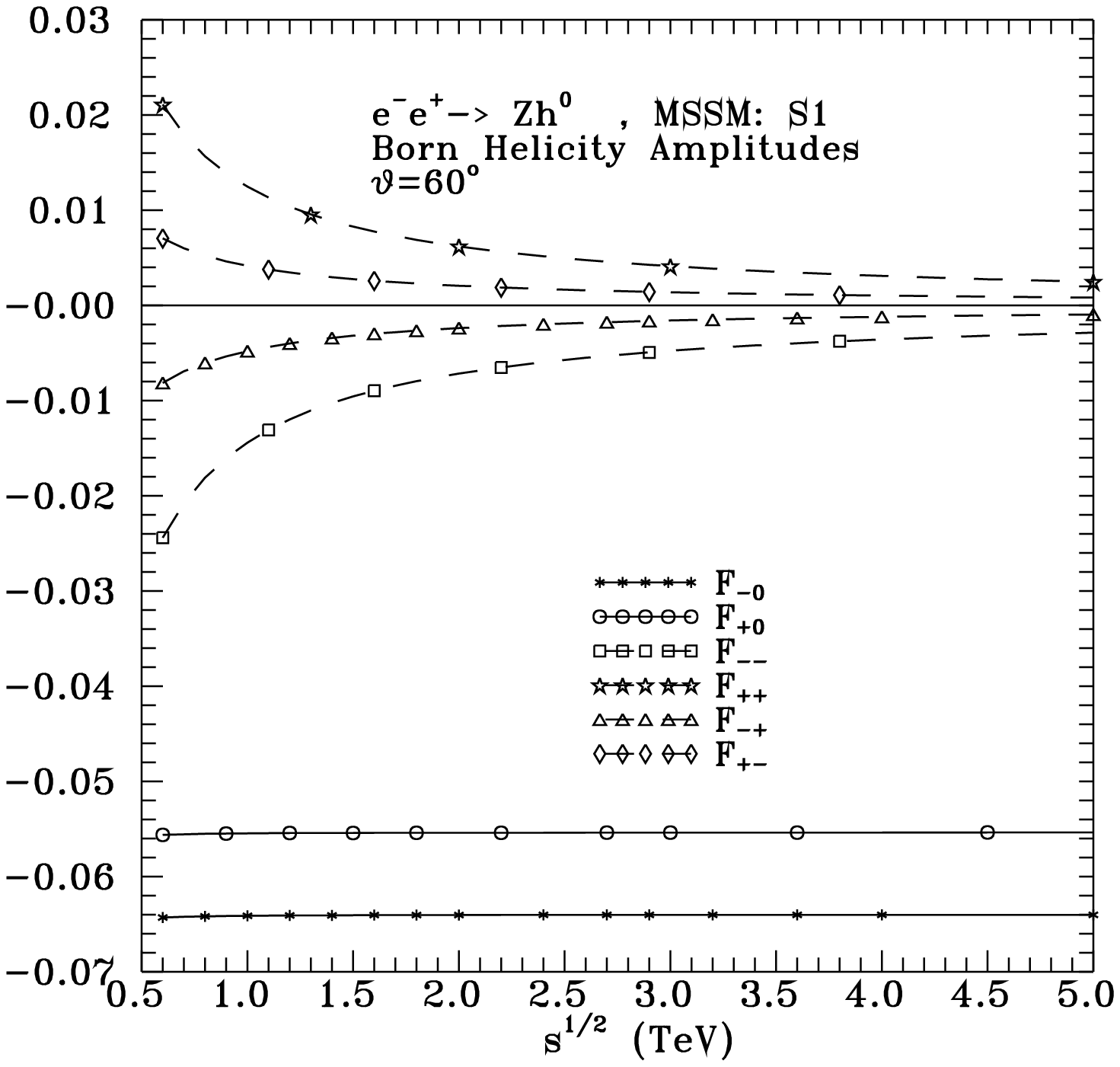,height=6.cm}
\]
\[
\epsfig{file=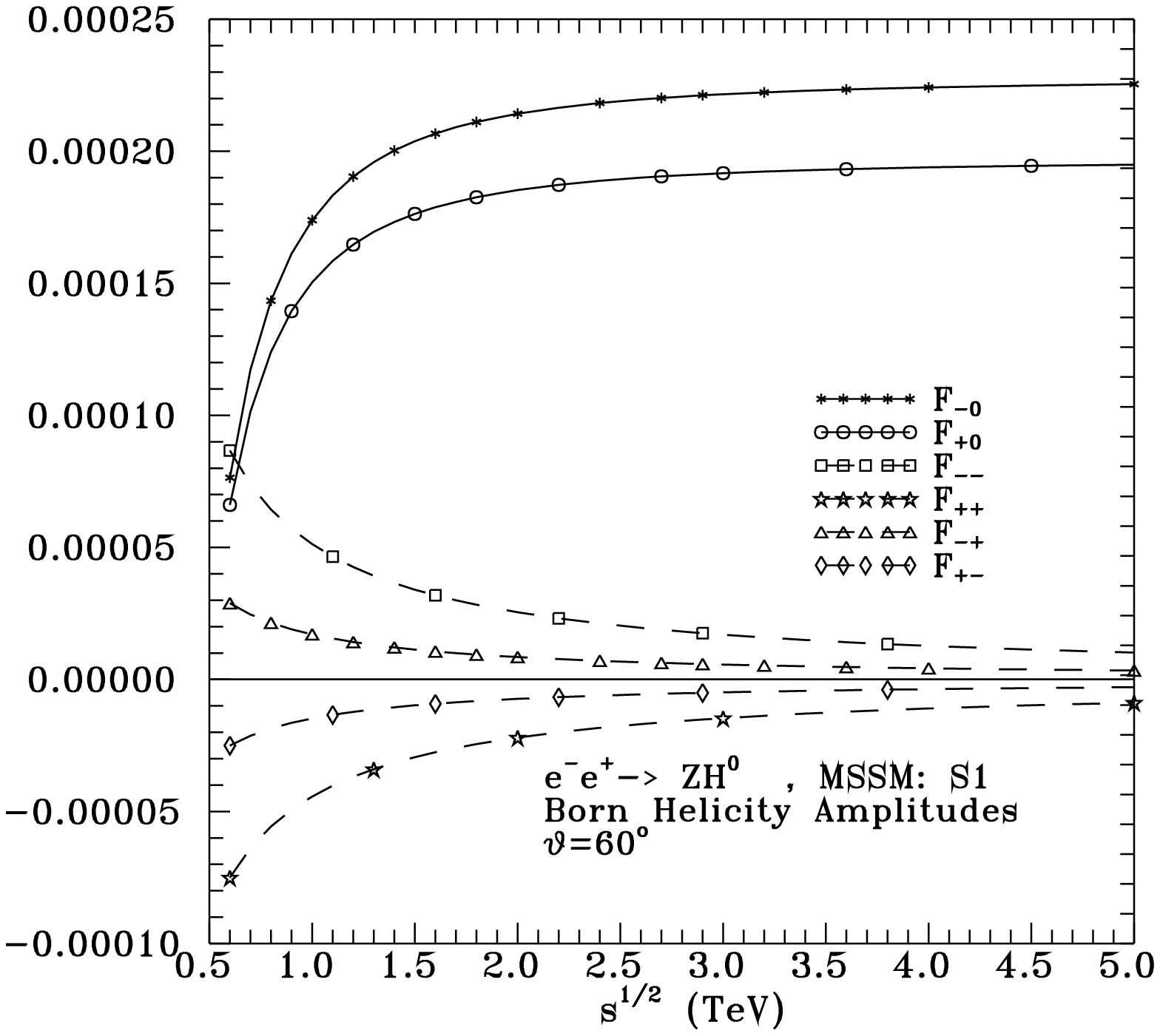, height=6.cm}
\]
\caption[1]{Energy dependence of the six Born helicity
amplitudes at $\theta=60^\circ$,
for the SM processes $e^-e^+\to Z H$ (upper left panel),
and the MSSM processes $e^-e^+\to Zh^0 ~, ZH^0$ (upper right  and the lower panels,  respectively).
For the SUSY cases we use the S1 MSSM benchmark of \cite{bench}.}
\label{Born-amp-fig}
\end{figure}

We first look at the  6 Born helicity amplitudes versus energy, at angle $60^\circ$.
Fig.\ref{Born-amp-fig} shows these Born amplitudes for $e^-e^+\to  HZ$ in SM,
and for $e^-e^+\to  h^0Z$ and $e^-e^+\to  H^0Z$ in the MSSM benchmark S1 mentioned above \cite{bench}.
As seen there, the HV amplitudes are suppressed  compared to the HC ones at high energies (above 1 TeV),
but they become comparable to them shortly below 1TeV.

Note that the $H^0$ SUSY amplitudes are very small for the S1 benchmark, due to the coupling factor $\cos(\beta-\alpha)\simeq 0$ in \cite{bench}.
This process will probably be unobservable
(cross sections will be about $10^{-5}$ times smaller than for $h^0$),
and it will not be useful to consider the one loop corrections.
Only strong anomalous couplings could lead to observable effects, in such a case.

\begin{figure}[p]
%\vspace{-1cm}
\[
\epsfig{file=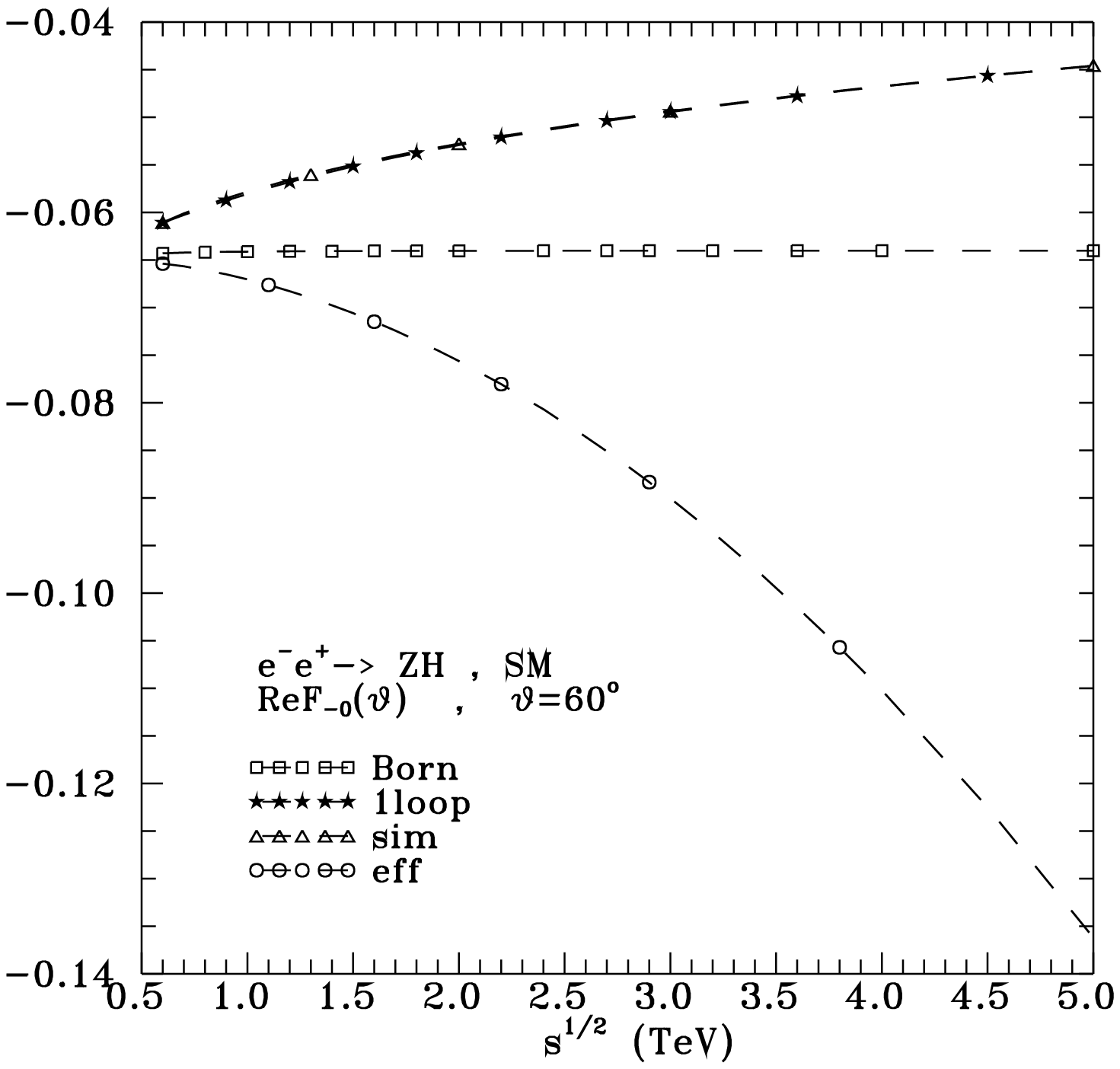, height=6.cm}\hspace{1.cm}
\epsfig{file=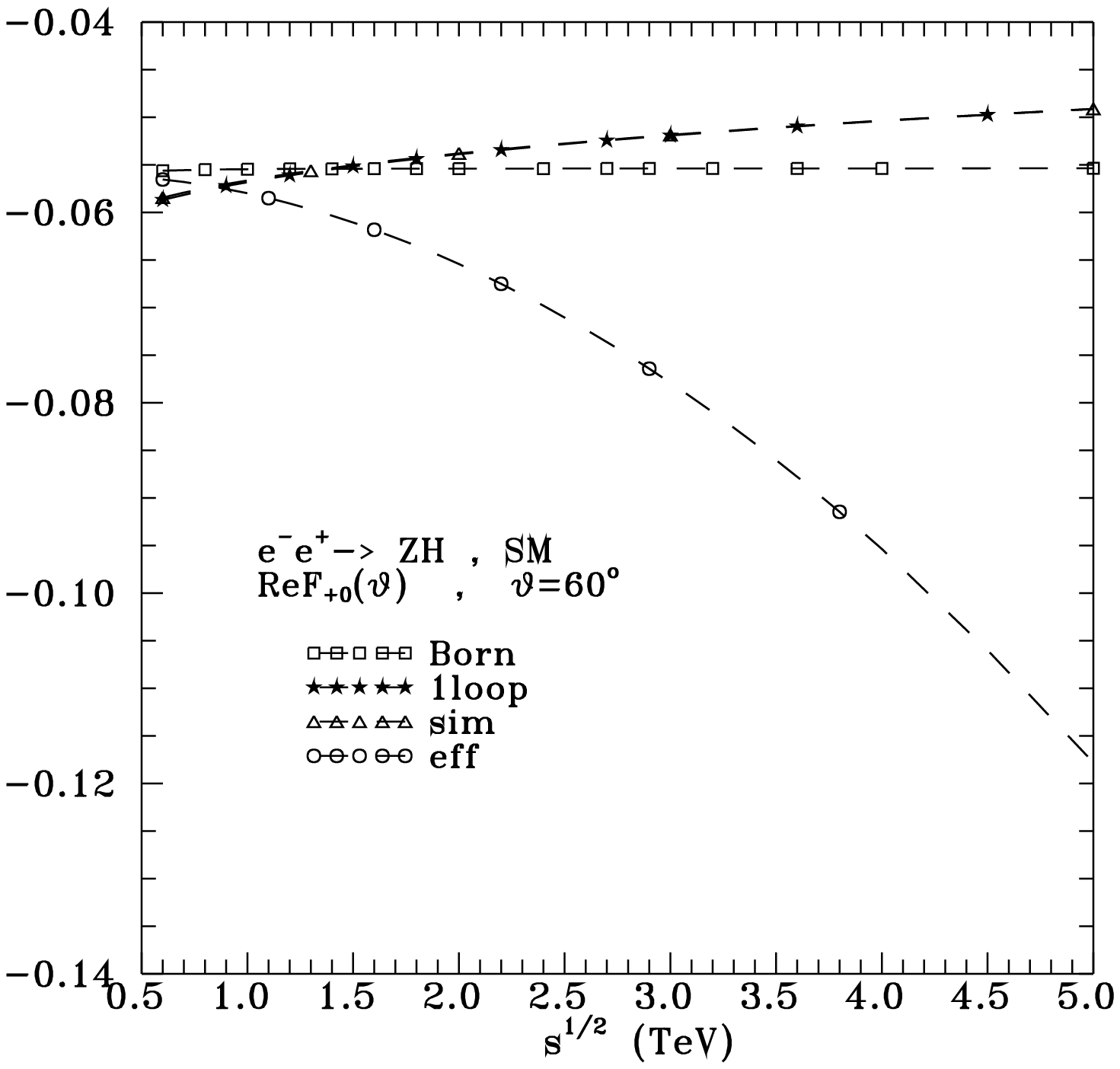,height=6.cm}
\]
\[
\epsfig{file=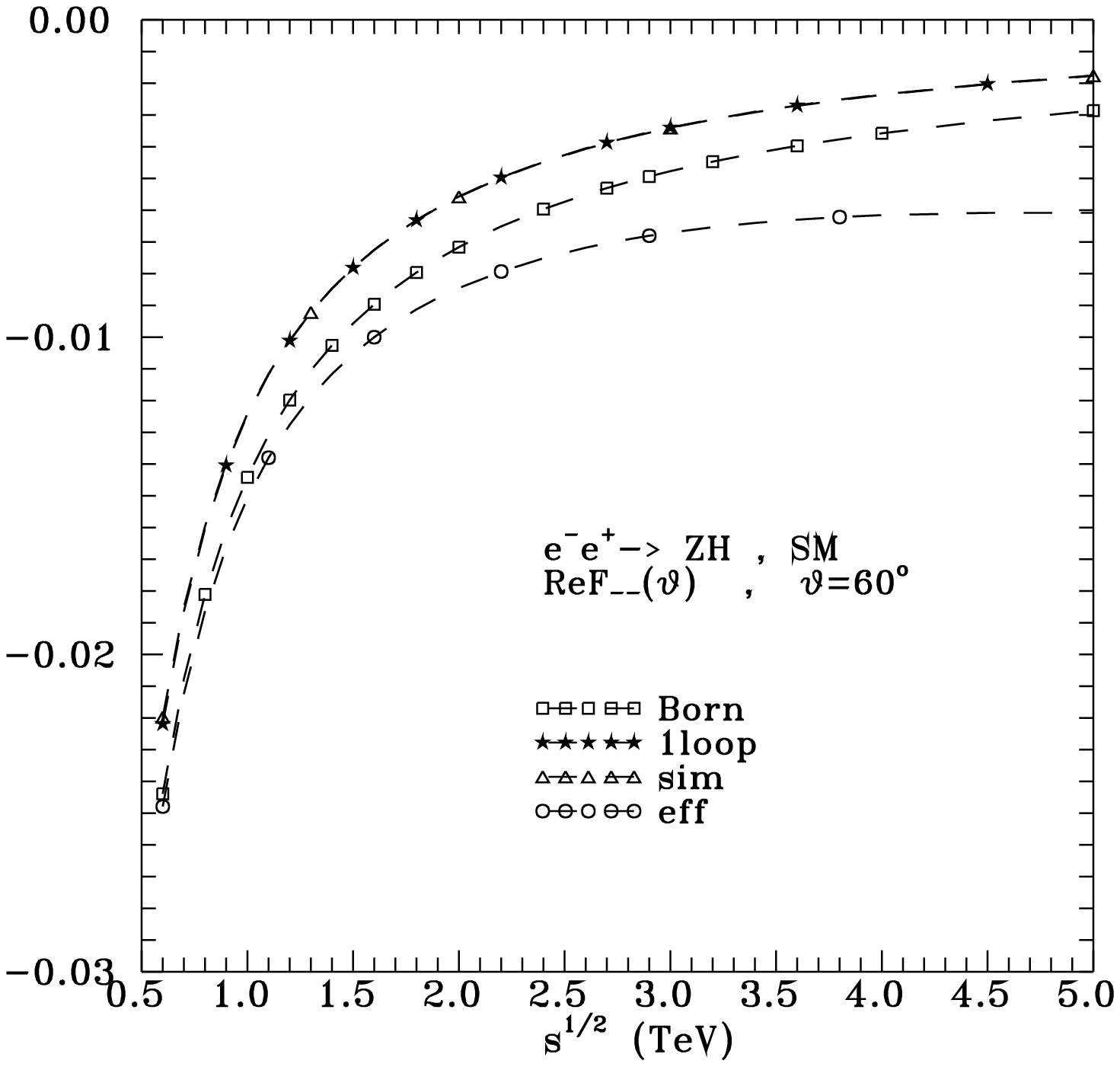, height=6.cm}\hspace{1.cm}
\epsfig{file=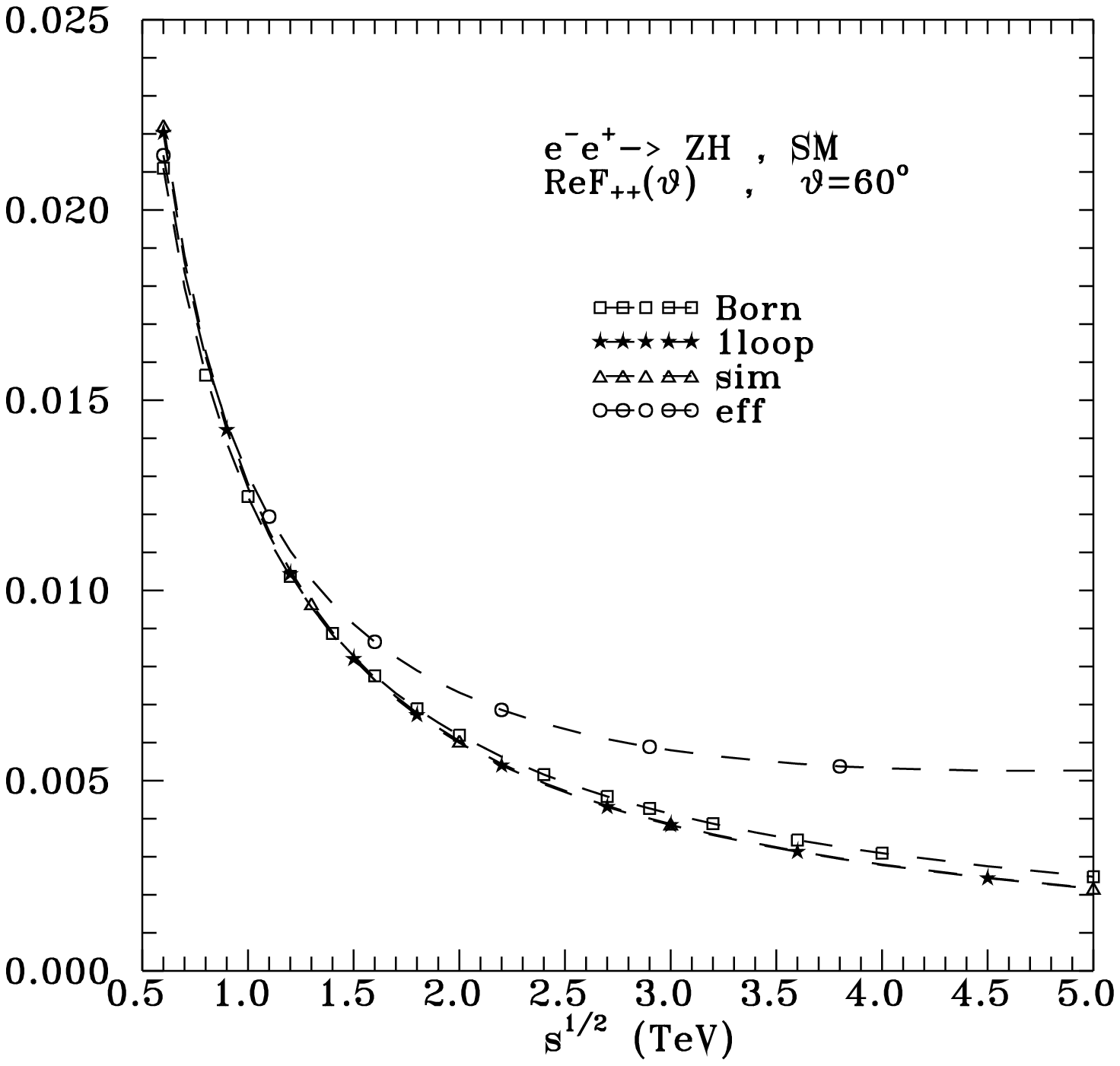,height=6.cm}
\]\[
\epsfig{file=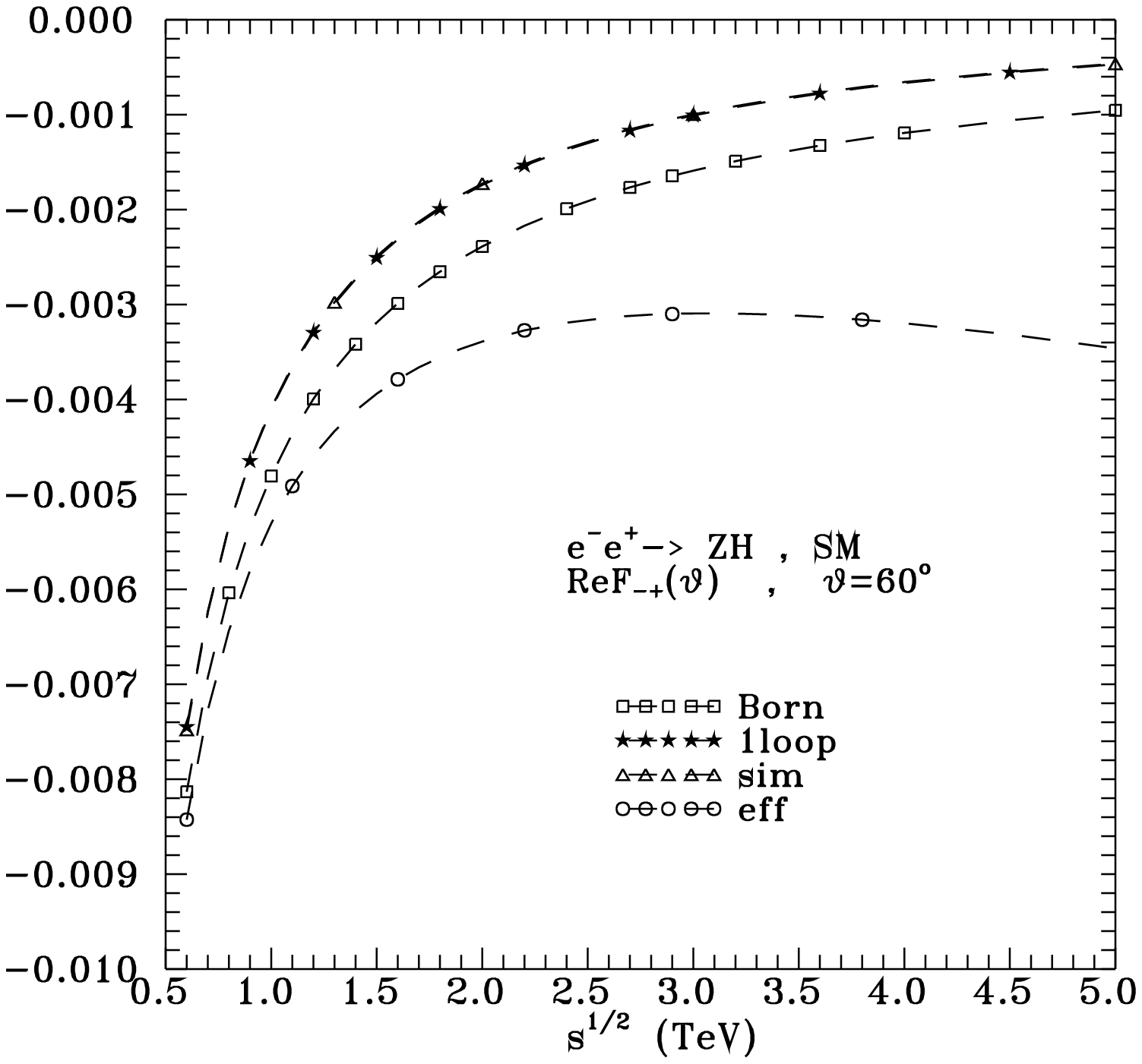, height=6.cm}\hspace{1.cm}
\epsfig{file=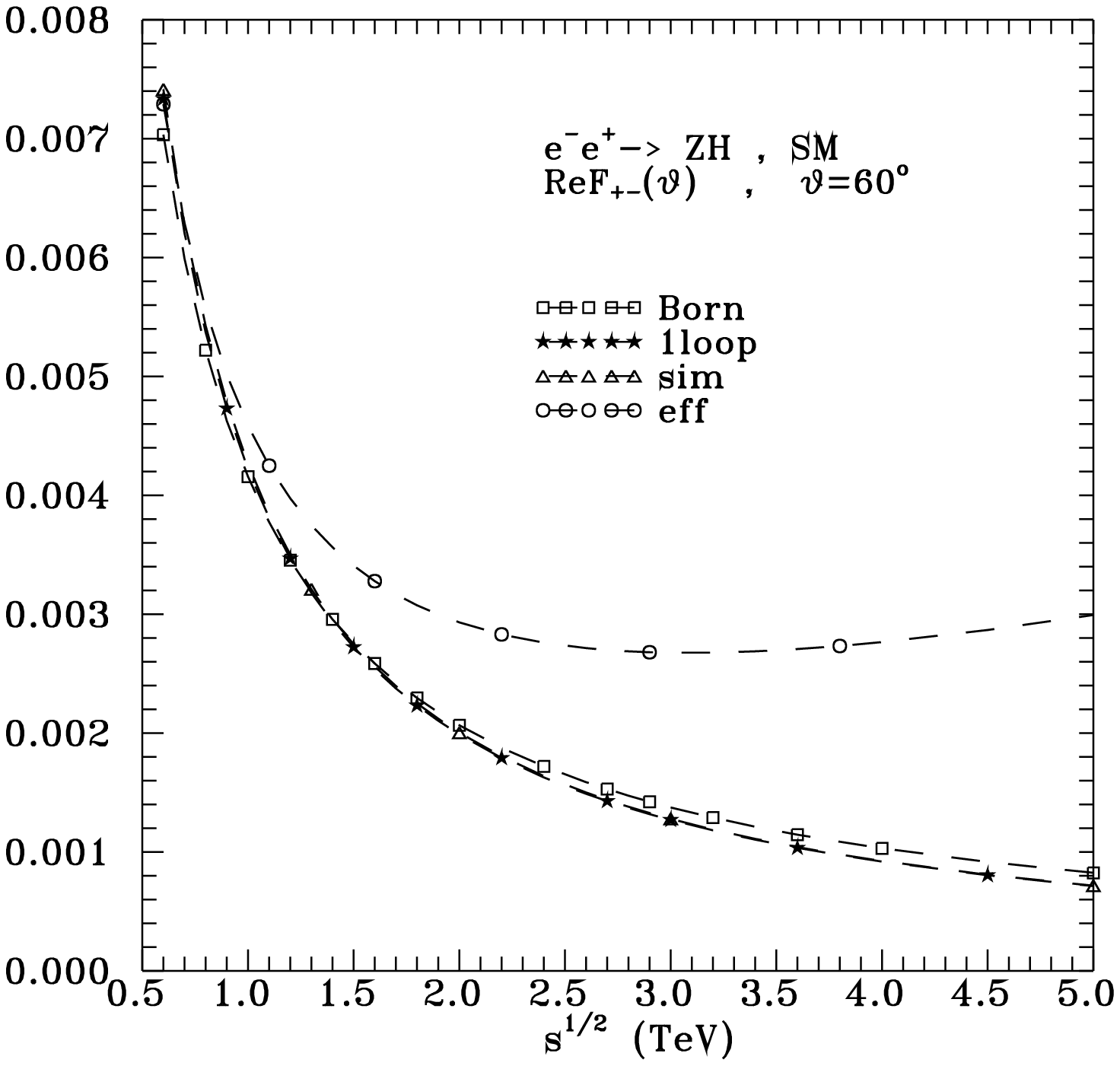,height=6.cm}
\]
\caption[1]{The real parts of the  six $e^-e^+\to Z H$ SM-amplitudes at $\theta=60^\circ$, in the Born,
1loop and sim approximations, together with amplitudes containing "eff" BSM contributions.
Upper row gives the HC amplitudes, while the HV ones are shown in the lower ones.}
\label{A2SM-amp-fig}
\end{figure}

\begin{figure}[p]
%\vspace{-1cm}
\[
\epsfig{file=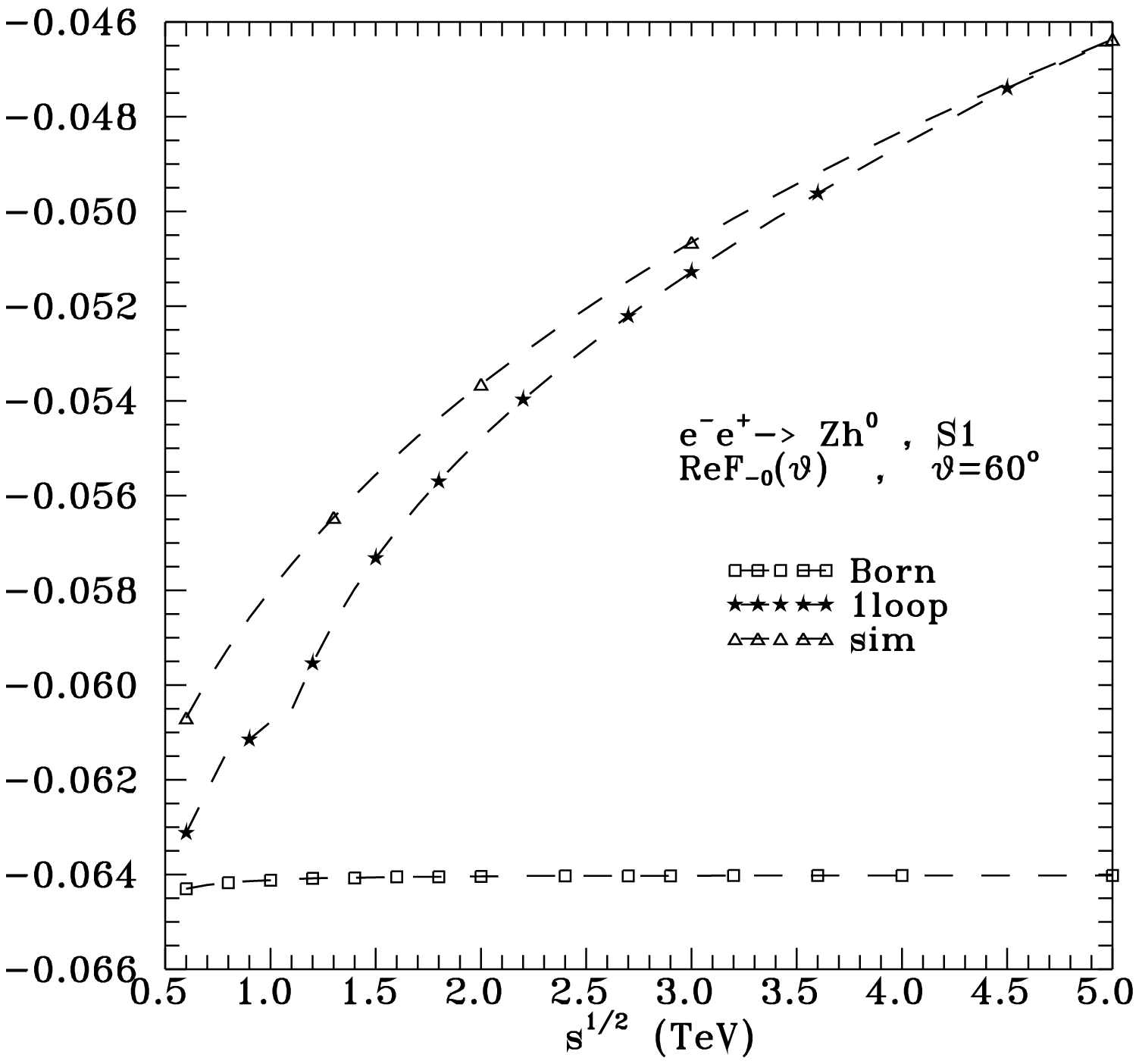, height=6.cm}\hspace{1.cm}
\epsfig{file=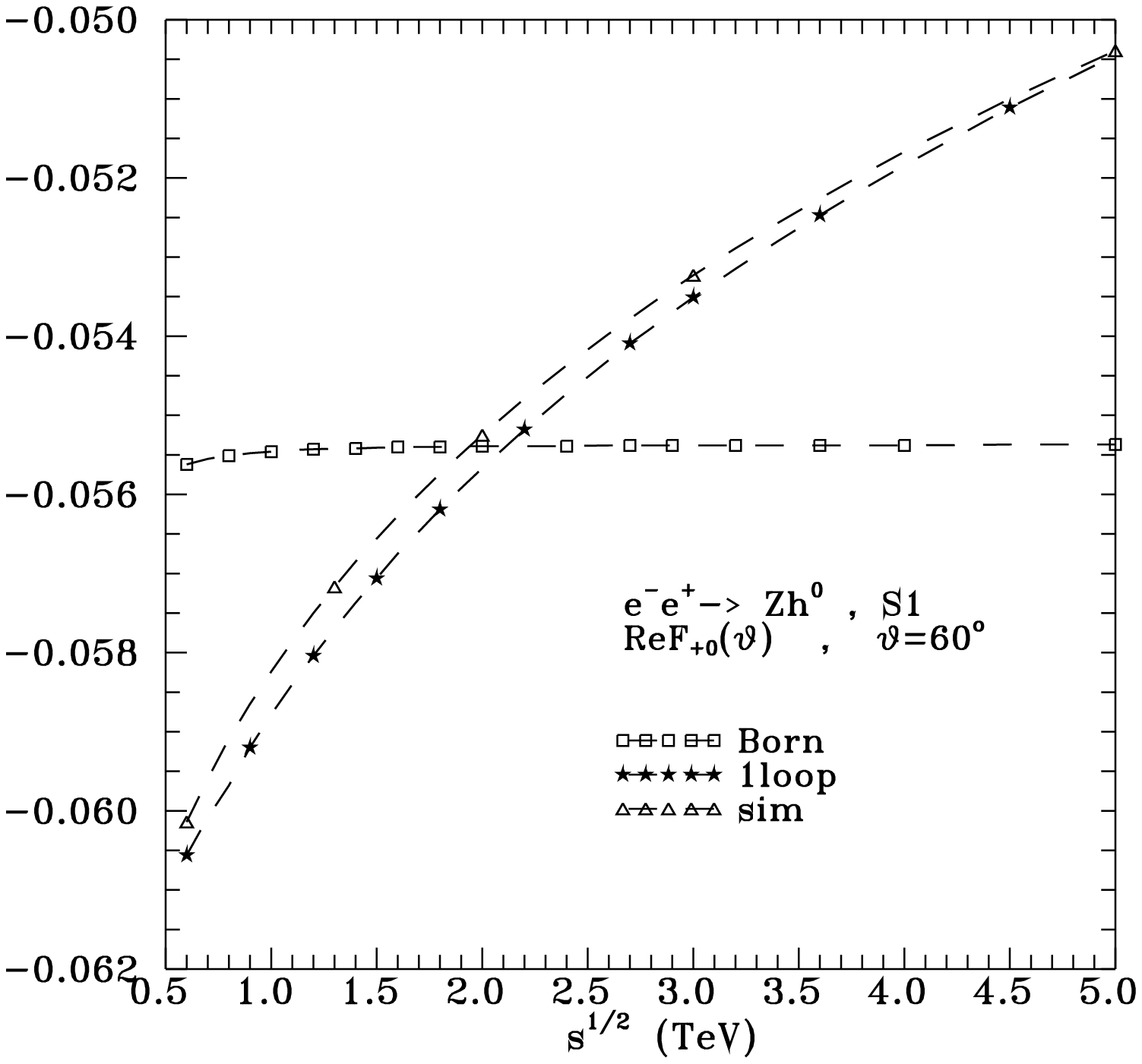,height=6.cm}
\]
\[
\epsfig{file=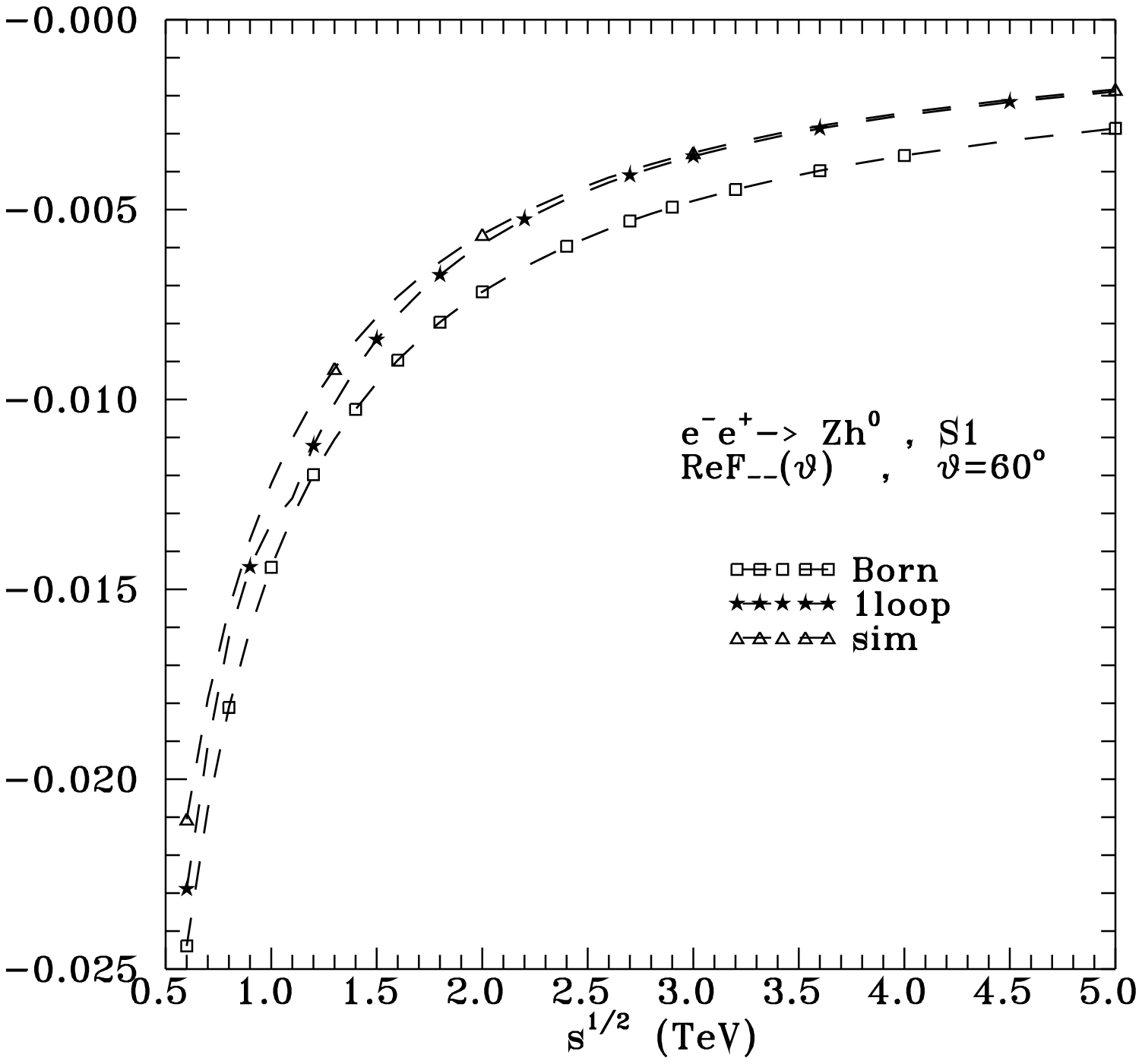, height=6.cm}\hspace{1.cm}
\epsfig{file=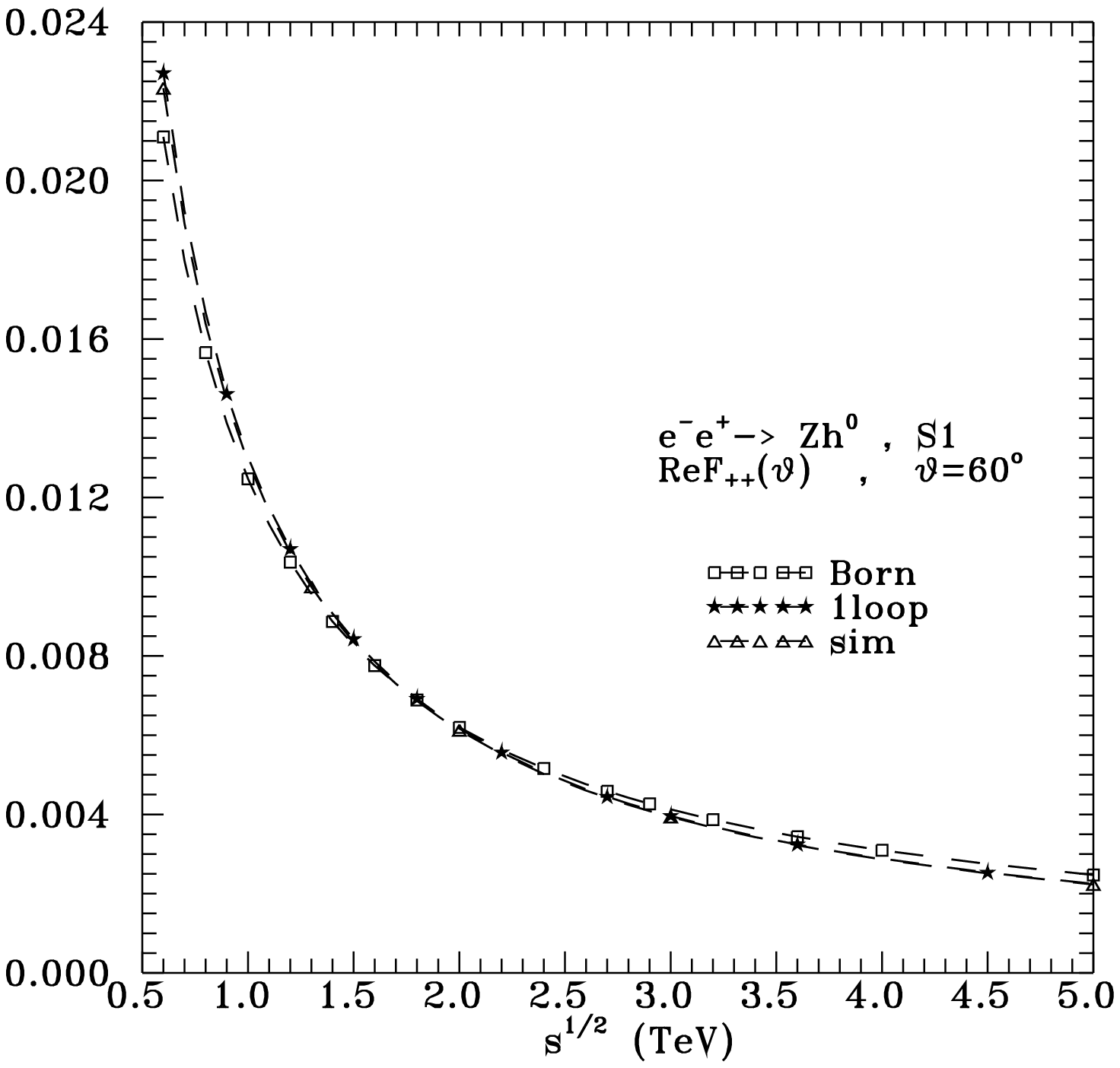,height=6.cm}
\]\[
\epsfig{file=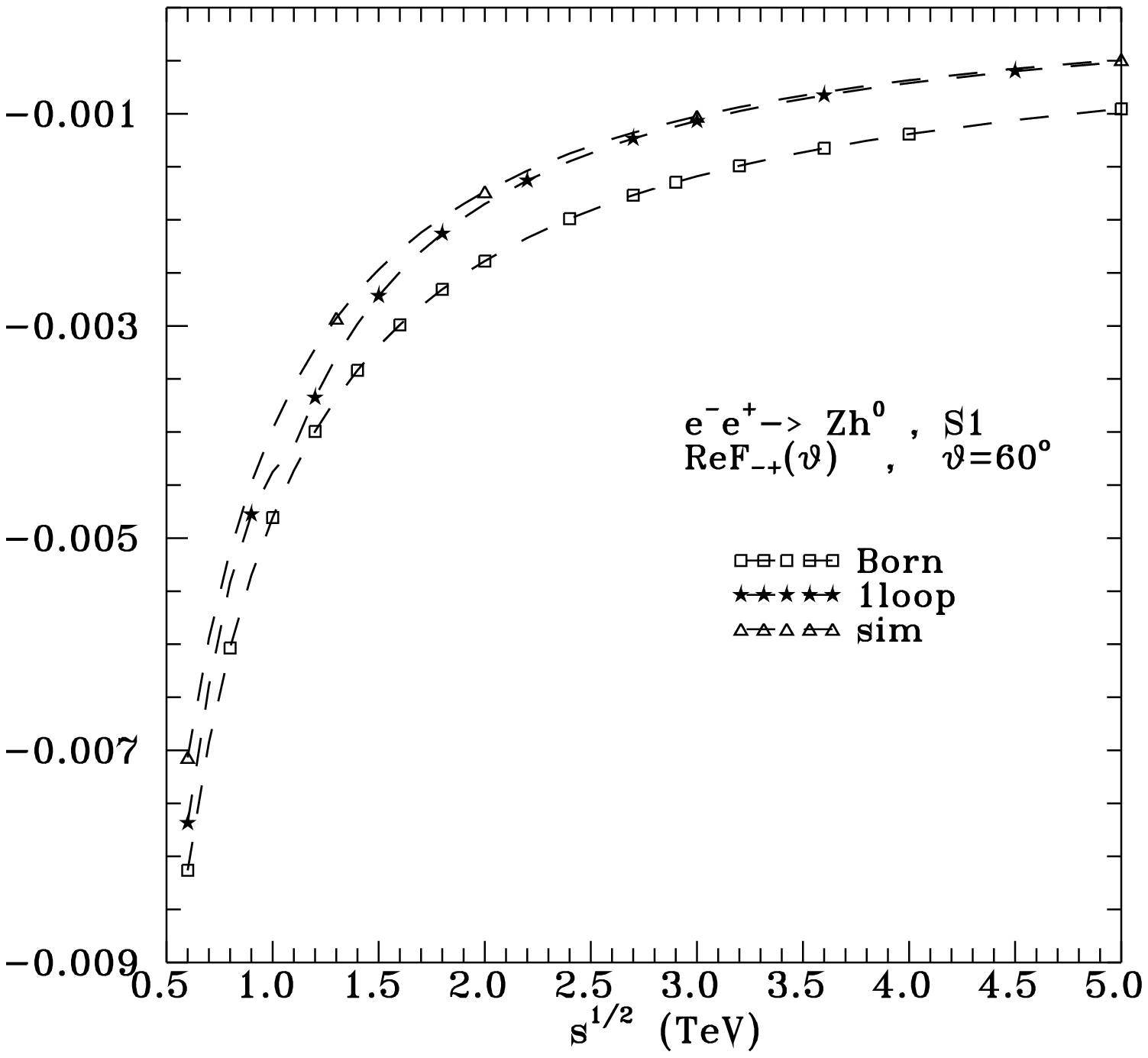, height=6.cm}\hspace{1.cm}
\epsfig{file=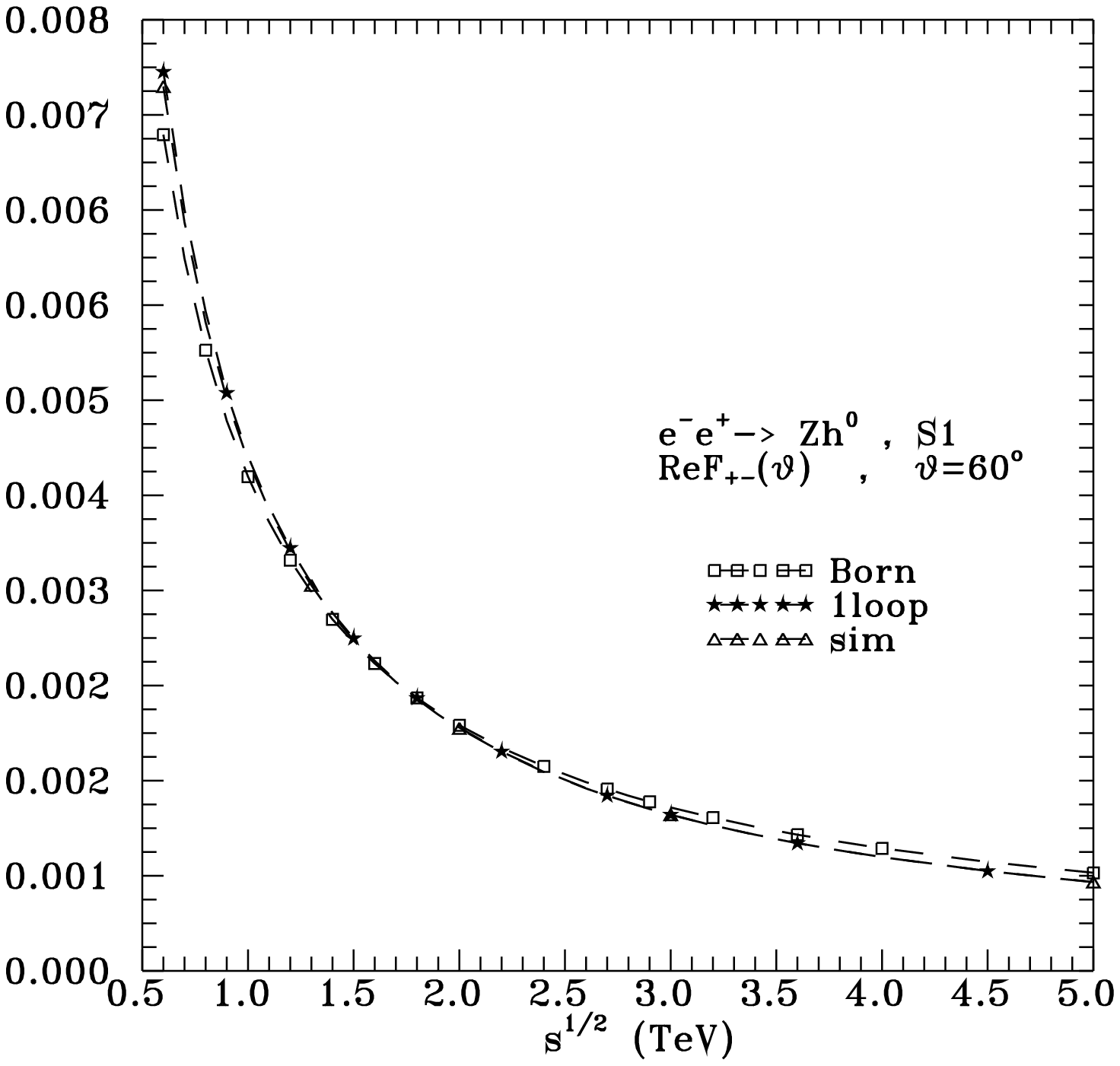,height=6.cm}
\]
\caption[1]{The real parts of the six MSSM $e^-e^+\to Z h^0$ amplitudes at $\theta=60^\circ$, for the S1 benchmark of \cite{bench}, in the Born, 1loop and sim approximations.
Upper row gives the HC amplitudes, while the HV ones are shown in the lower ones.}
\label{A2MSSM-amp-fig}
\end{figure}

We then present the one loop results for  the six amplitudes in  $e^-e^+\to  ZH$ (Fig.\ref{A2SM-amp-fig}) for  SM, and  $e^-e^+\to  Zh^0$ (Fig.\ref{A2MSSM-amp-fig} ) for MSSM. In both cases, the results are plotted versus energy at an angle $60^\circ$, and  compared to their sim and Born approximations. In all cases only the real parts of the amplitudes are shown; imaginary parts are much smaller.   In the SM case (Fig.\ref{A2SM-amp-fig}),
we also show how the various amplitudes would look like in the case that an additional
effective anomalous coupling of 0.0005, as described in Sect.5,  also exists.

Comparing the upper rows of Figs.\ref{A2SM-amp-fig}, \ref{A2MSSM-amp-fig}, with the middle and lowest ones,
one recognizes the large logarithmic (Sudakov) one-loop corrections to the Born amplitudes in the HC cases;
 for the HV amplitudes, these corrections  are relatively smaller.
In the SM case (Fig.\ref{A2SM-amp-fig}), the sim accuracy is quite  good, already at around 1 TeV.
In the $h^0$ case (Fig.\ref{A2MSSM-amp-fig}) though, the accuracy is only at the few percent level at 1 TeV, but it improves quickly  as the various supersymmetric thresholds are overpassed and the energy approaches 5 TeV.

As seen in Fig.\ref{A2SM-amp-fig}, such a sim  accuracy may be sufficient for discriminating the SM contributions to both the the HC and HV amplitudes, from a BSM contribution, like the one described by the anomalous couplings in Sect.5. For achieving this, it assumed of course that an amplitude analysis of the
experimental data, like the one sketched  Sect.6, is successfully performed.
In our illustrations we use constant effective couplings, leading to BSM contributions strongly
growing with the energy. Obviously such contributions  could be tempered above  some given basic scale,
by form factors. \\

We now turn to the various  observable introduced in Sect. 6
and study their  effects in SM, BSM and MSSM.

\subsection{ Unpolarized differential cross sections}

\begin{figure}[p]
%\vspace{-1cm}
\[
\epsfig{file=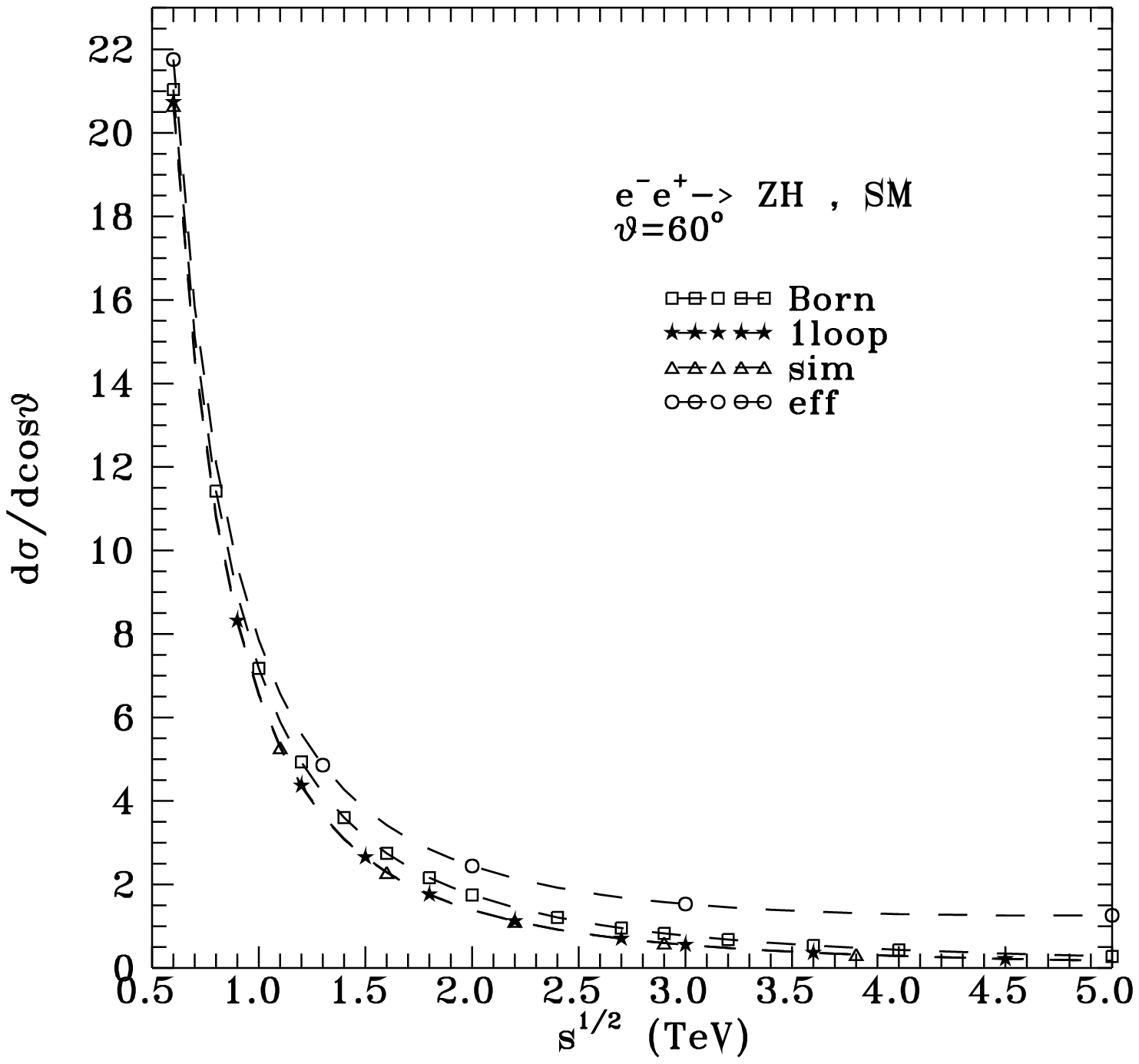, height=6.cm}\hspace{1.cm}
\epsfig{file=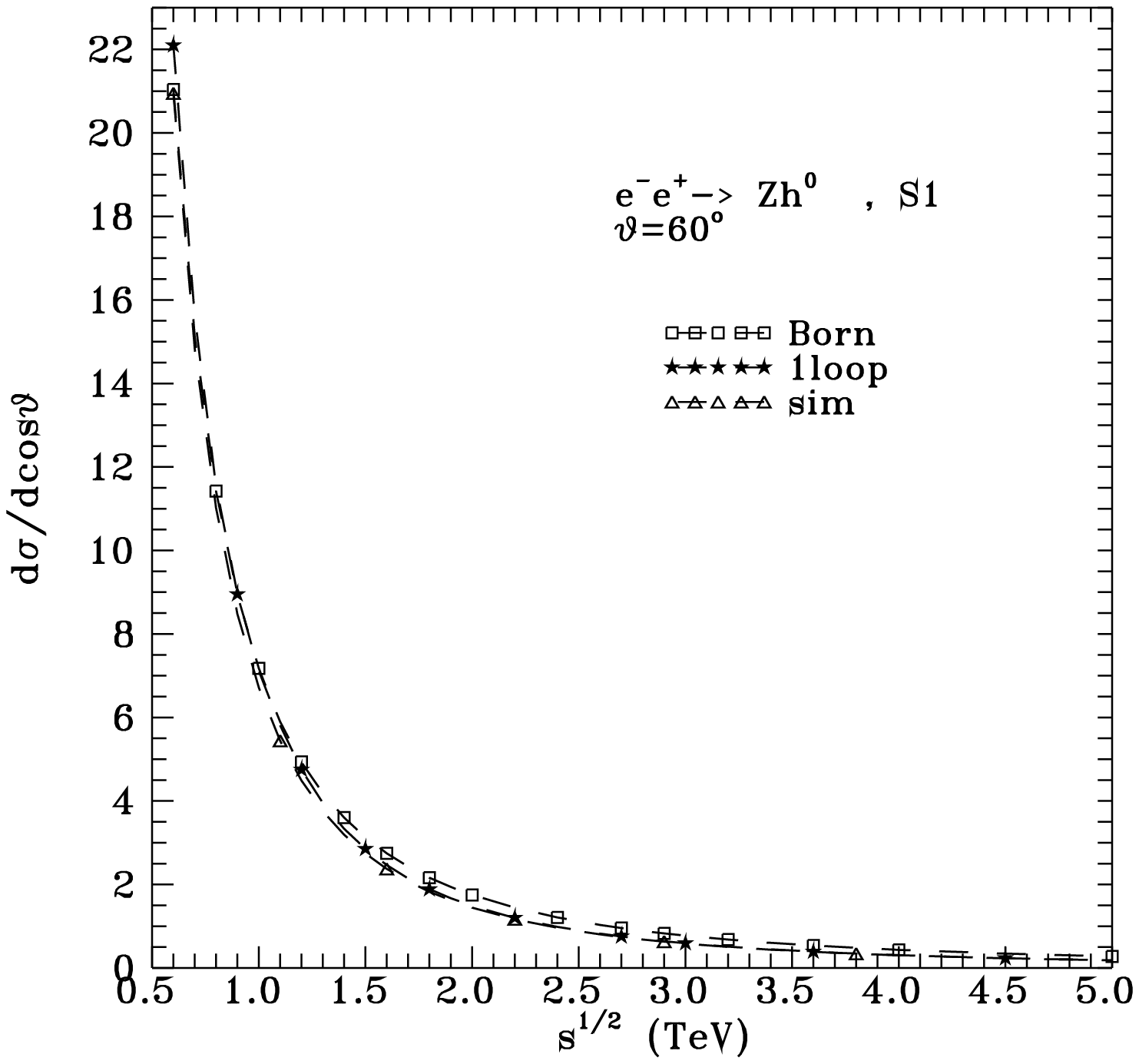,height=6.cm}
\]
\[
\epsfig{file=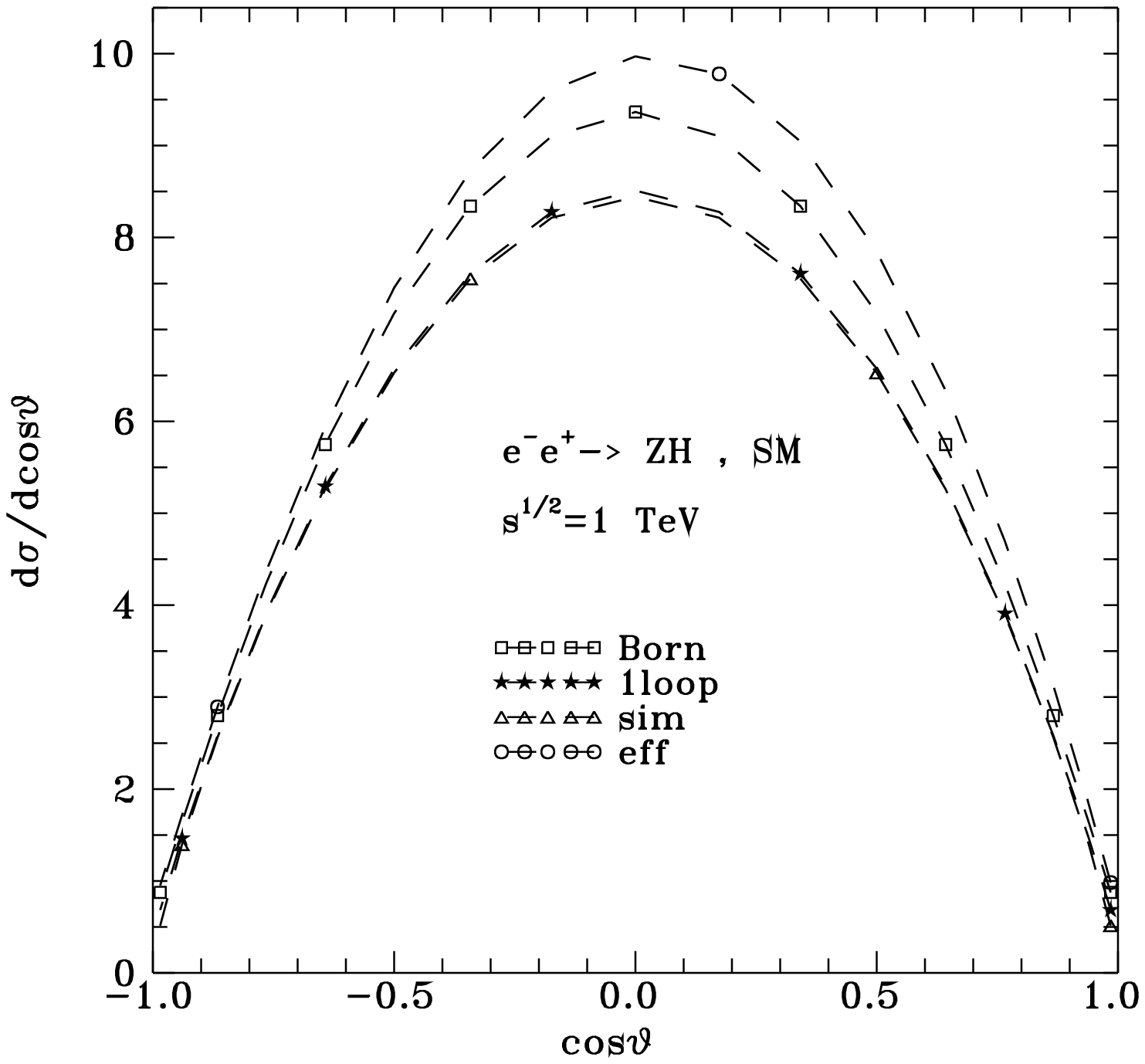, height=6.cm}\hspace{1.cm}
\epsfig{file=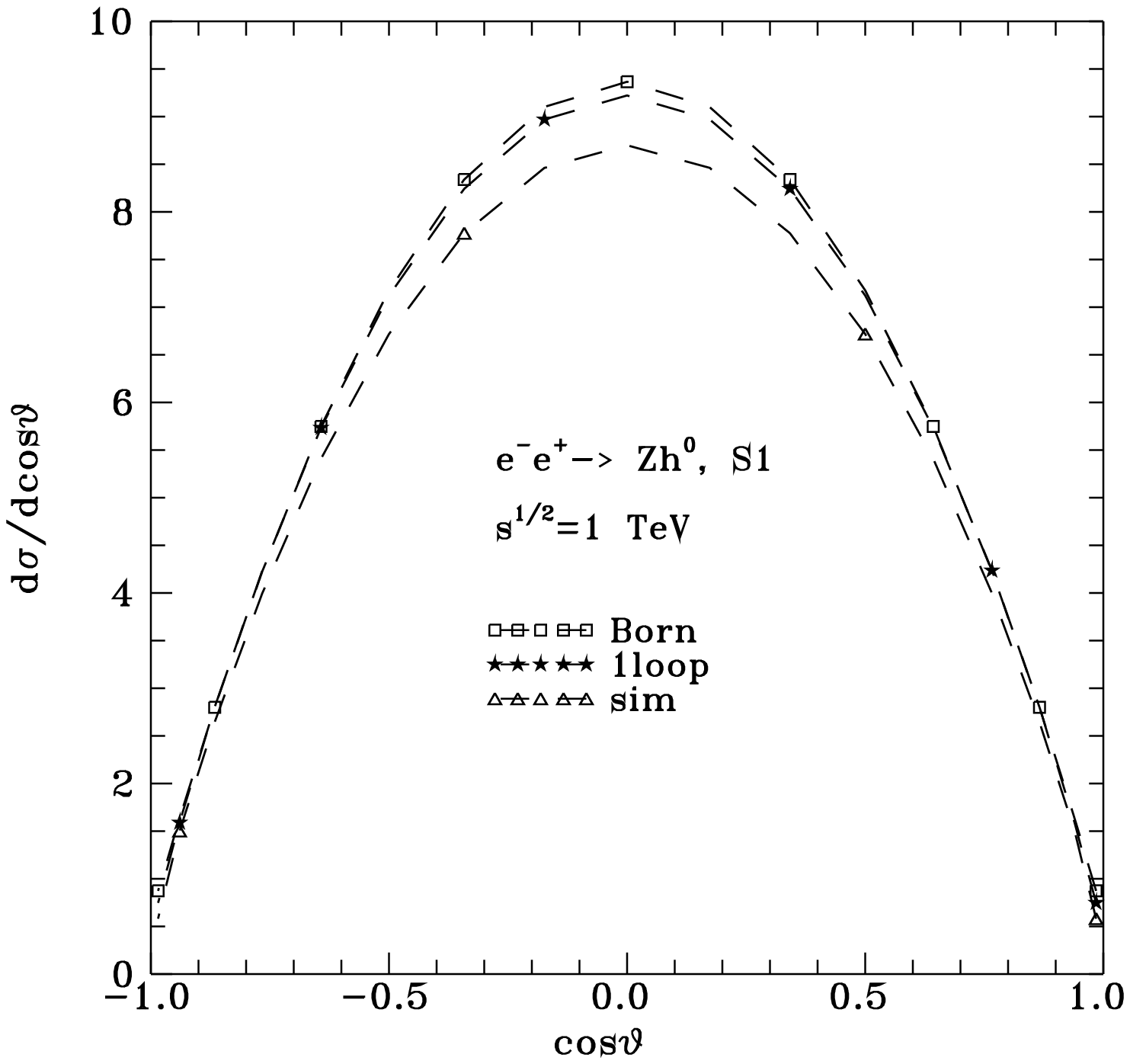,height=6.cm}
\]
\[
\epsfig{file=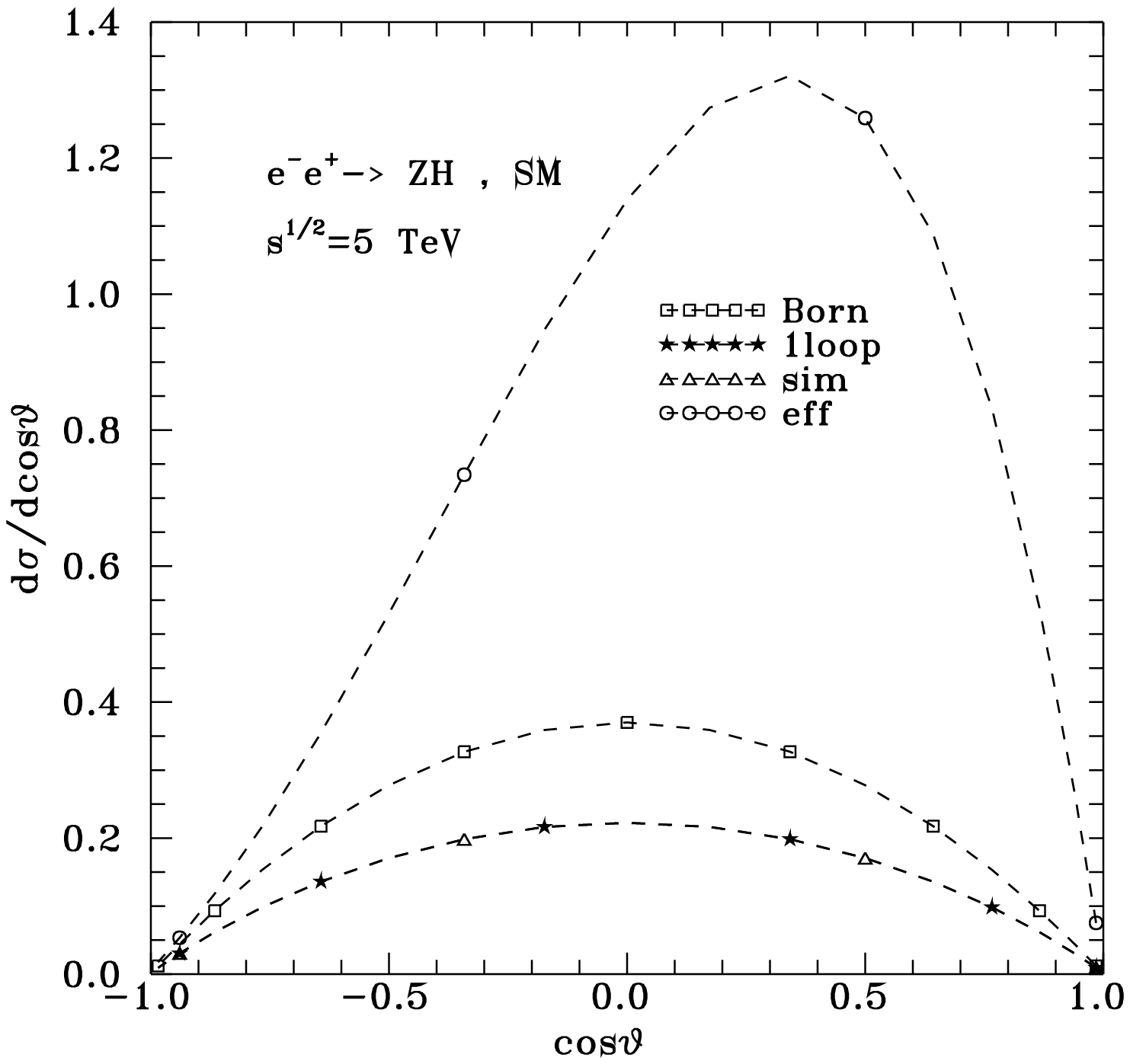, height=6.cm}\hspace{1.cm}
\epsfig{file=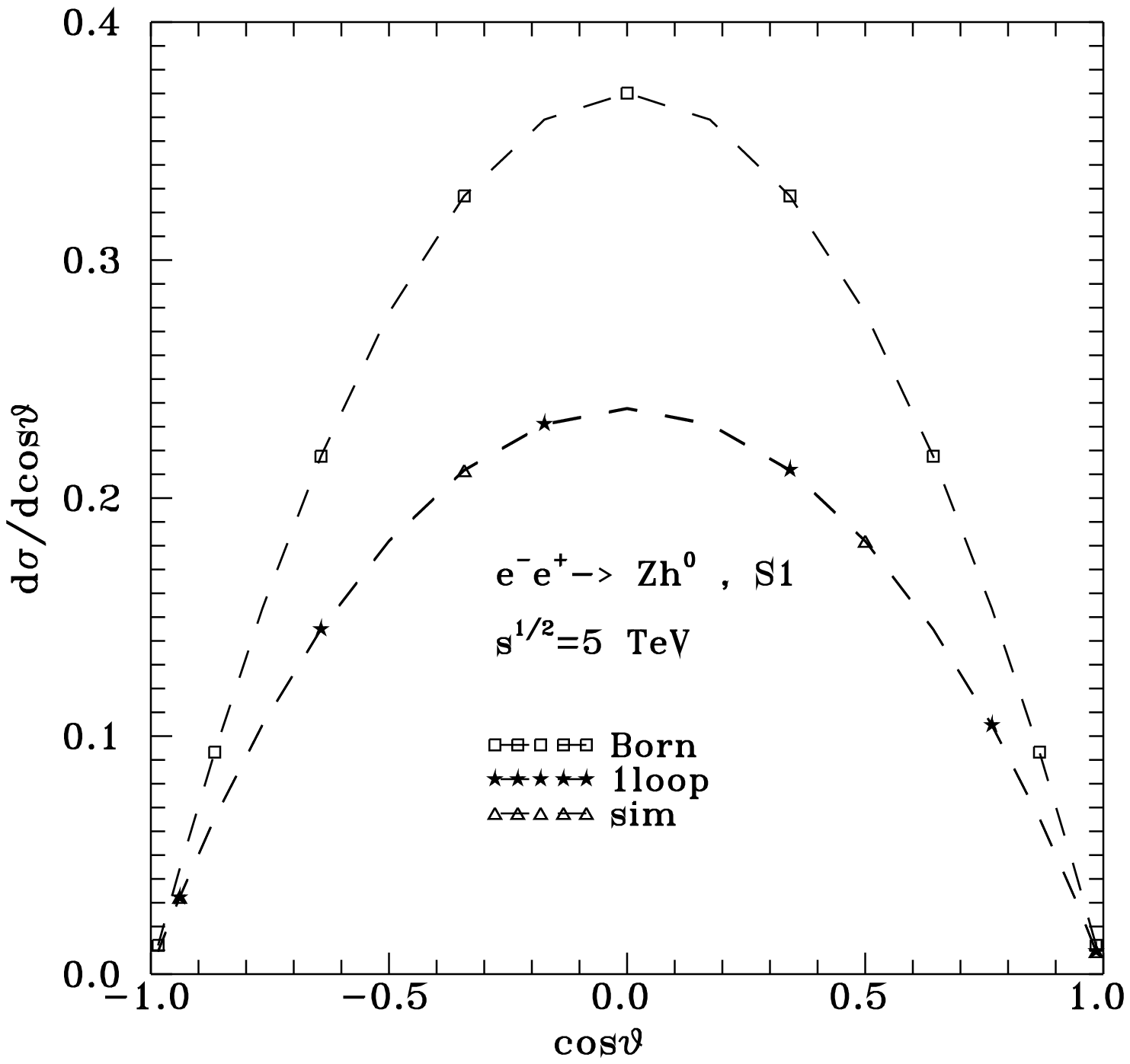,height=6.cm}
\]
\caption[1]{The unpolarized differential cross sections in SM (left panels) and and the S1 MSSM benchmark \cite{bench} (right panels). Upper row describes energy dependence at $\theta =60^\circ$; middle (lowest) panels  give the angular distributions at $\sqrt{s}=1~ {\rm TeV}$ ($\sqrt{s}=5~ {\rm TeV}$).}
\label{sigmas-fig}
\end{figure}

Fig.\ref{sigmas-fig} shows the unpolarized differential cross section for $e^-e^+\to Z H$
in SM (left panels) and for $e^-e^+\to Z h^0$ in MSSM (right panels),
first versus the energy at $60^\circ$ (upper panels),
and then versus $\cos\theta$ at  1 TeV (middle) and at 5 TeV (lowest) panels.
The various panels intend to compare  the Born,  exact one loop and the sim approximation.
In the left panels the possible effect of an additional effective anomalous interaction is also shown.

Note that cross sections become largest at angles around $90^\circ$, due to the $\sin\theta$
dependence of the leading Born HC amplitudes in (\ref{FBorn-Long}).
 The 1 loop contribution is of the order of 20\%
at 1 TeV, and increasing with the energy. The sim approximation is already good
at 1 TeV and becomes better at higher energies.

\subsection{Differential Asymmetries}

\begin{figure}[p]
%\vspace{-1cm}
\[
\epsfig{file=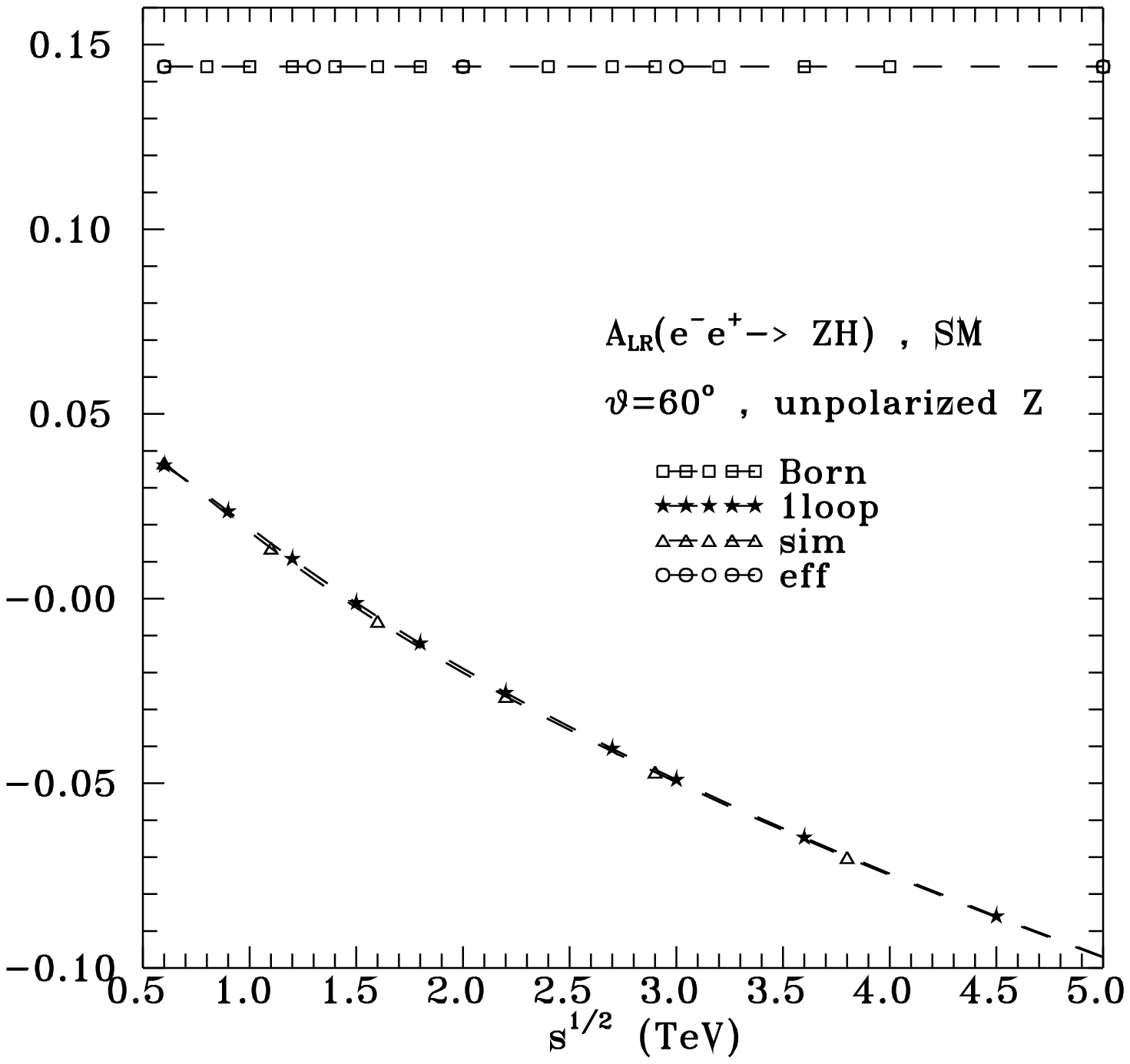, height=6.cm}\hspace{1.cm}
\epsfig{file=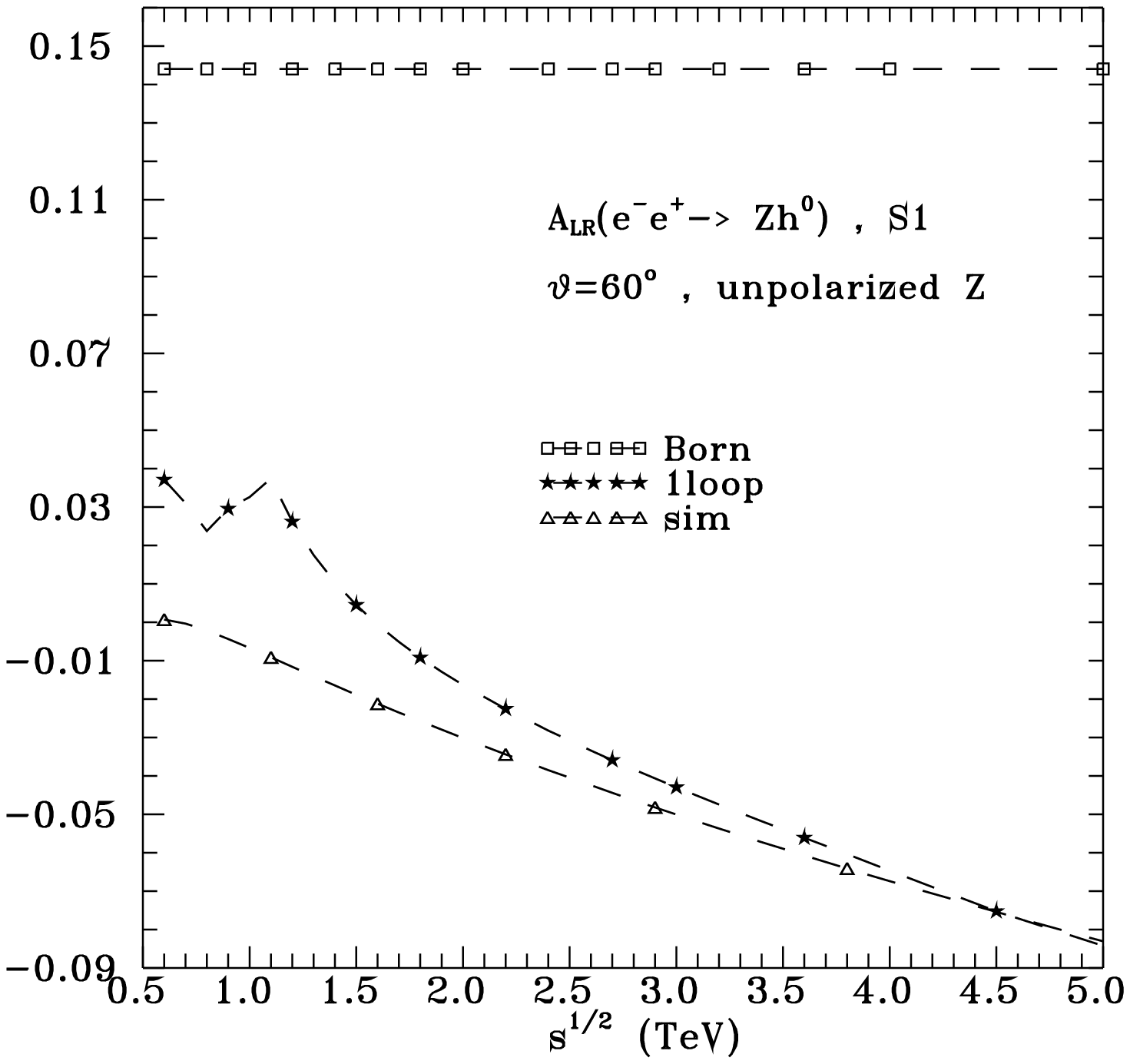,height=6.cm}
\]
\[
\epsfig{file=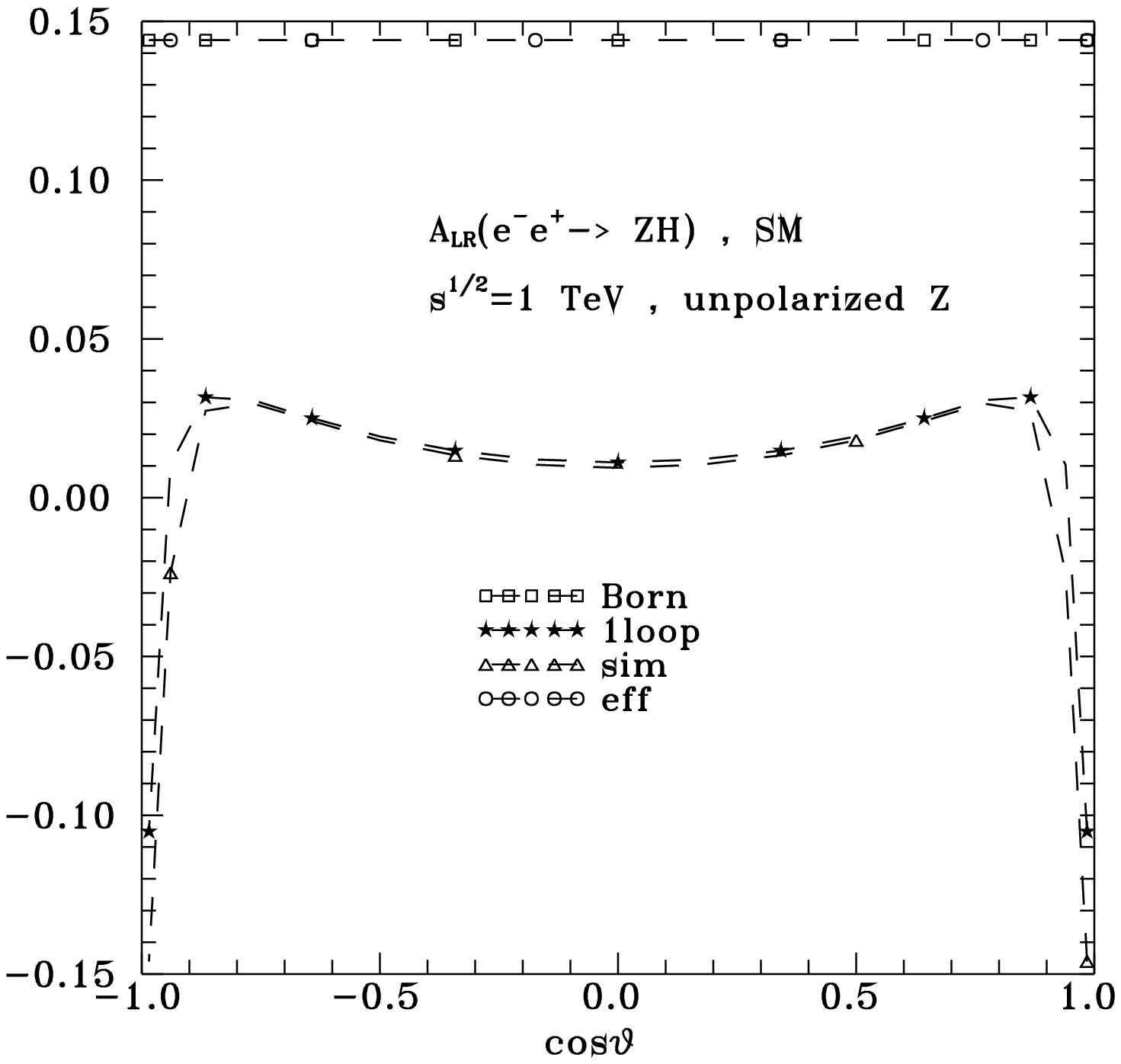, height=6.cm}\hspace{1.cm}
\epsfig{file=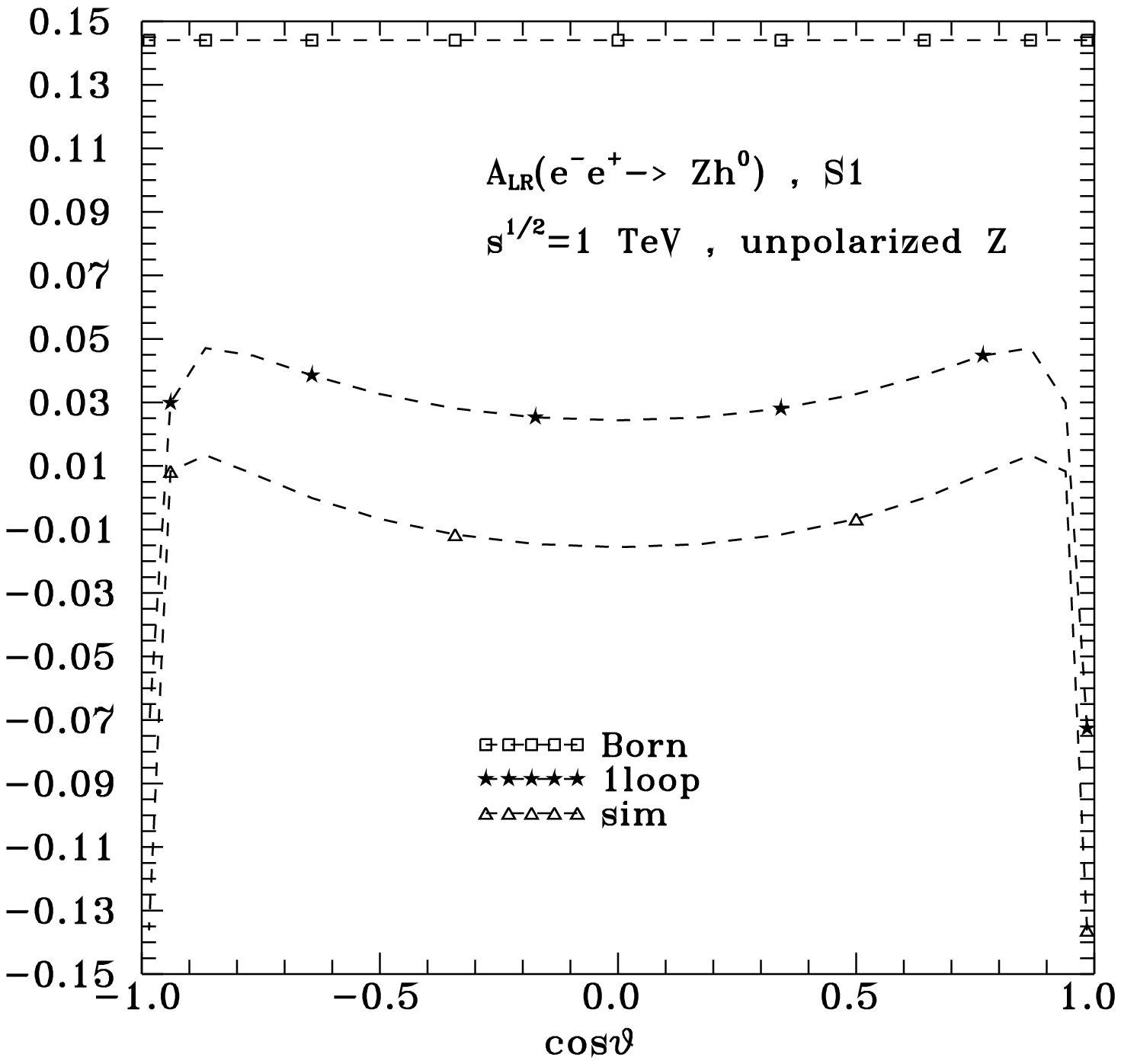,height=6.cm}
\]
\[
\epsfig{file=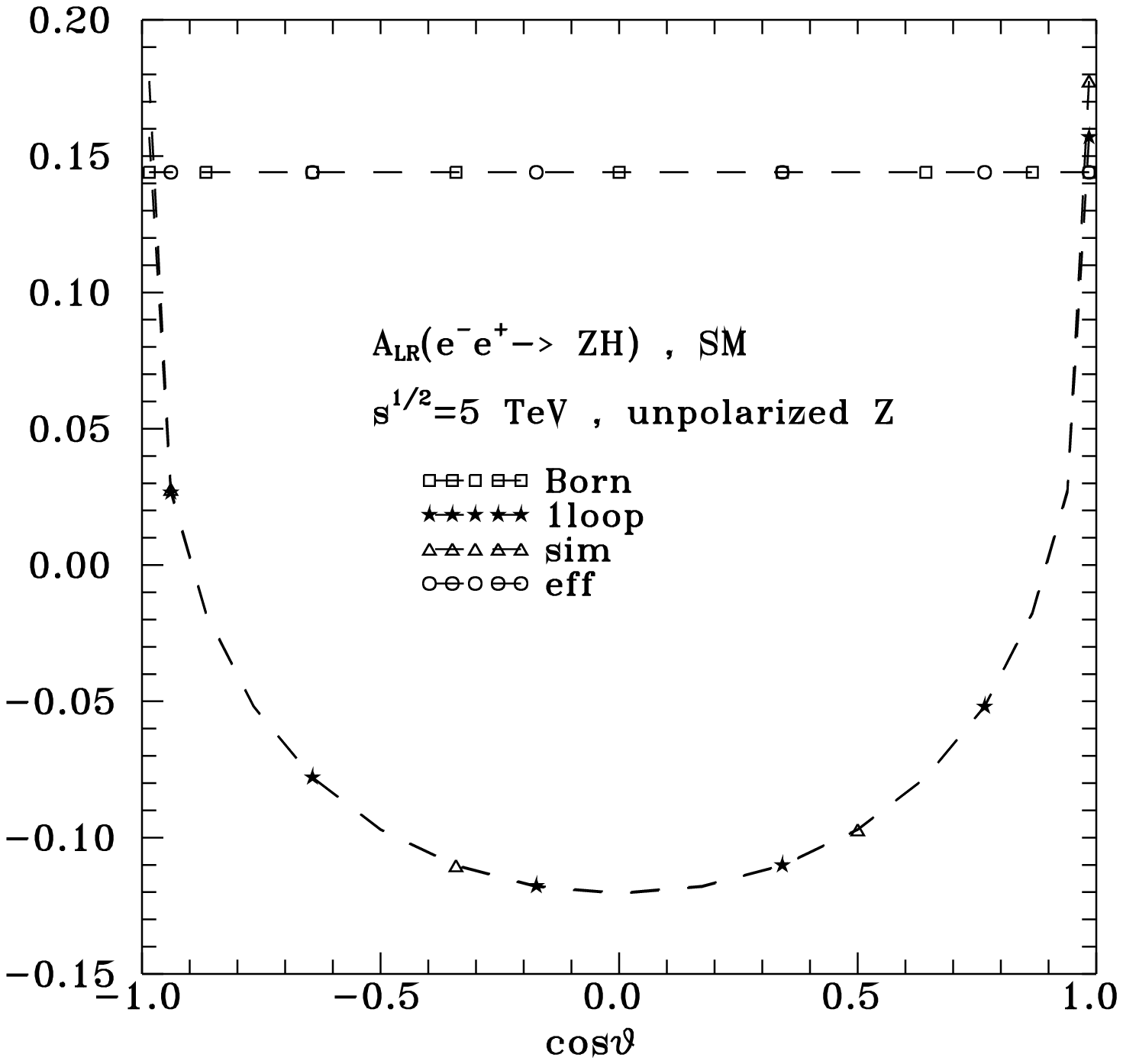, height=6.cm}\hspace{1.cm}
\epsfig{file=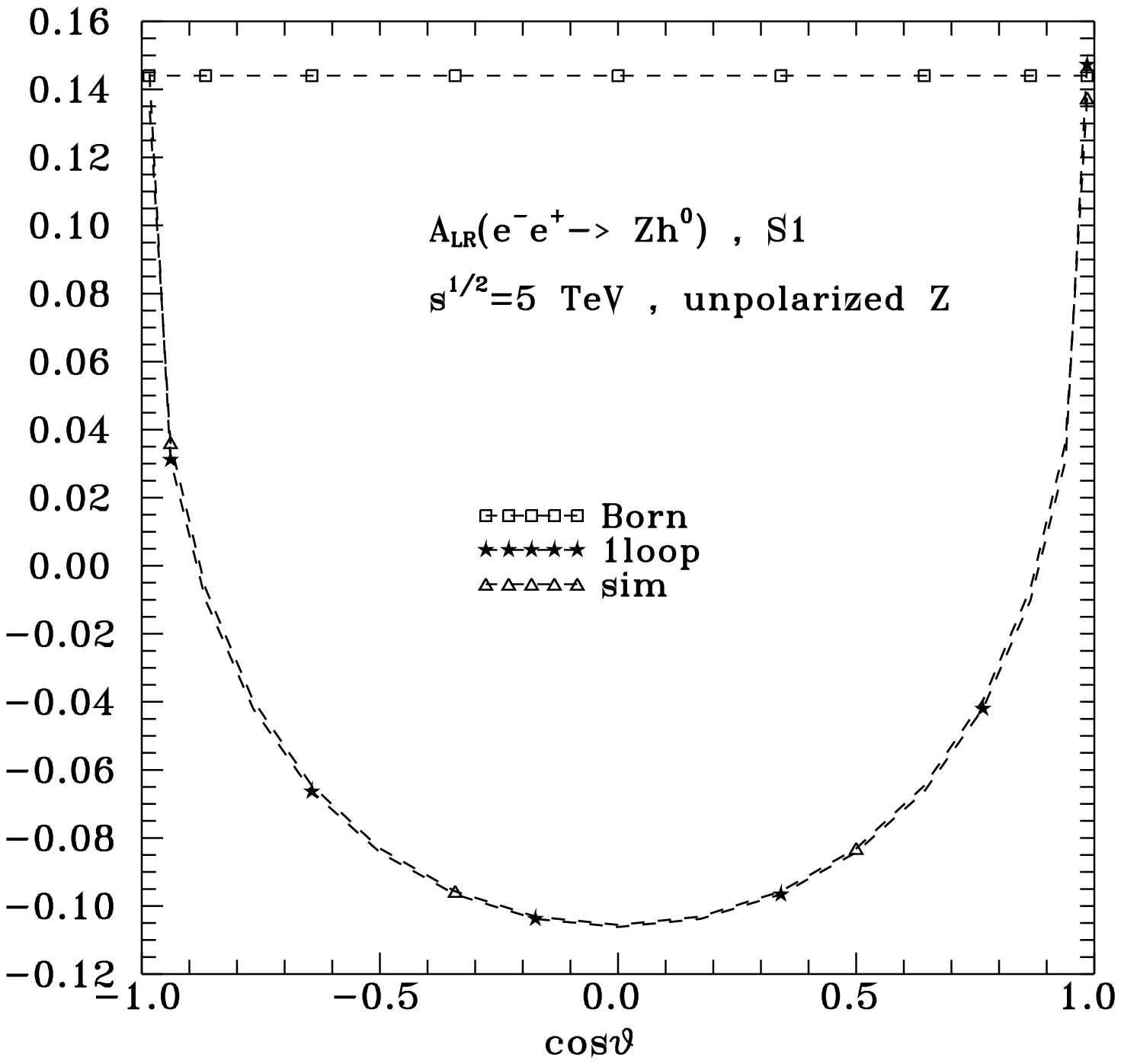,height=6.cm}
\]
\caption[1]{The $A_{LR}$ asymmetries for unpolarized Z, as  defined in (\ref{ALR-unpolZ}).
See caption   in Fig.\ref{sigmas-fig}.}
\label{ALR-unpolarizedZ-fig}
\end{figure}

Fig.\ref{ALR-unpolarizedZ-fig} shows $A_{LR}$, defined in (\ref{ALR-unpolZ})  using polarized
$e^\mp$ beams,  for  $e^-e^+\to  ZH$ in SM (left panels) and  $e^-e^+\to  Z h^0$ in MSSM (right panels), for the cases that  the $Z$ polarization is not looked at. Upper panels describe the energy dependence  at $60^\circ$, while the middle and lower panels show the angular distributions
 at  1 and  5 TeV respectively.

At the Born level,  $A_{LR}^{\rm Born}\simeq 0.14$, with the actual constant value determined by the factor
$[(g^{Z}_{eL})^2-(g^{Z}_{eR})^2]/ [(g^{Z}_{eL})^2+(g^{Z}_{eR})^2]$.
In all cases, the one loop contribution is  quite large, as compared to
Born one, for all energies and angles.
The sim approximation is good at high energy, but becomes worse as the energy  decreases.\\

\begin{figure}[p]
%\vspace{-1cm}
\[
\epsfig{file=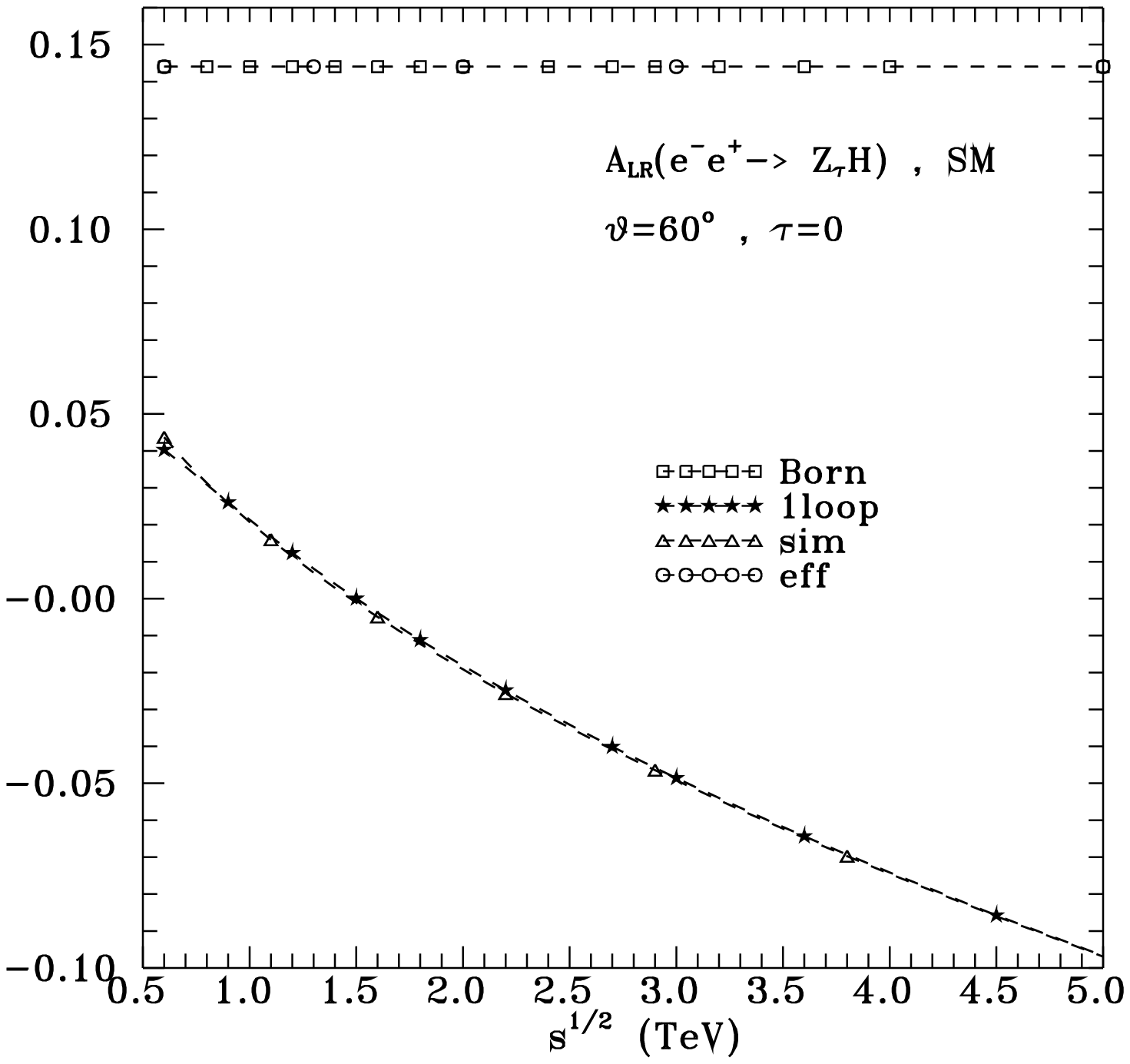, height=6.cm}\hspace{1.cm}
\epsfig{file=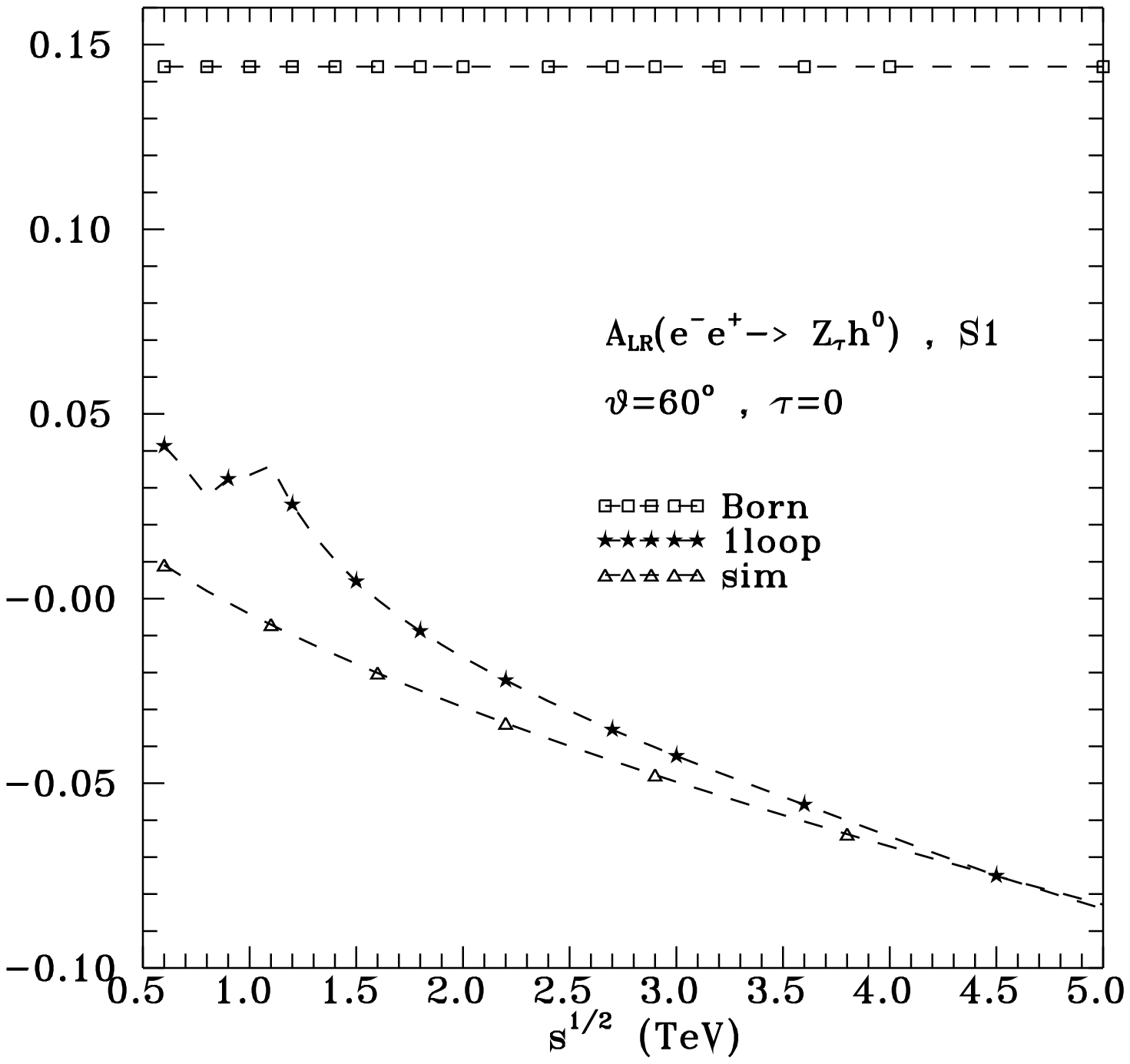,height=6.cm}
\]
\[
\epsfig{file=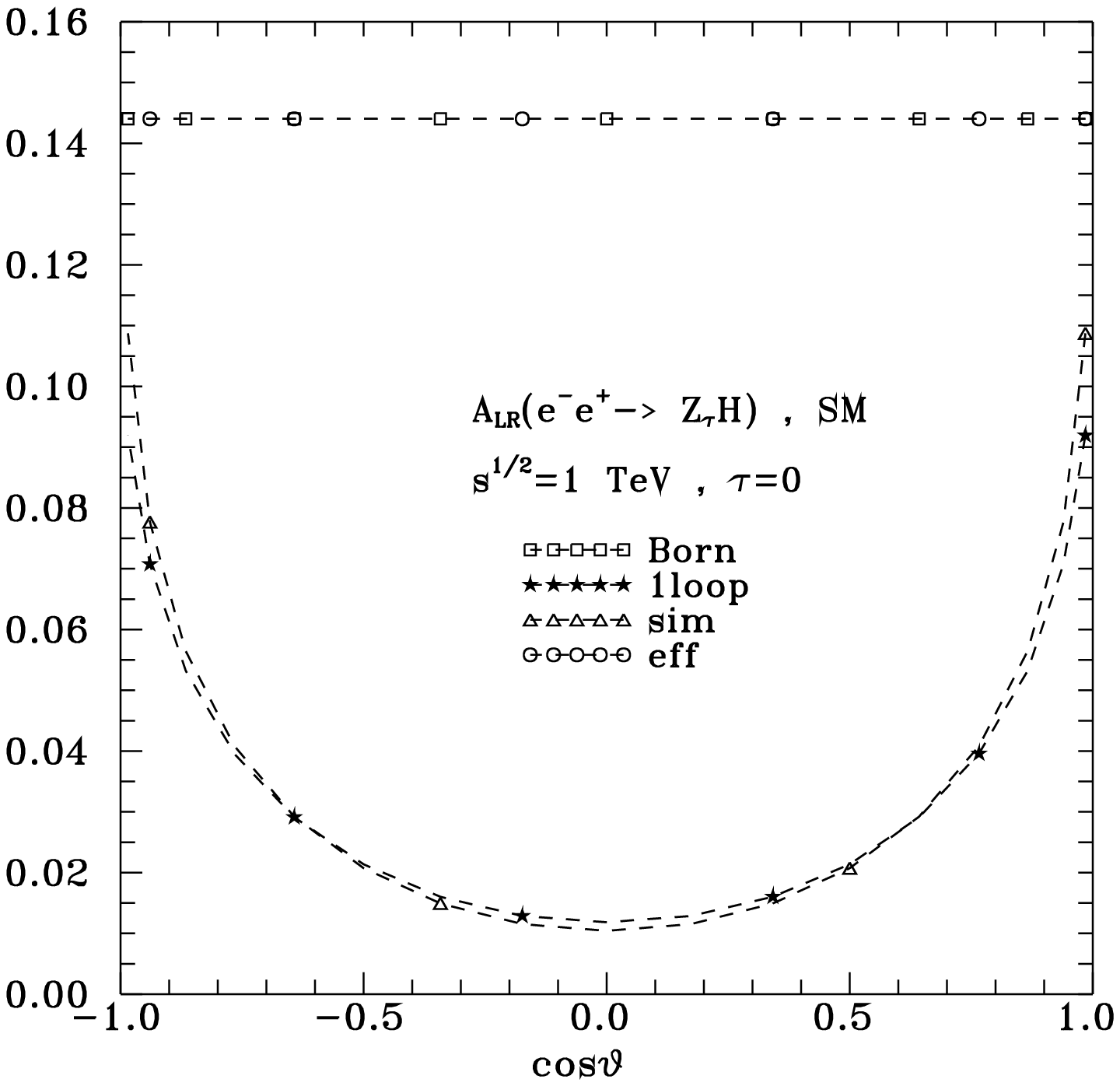, height=6.cm}\hspace{1.cm}
\epsfig{file=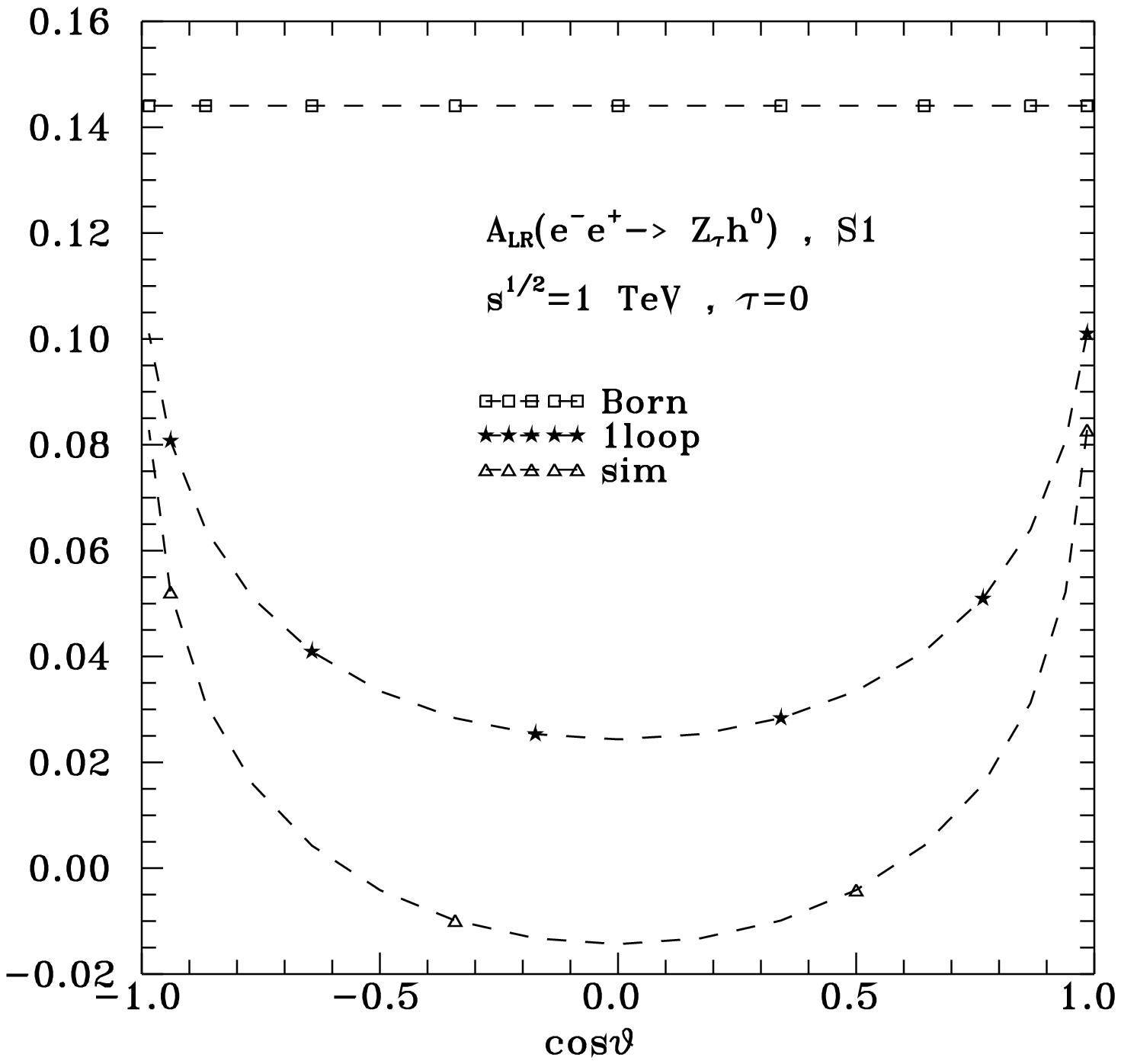,height=6.cm}
\]
\[
\epsfig{file=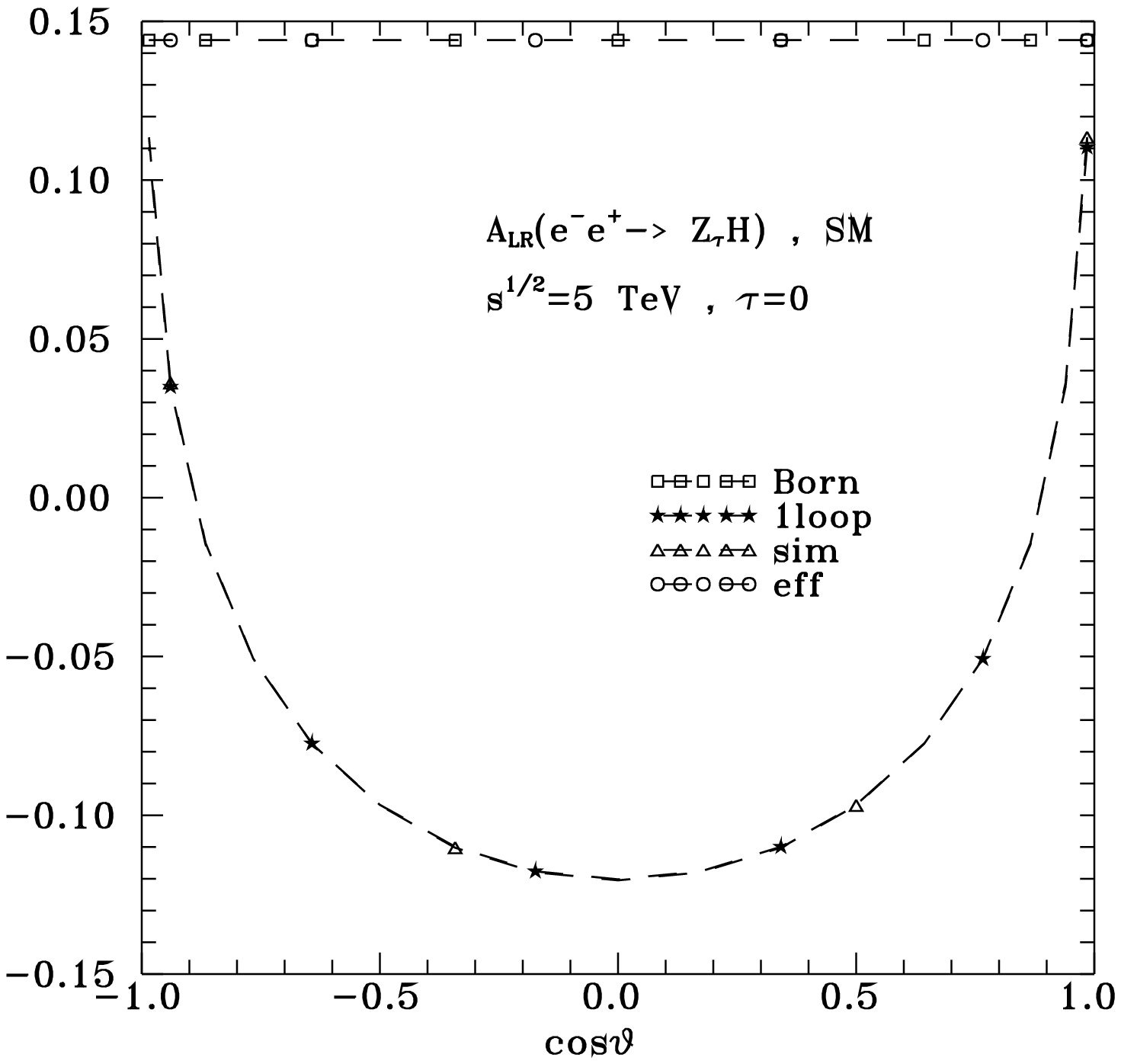, height=6.cm}\hspace{1.cm}
\epsfig{file=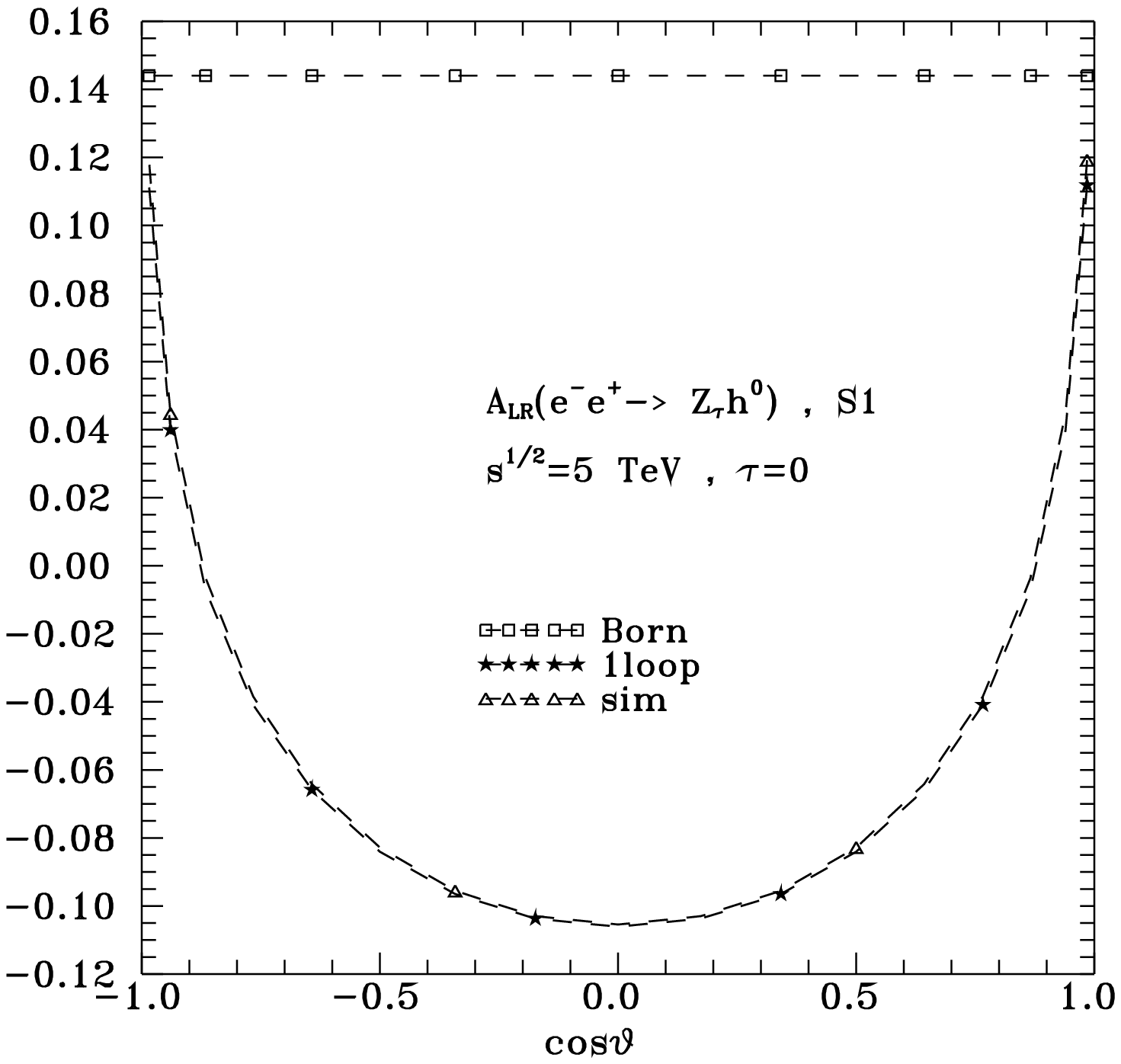,height=6.cm}
\]
\caption[1]{ The $A_{LR}$ asymmetries defined in (\ref{ALR-tau}), for a final $Z$ of helicity $\tau=0$.
See caption in  Fig.\ref{sigmas-fig}.}
\label{C2-tau0-fig}
\end{figure}

\begin{figure}[p]
%\vspace{-1cm}
\[
\epsfig{file=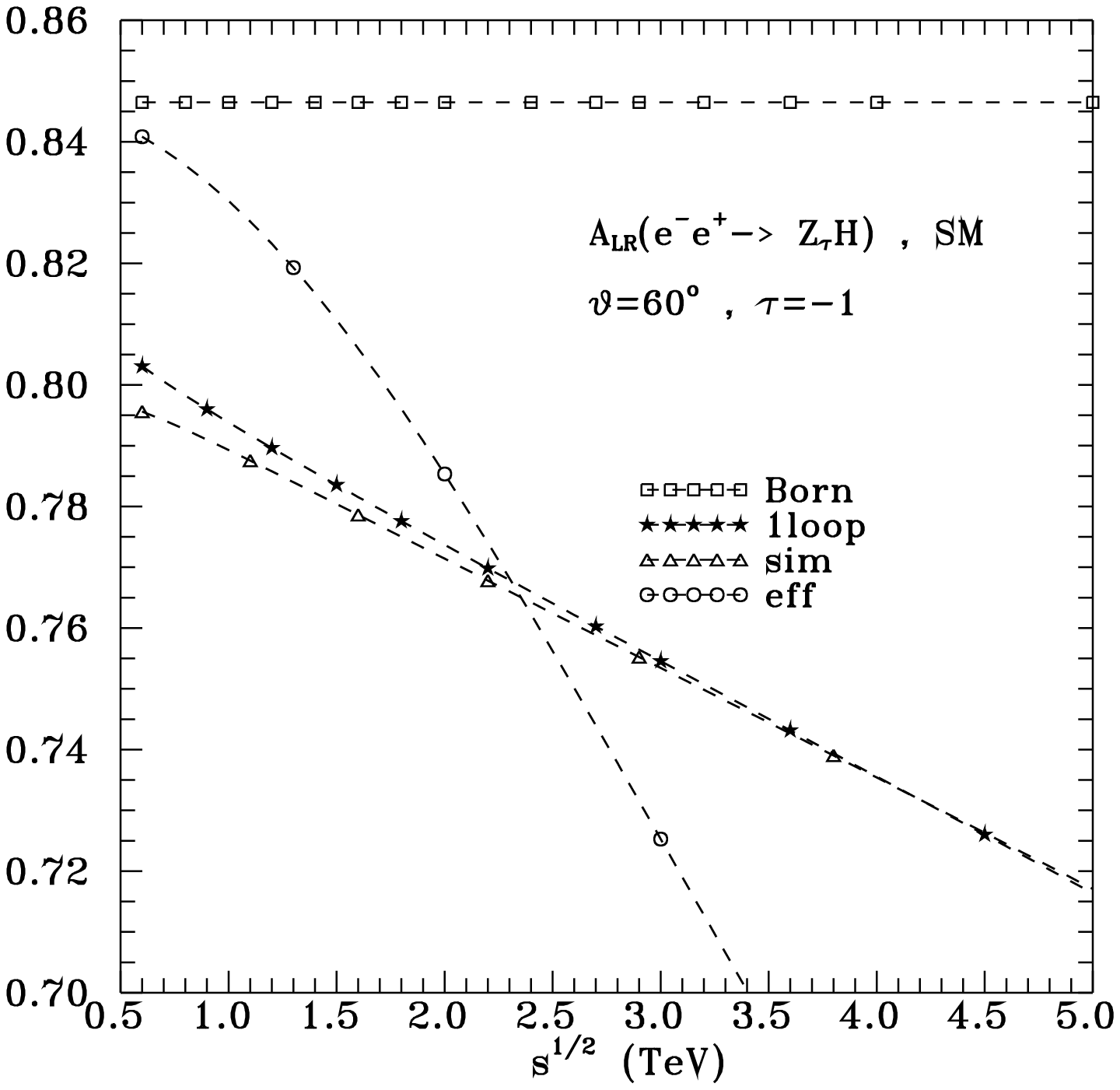, height=6.cm}\hspace{1.cm}
\epsfig{file=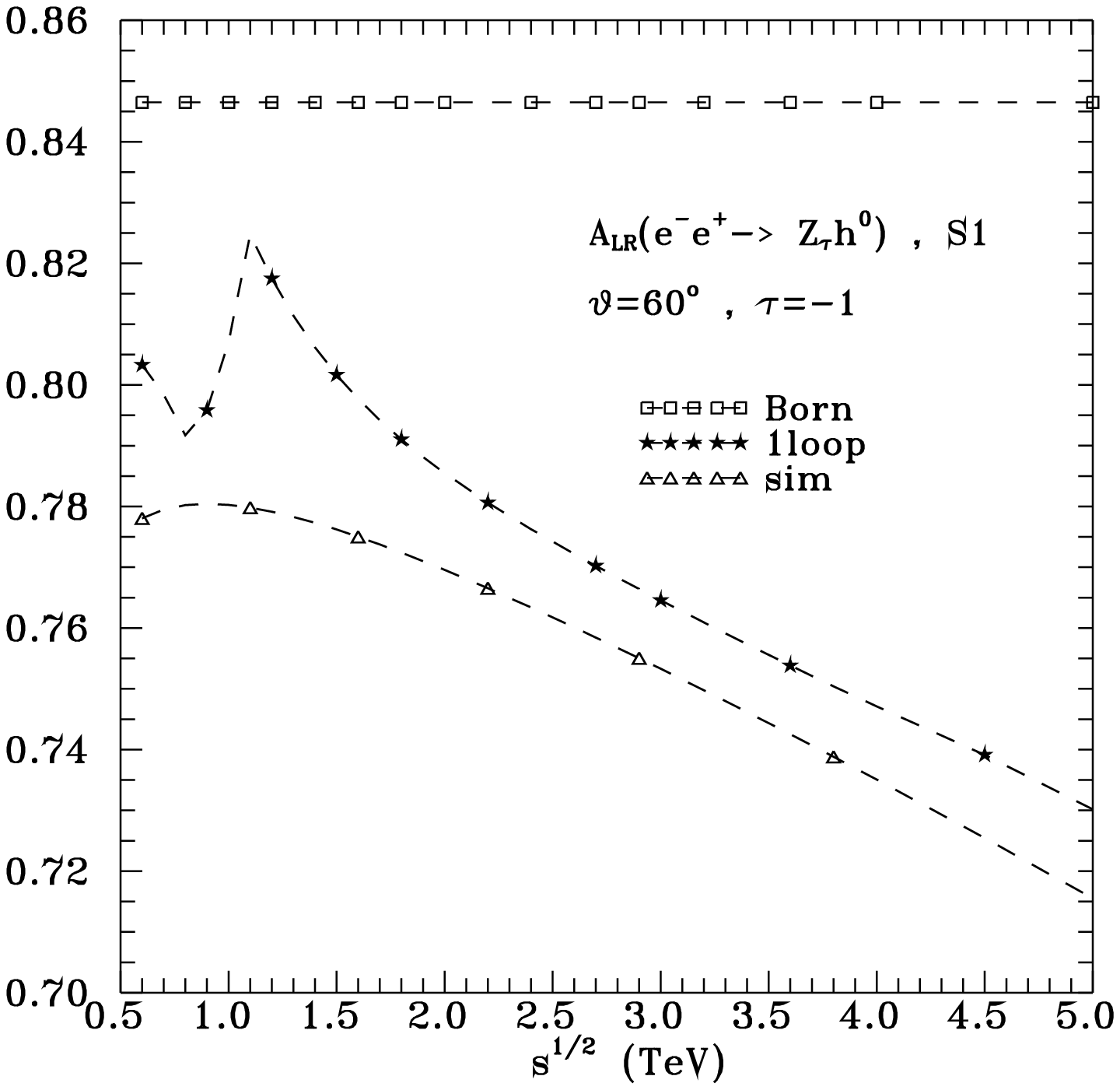,height=6.cm}
\]
\[
\epsfig{file=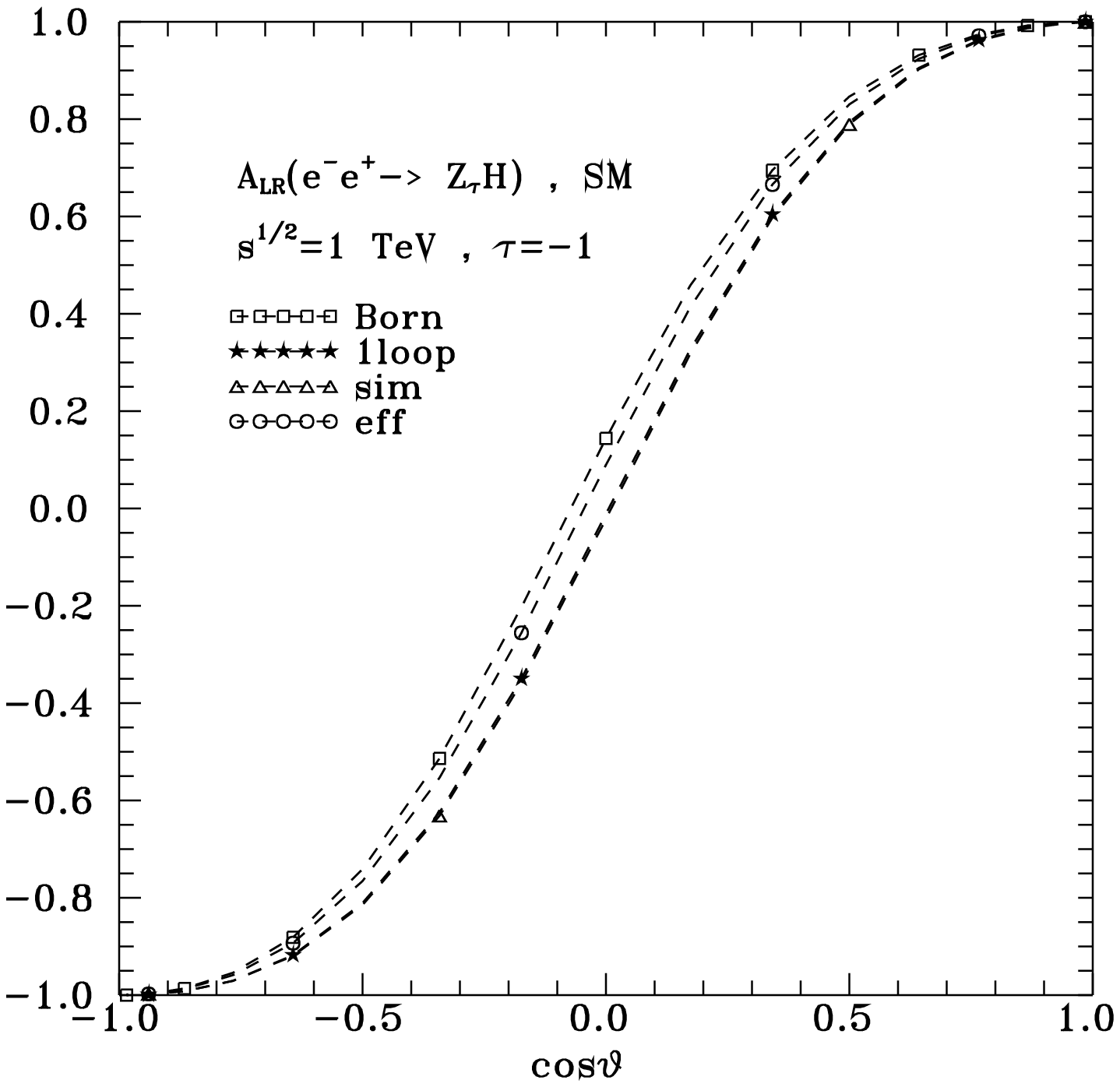, height=6.cm}\hspace{1.cm}
\epsfig{file=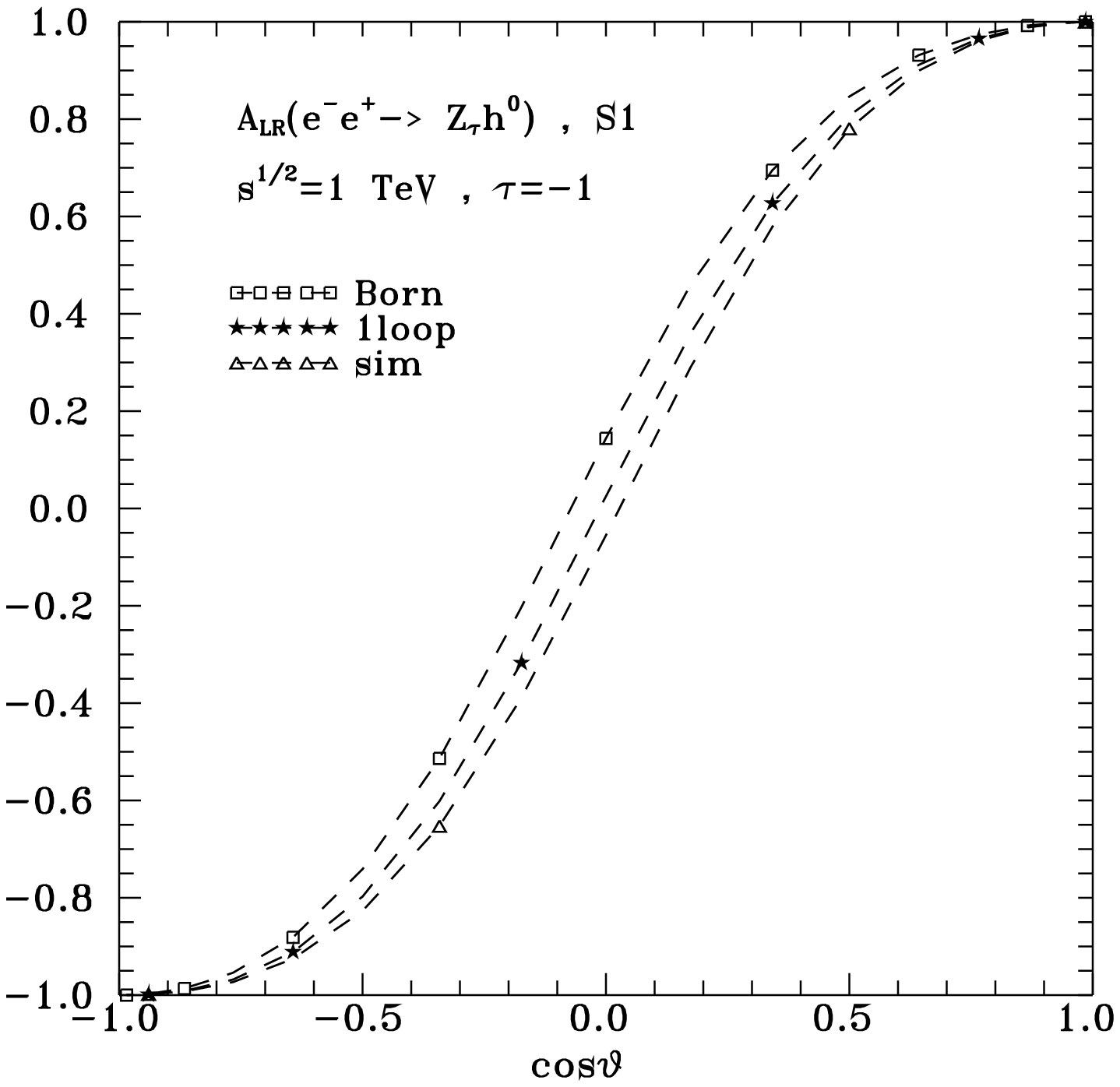,height=6.cm}
\]
\[
\epsfig{file=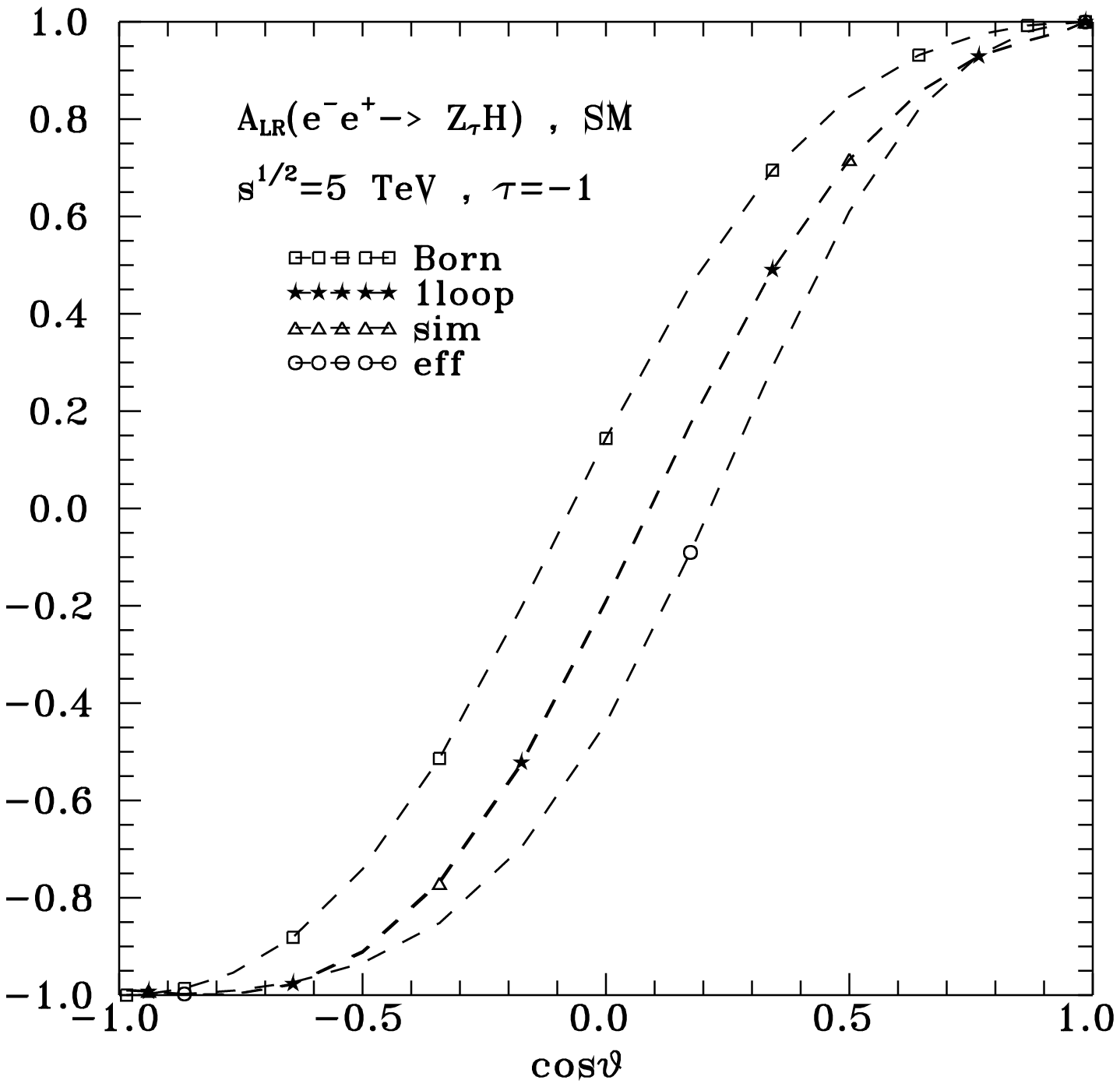, height=6.cm}\hspace{1.cm}
\epsfig{file=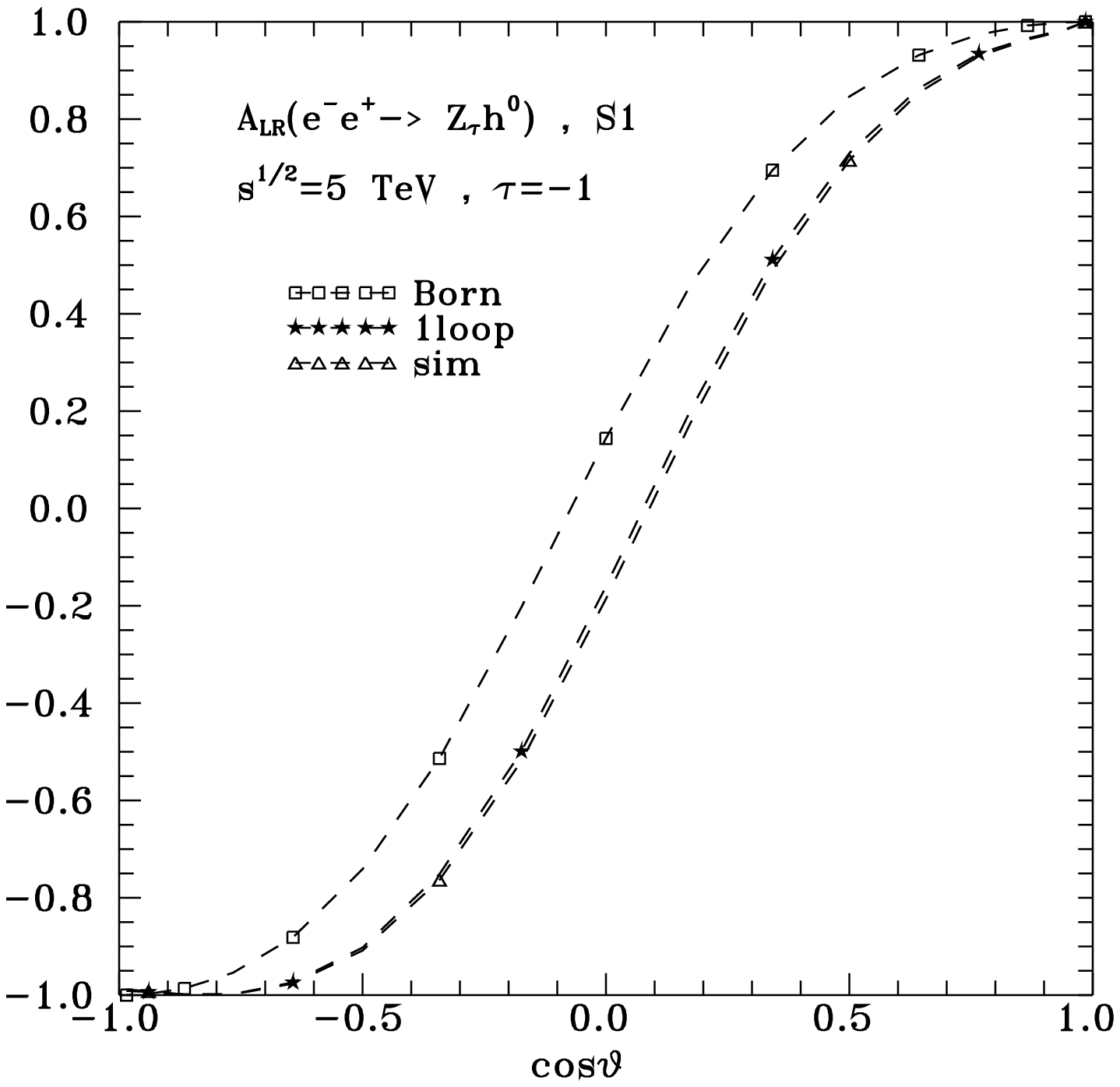,height=6.cm}
\]
\caption[1]{ The $A_{LR}$ asymmetries defined in (\ref{ALR-tau}), for a final $Z$ of helicity $\tau=-1$.
See caption in Fig.\ref{sigmas-fig}}
\label{C2-taum-fig}
\end{figure}

\begin{figure}[p]
%\vspace{-1cm}
\[
\epsfig{file=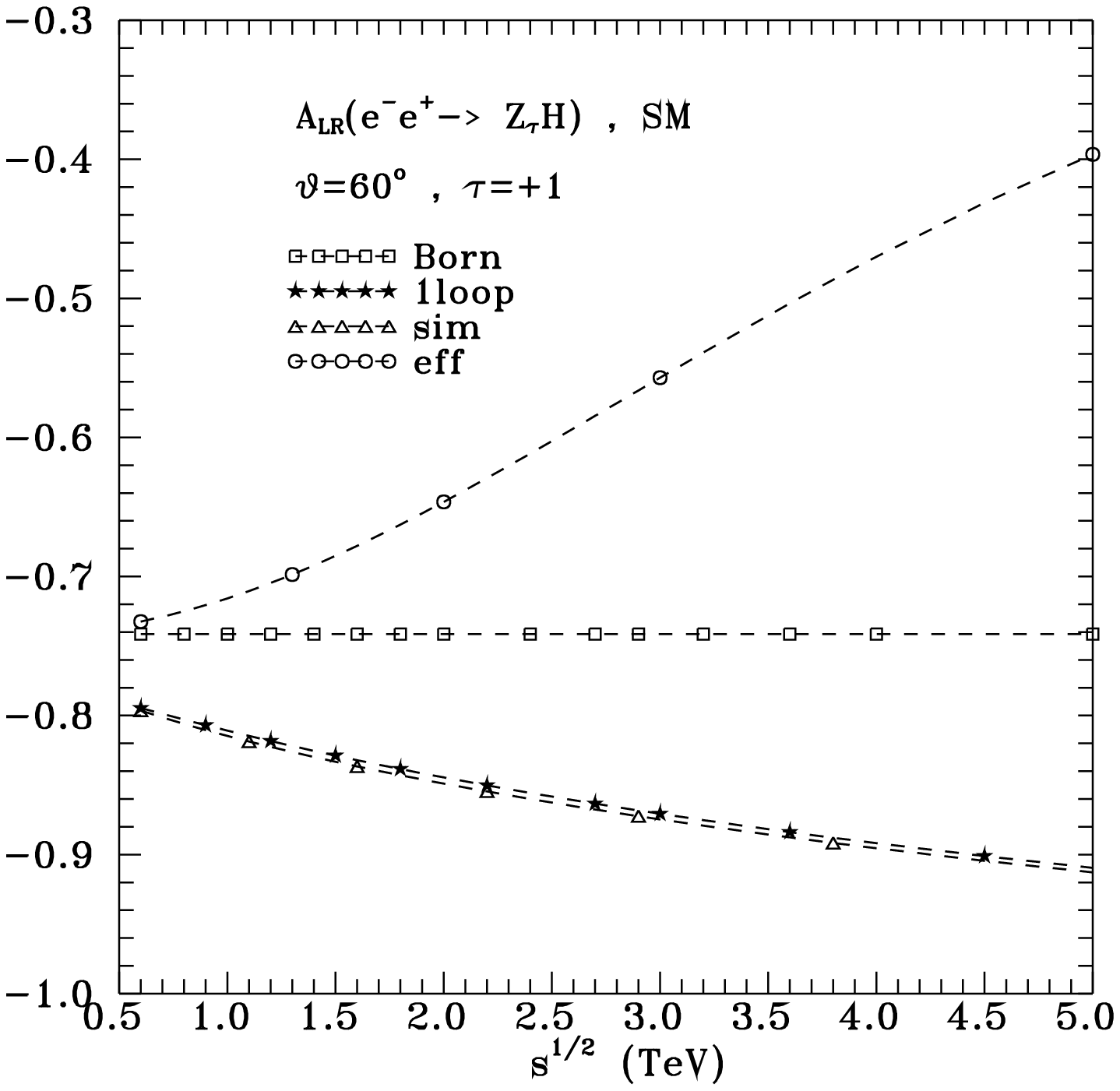, height=6.cm}\hspace{1.cm}
\epsfig{file=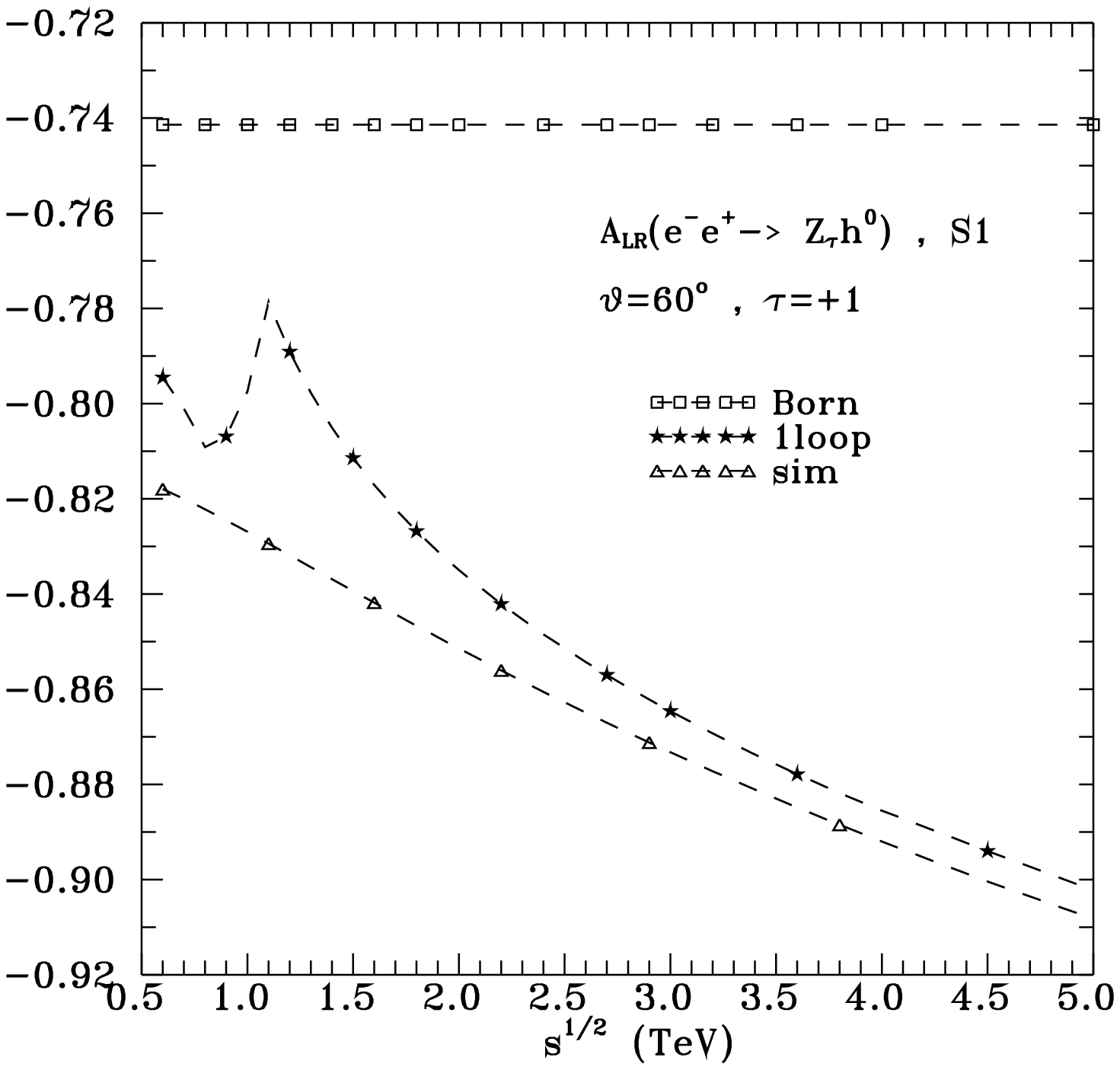,height=6.cm}
\]
\[
\epsfig{file=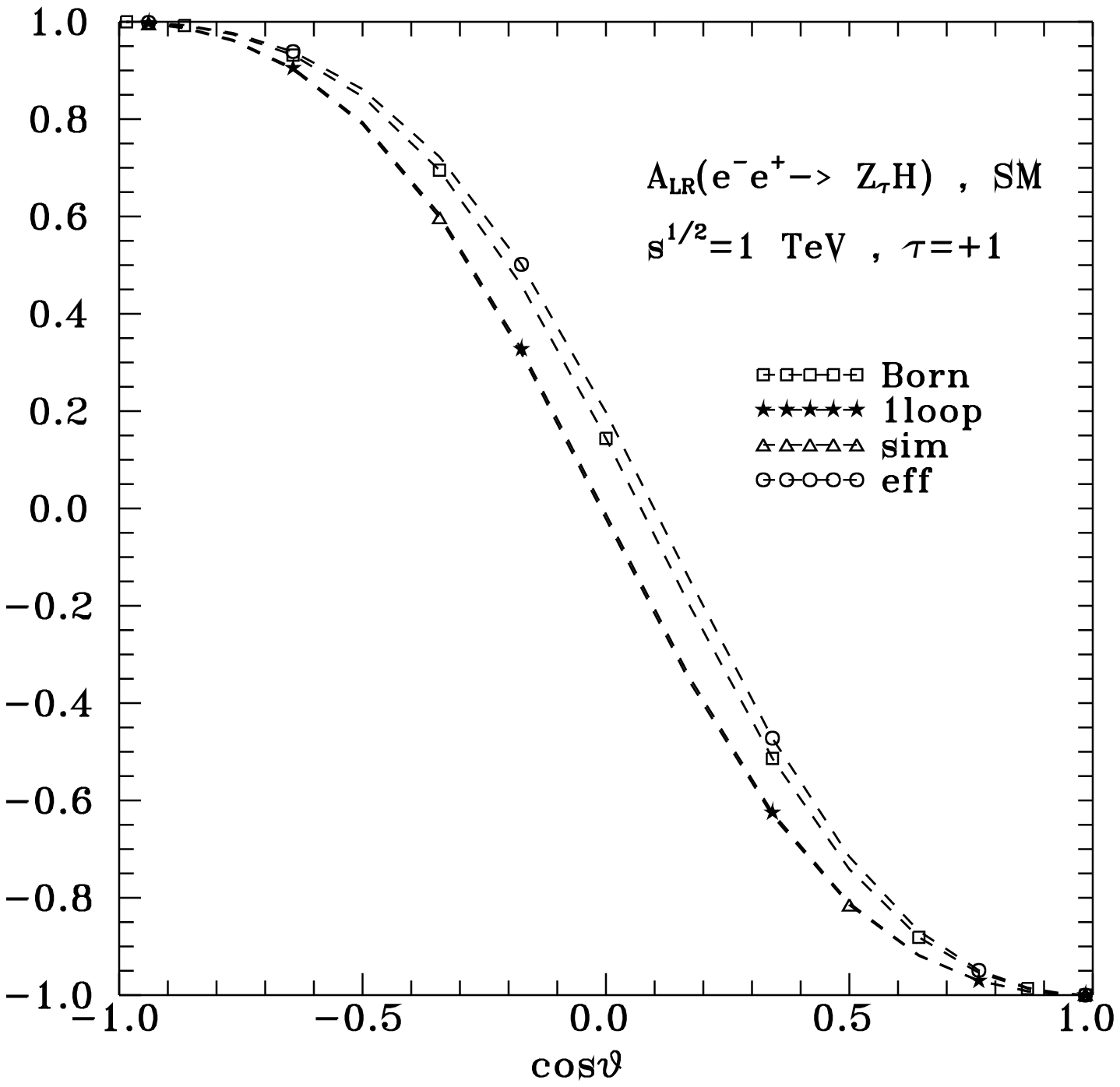, height=6.cm}\hspace{1.cm}
\epsfig{file=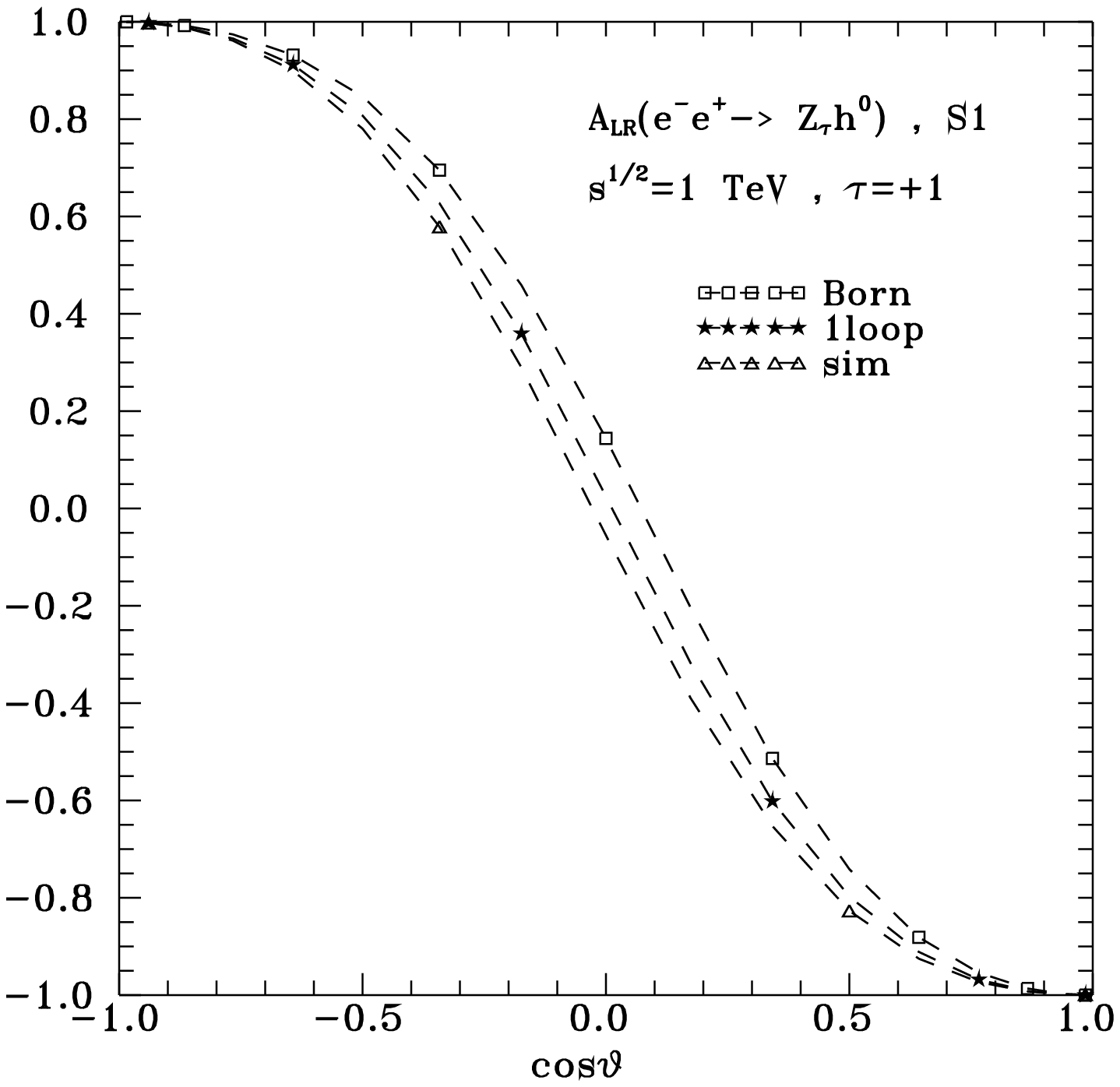,height=6.cm}
\]
\[
\epsfig{file=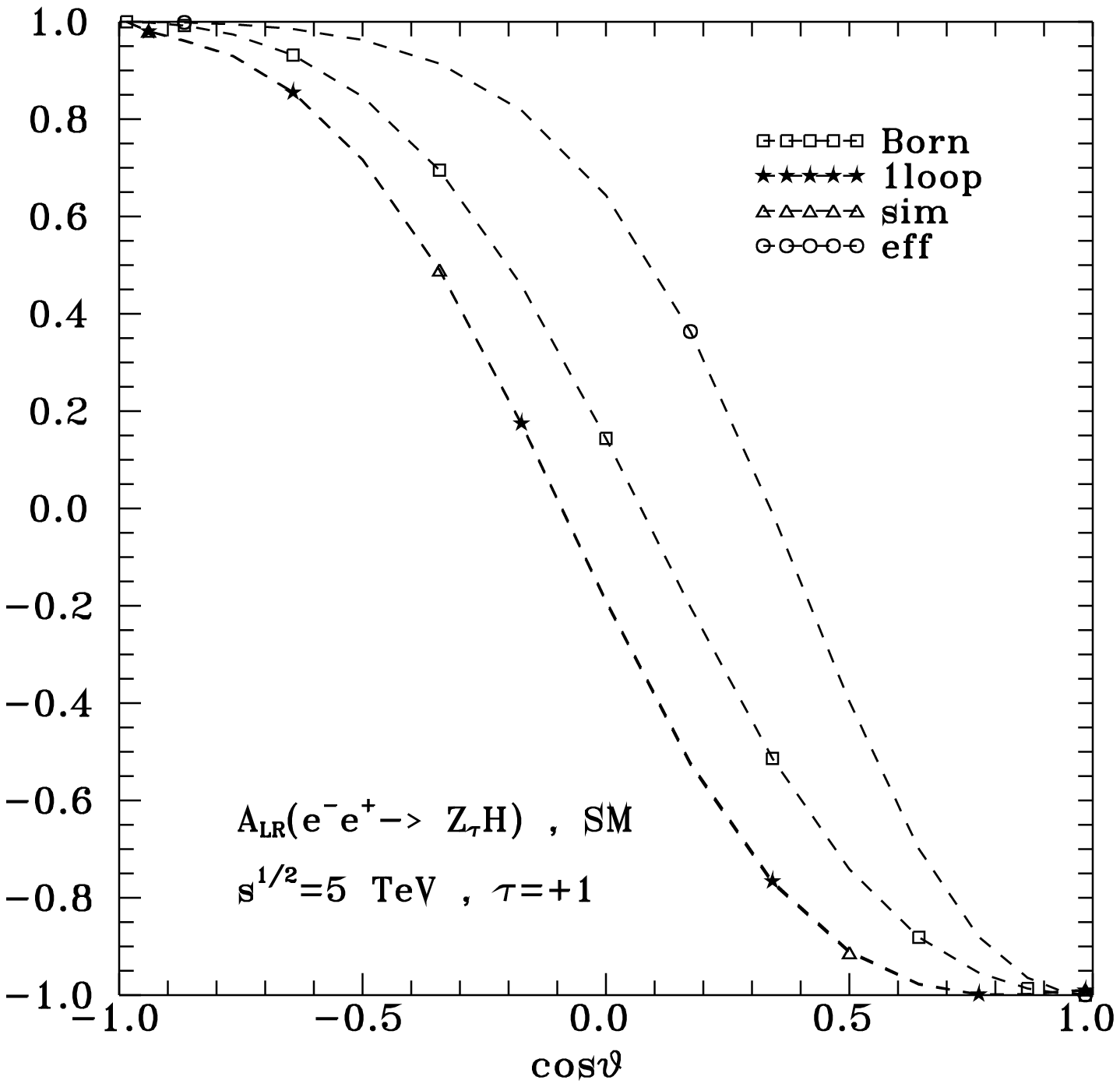, height=6.cm}\hspace{1.cm}
\epsfig{file=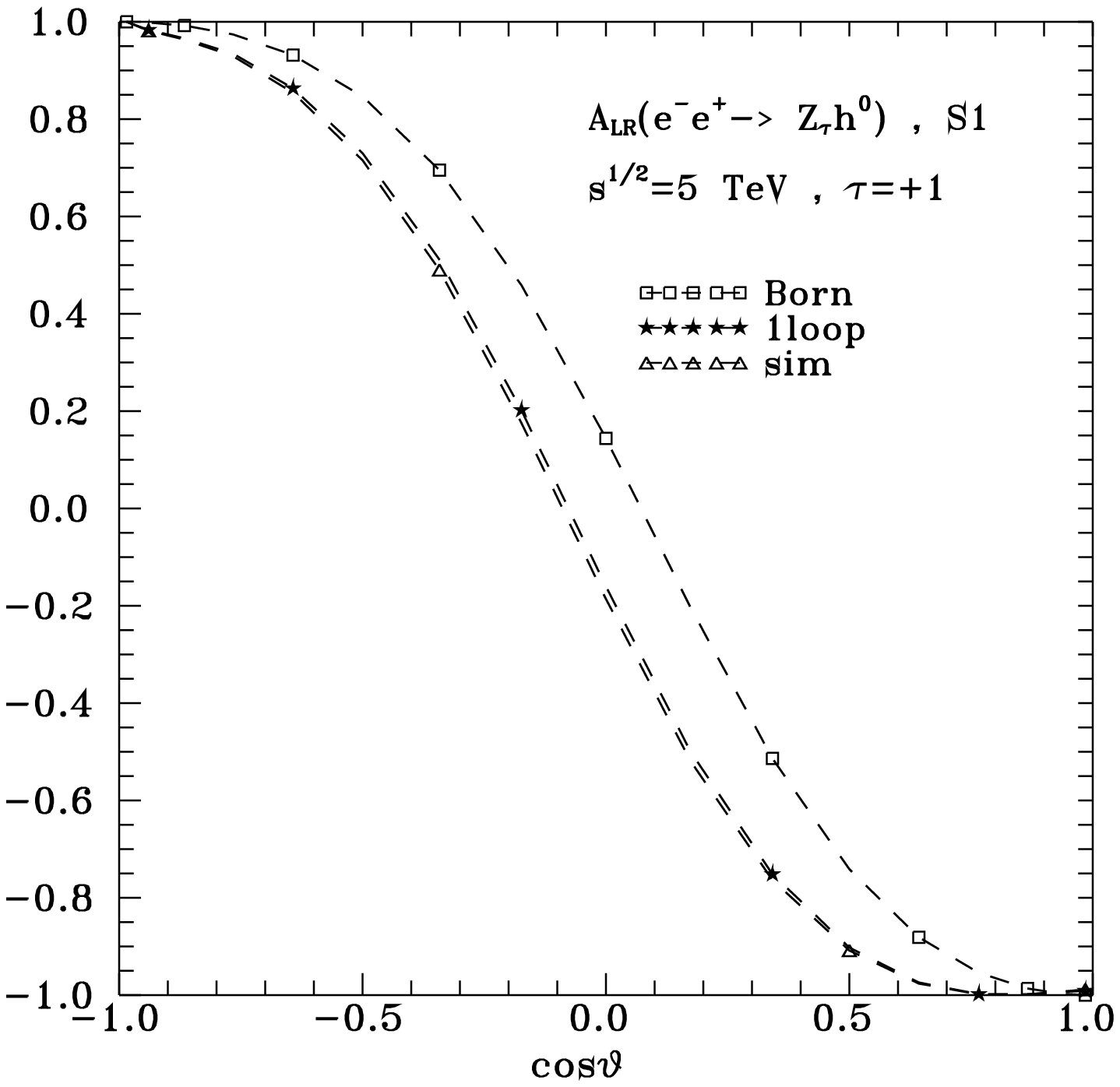,height=6.cm}
\]
\caption[1]{ The $A_{LR}$ asymmetries defined in (\ref{ALR-tau}), for a final $Z$ of helicity $\tau=1$.
See caption  in Fig.\ref{sigmas-fig}.}
\label{C2-taup-fig}
\end{figure}

We next turn to  $A_{LR}$ defined in (\ref{ALR-tau}),
where, in addition to using polarized $e^\mp$-beams, the  $Z$ helicity is also measured. Figs.\ref{C2-tau0-fig},\ref{C2-taum-fig},\ref{C2-taup-fig} describe
 $e^-e^+\to  Z H$ in SM (left panels) and $e^-e^+\to  Z h^0$ in MSSM (right panels), for
  Z-helicities $\tau=0, -1, +1$ respectively.
The upper panels give the  energy dependencies at $60^0$, while the middle and lower panels give
  the angular dependencies  at 1 and  5 TeV respectively. Comparing the complete one-loop results with the Born contributions, we realize that the one-loop corrections are important.

In the $\tau=0$ case shown in Fig.\ref{C2-tau0-fig}, the Born value of $A_{LR}$  is again $\simeq 0.14$, and  strongly affected by one loop corrections. In agreement with the CP rule (\ref{ALRFB-tau}), the angular dependence is forward-backward symmetric.

In the $\tau=\mp 1$ cases shown in Figs.\ref{C2-taum-fig} and \ref{C2-taup-fig}, the Born contributions  appear almost forward-backward antisymmetric; see middle and lower panels.
This is due to the terms $(1\pm\cos\theta)$ in (\ref{FBorn-T}) and to their coefficients $g^{Z}_{eL}$ and $g^{Z}_{eR}$ being  almost opposite.
We also note that for  CP invariant interactions  $A_{LR}(Z^+)$ and $A_{LR}(Z^-)$ are Forward-Backward
symmetric to each other, see (\ref{ALRFB-tau}).

When the $Z$-polarization is not looked at, all these $\tau=0$ and  $\tau=\mp 1$ effects of course disappear, and the global $A_{LR}$  appears forward-backward symmetric; see Fig.\ref{ALR-unpolarizedZ-fig}.

These implications of the different $A_{LR}$ measurements would produce interesting tests of the nature
of the various contributions. The accuracy of sim is again good for high energies above 1 TeV.

\subsection{$A^{{\rm pol}~Z}$ and $A^{{\rm pol}~Z}(\lambda)$ asymmetries.}

\begin{figure}[p]
%\vspace{-1cm}
\[
\epsfig{file=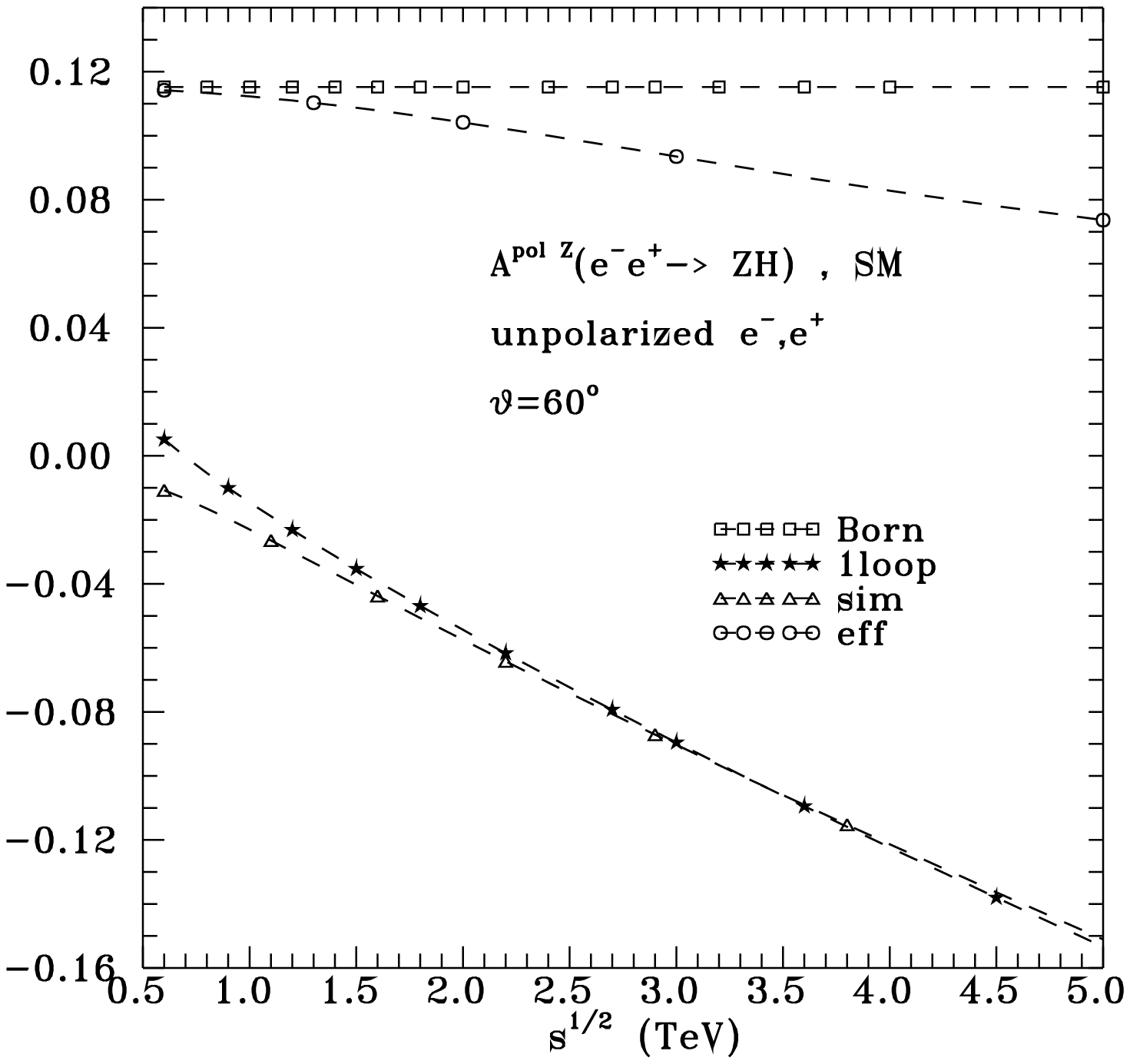, height=6.cm}\hspace{1.cm}
\epsfig{file=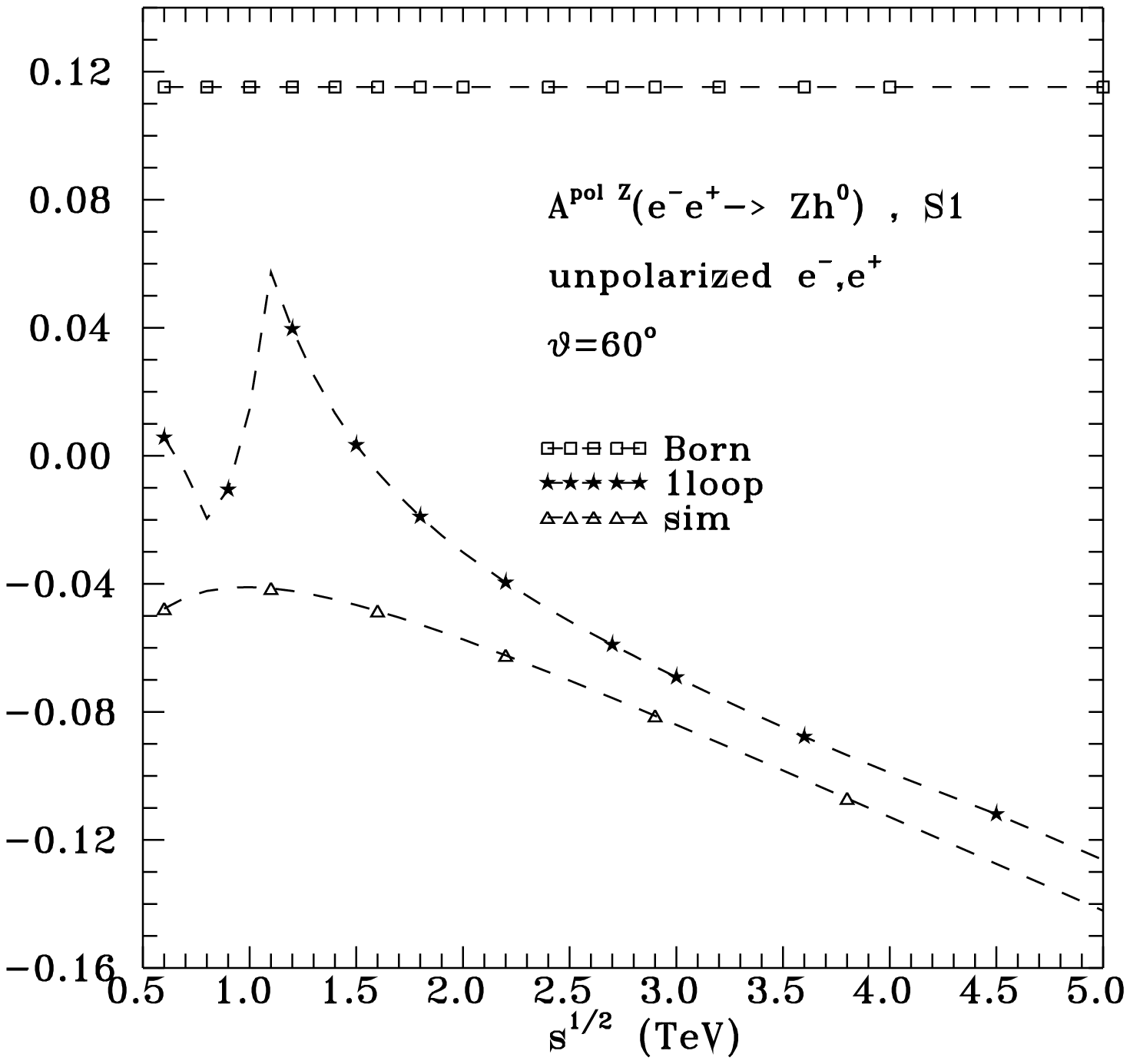,height=6.cm}
\]
\[
\epsfig{file=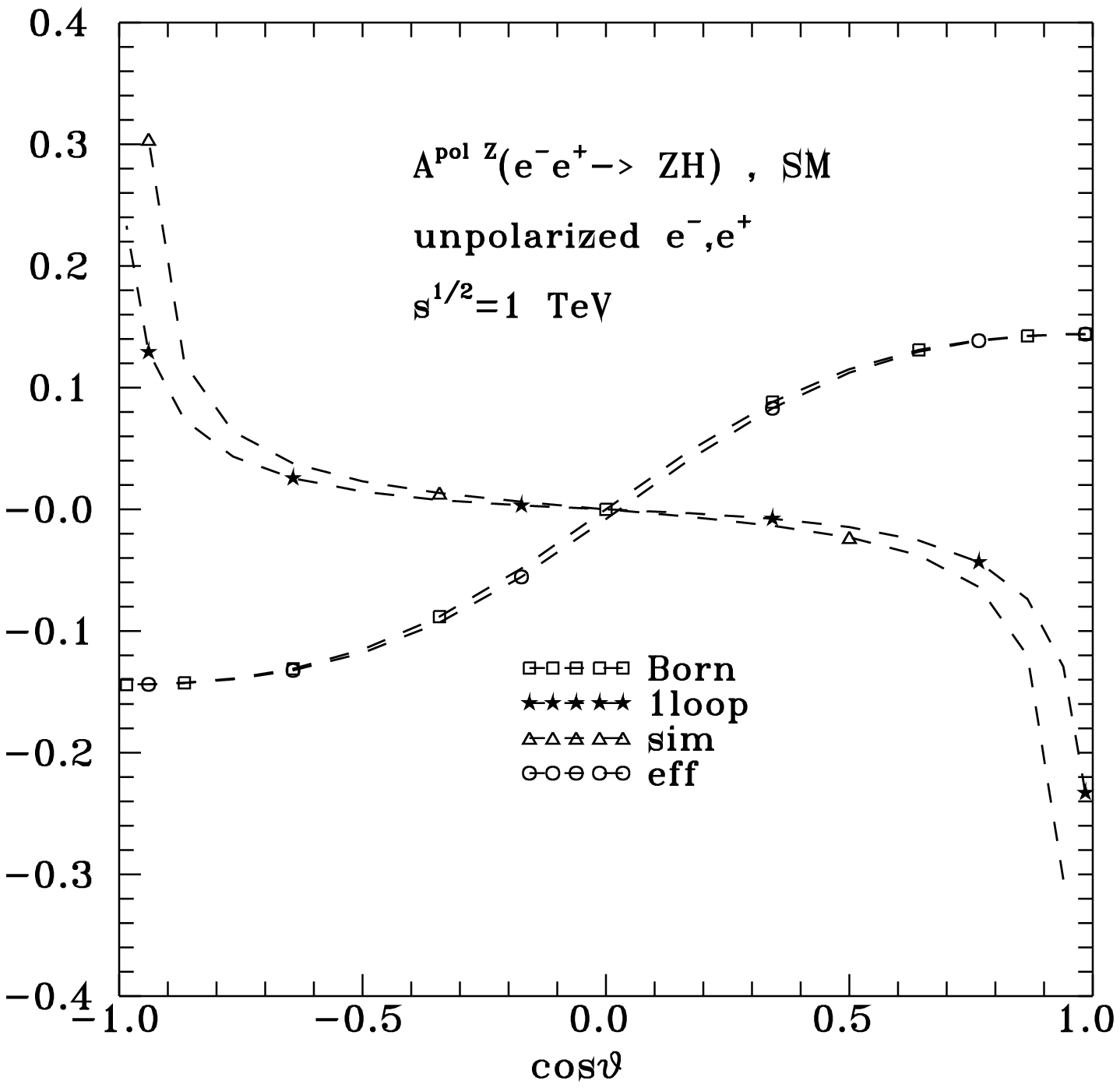, height=6.cm}\hspace{1.cm}
\epsfig{file=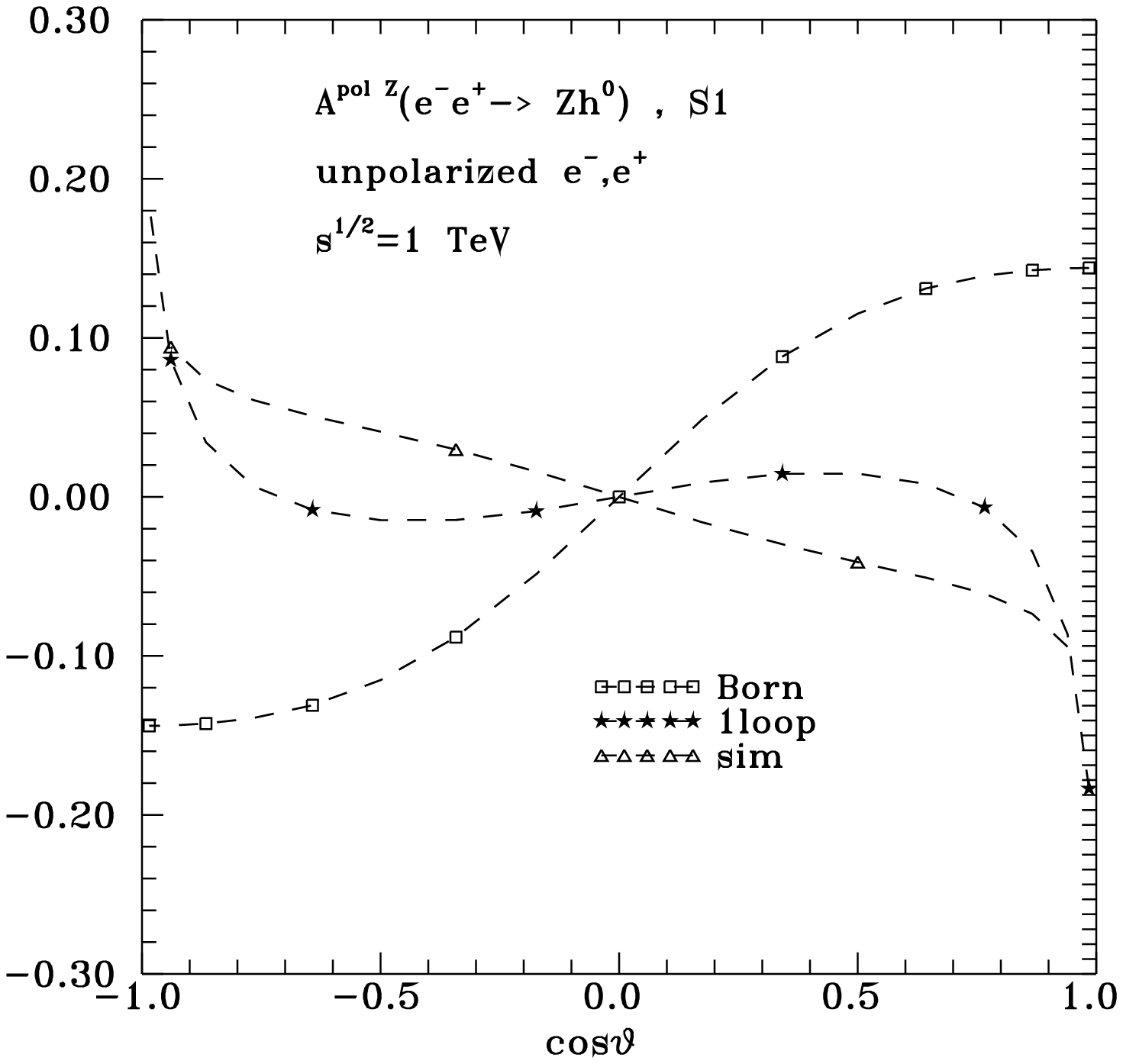,height=6.cm}
\]
\[
\epsfig{file=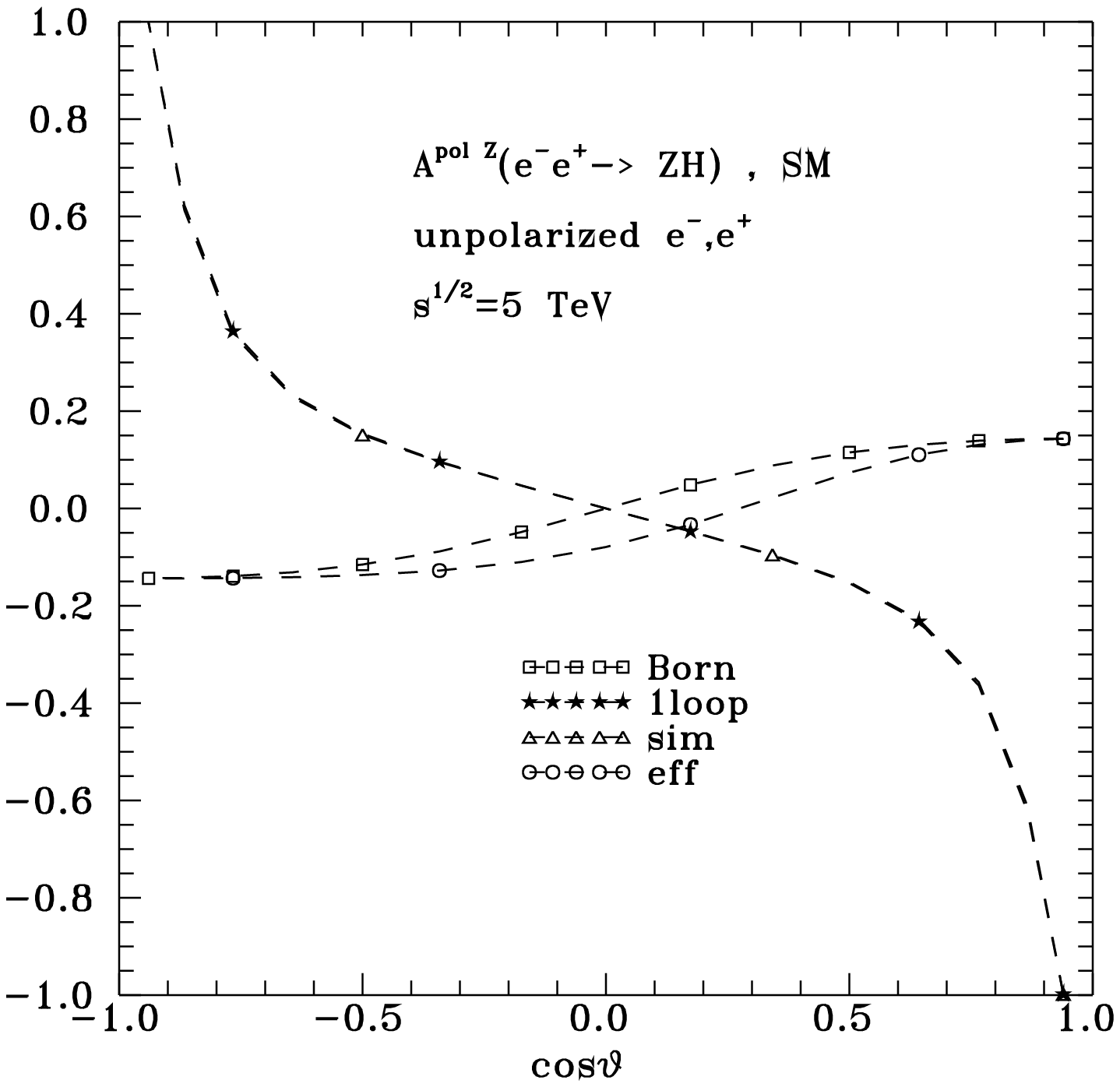, height=6.cm}\hspace{1.cm}
\epsfig{file=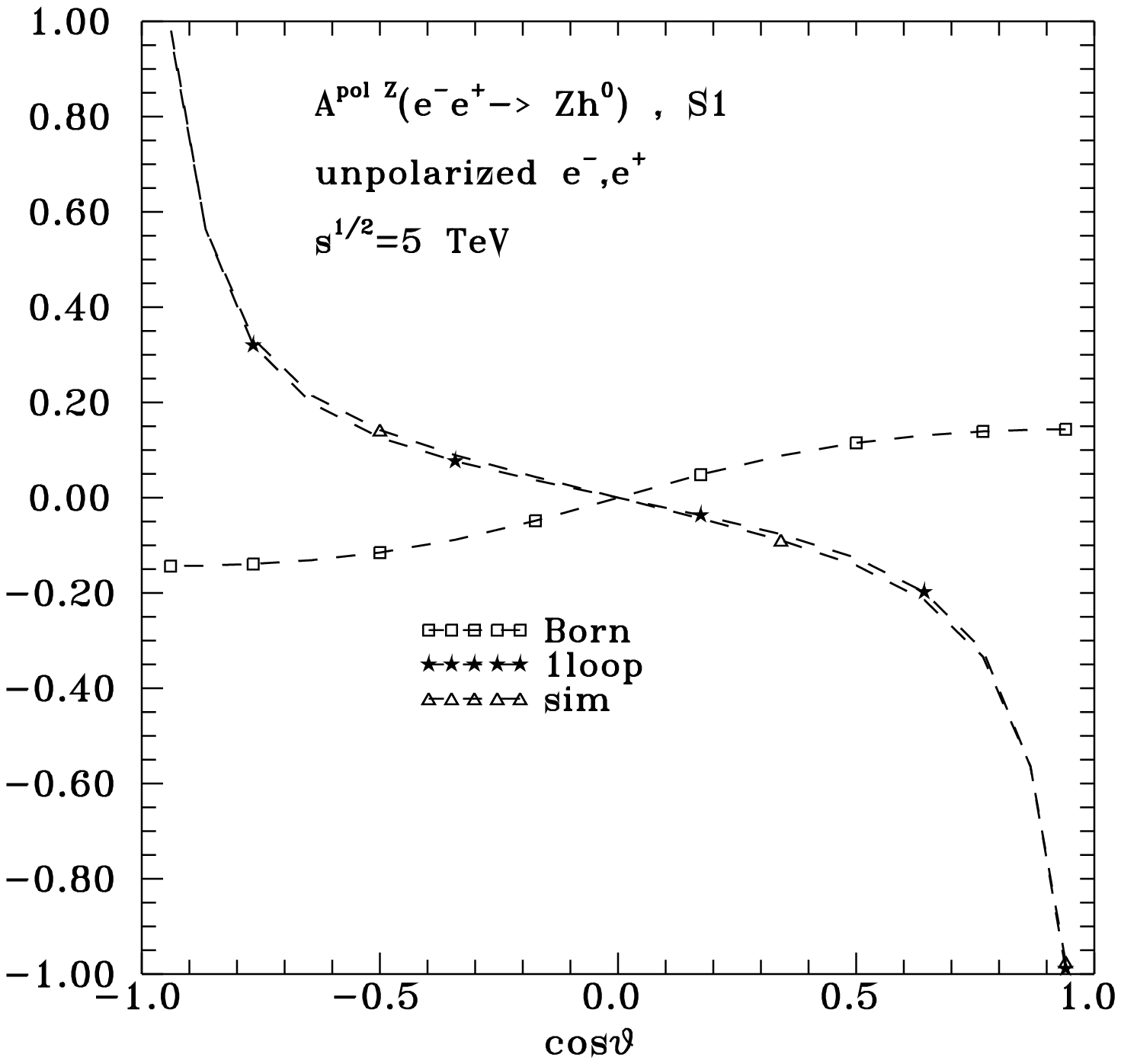,height=6.cm}
\]
\caption[1]{$A^{{\rm pol}~ Z}$ asymmetries defined in (\ref{A-polZ}),  for unpolarized $(e^-,~ e^+)$-beams. See caption  in Fig.\ref{sigmas-fig}. }
\label{C3-fig}
\end{figure}

\begin{figure}[p]
%\vspace{-1cm}
\[
\epsfig{file=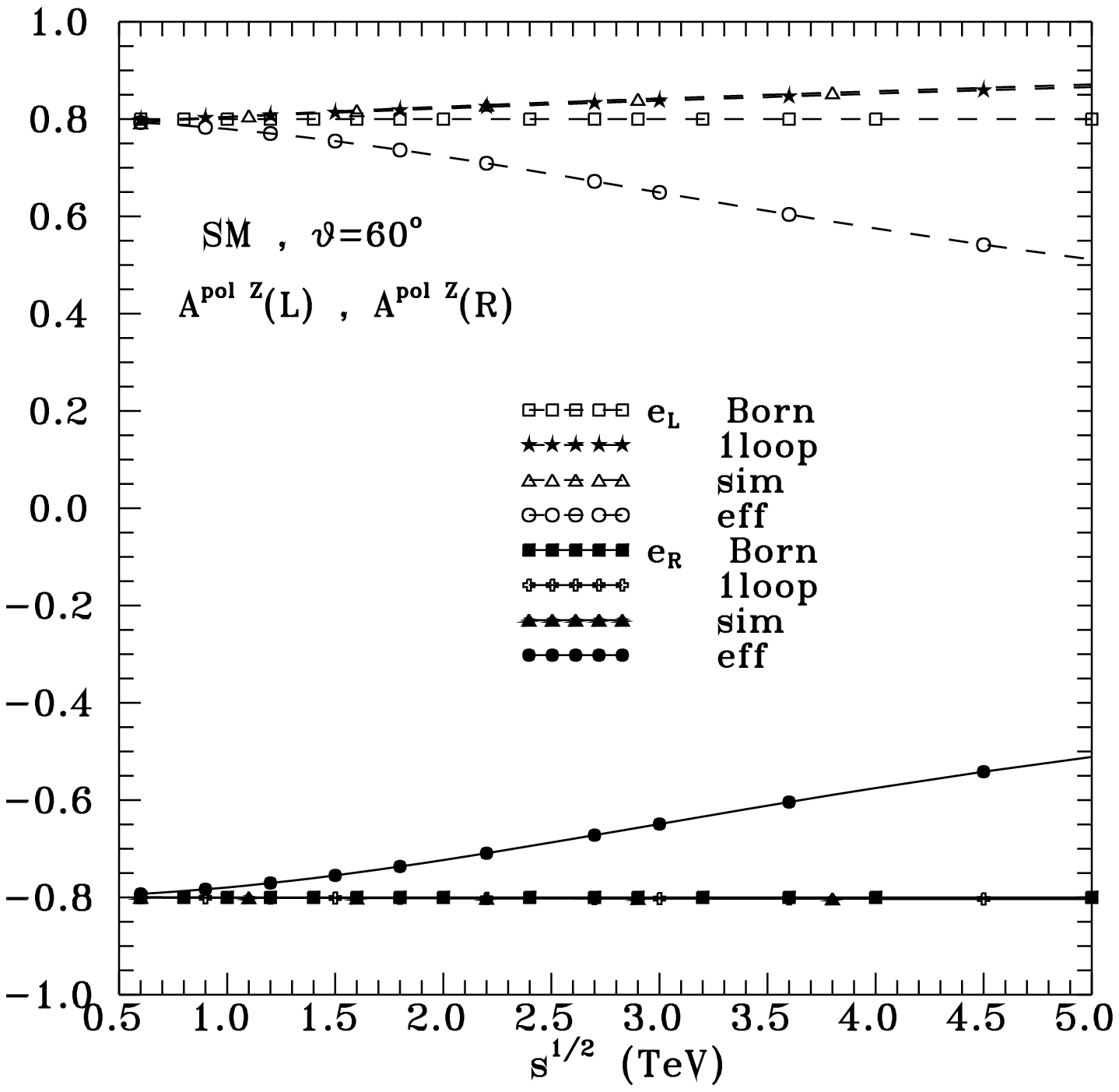, height=6.cm}\hspace{1.cm}
\epsfig{file=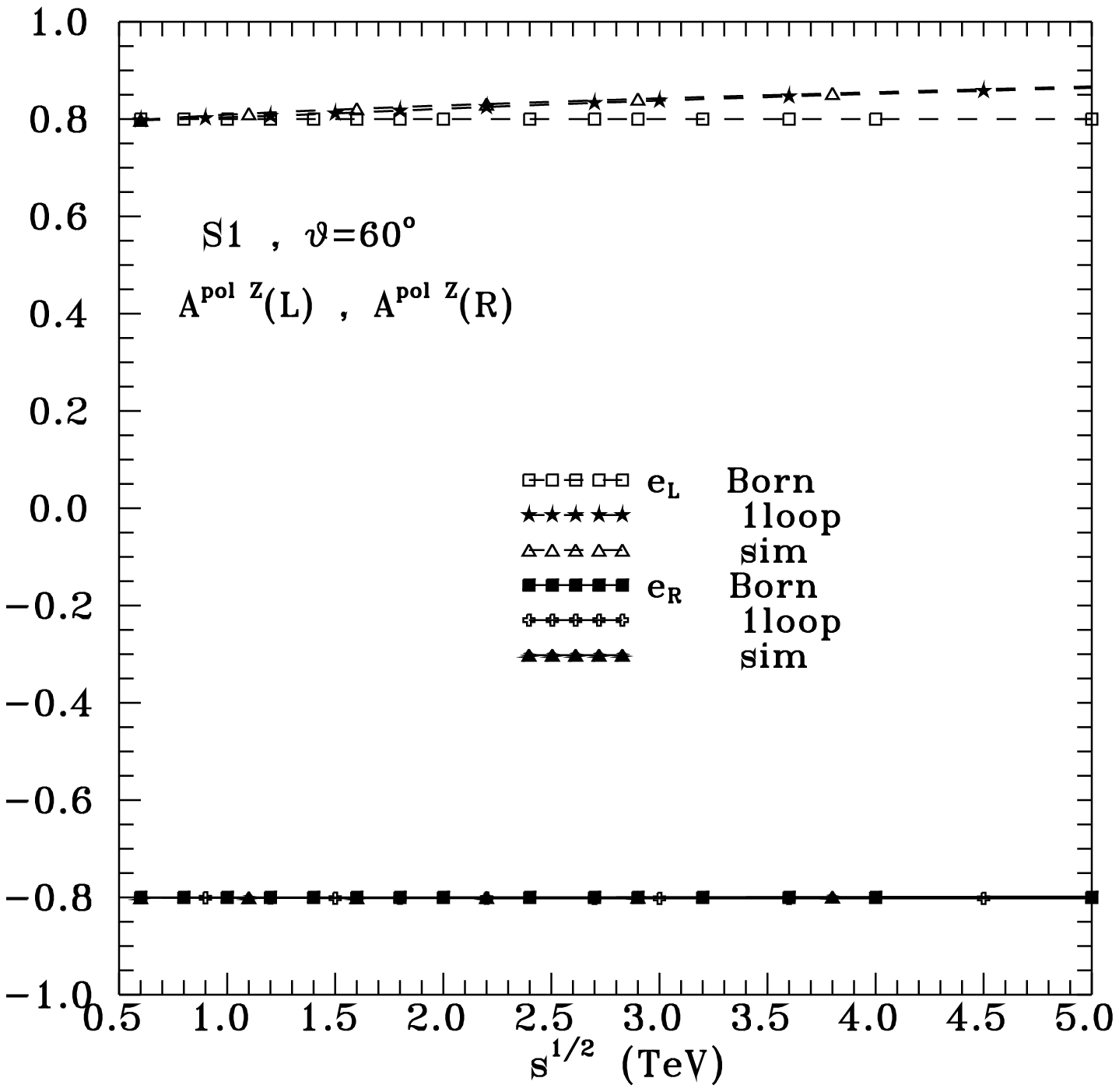,height=6.cm}
\]
\[
\epsfig{file=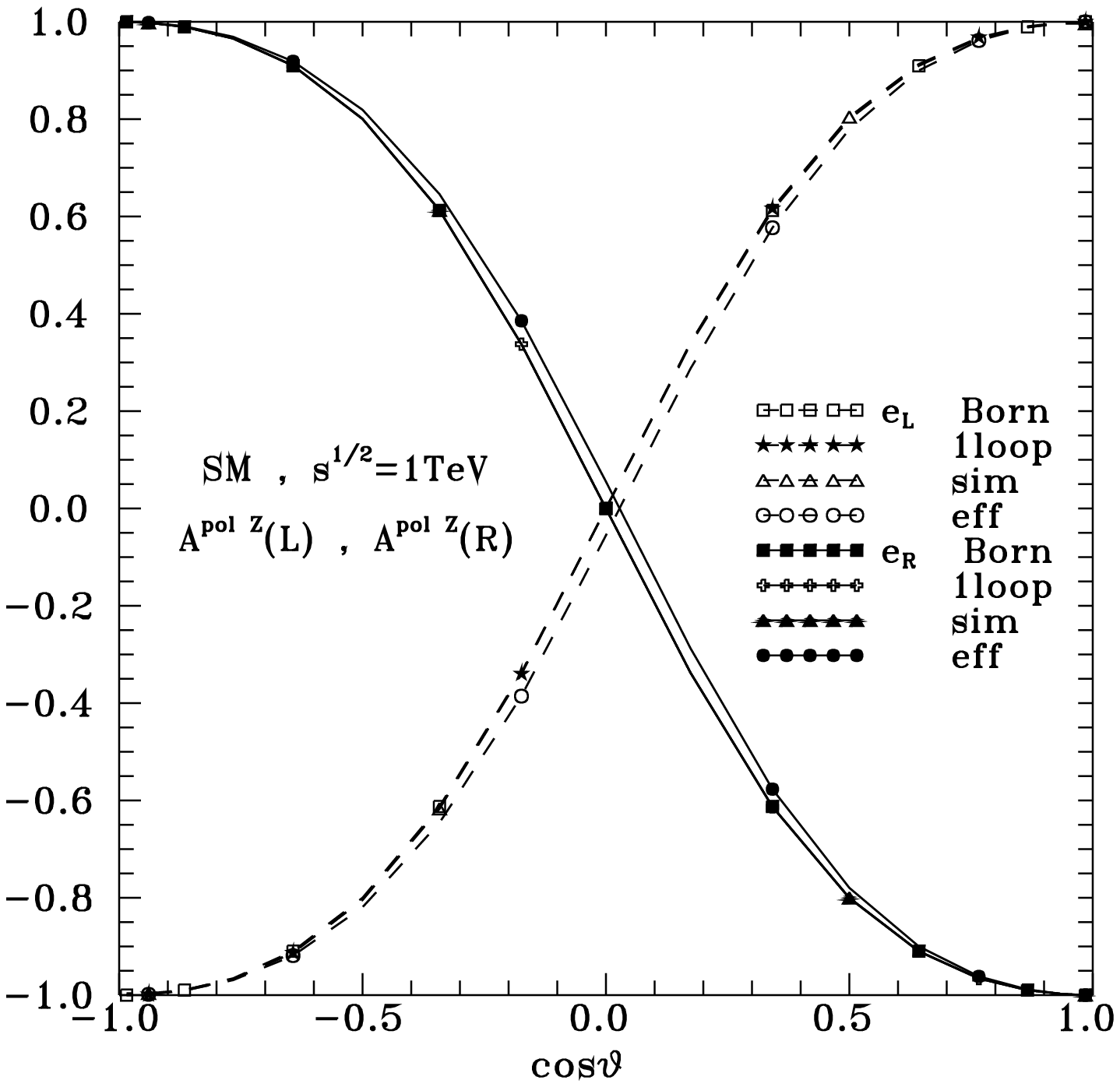, height=6.cm}\hspace{1.cm}
\epsfig{file=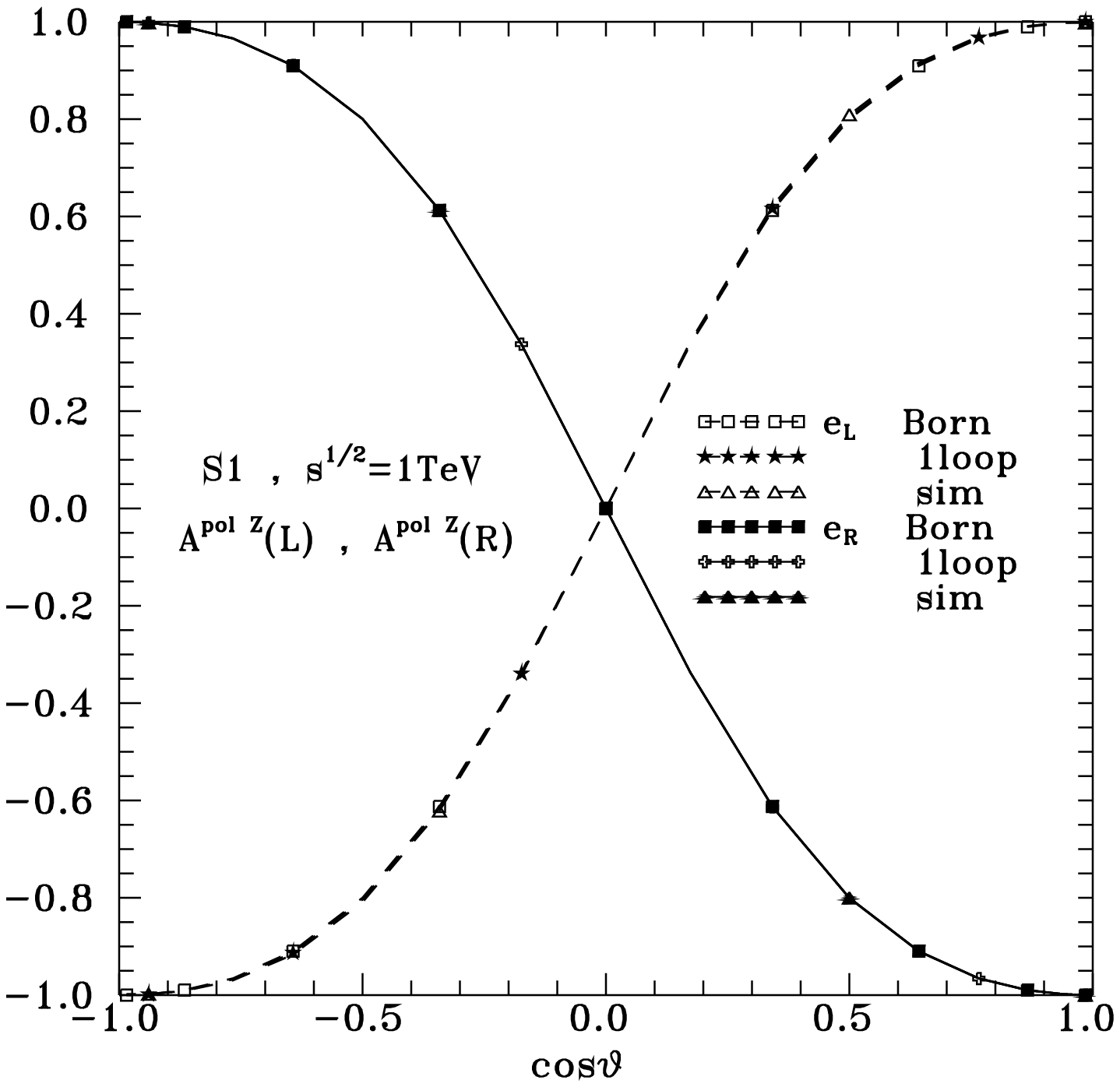,height=6.cm}
\]
\[
\epsfig{file=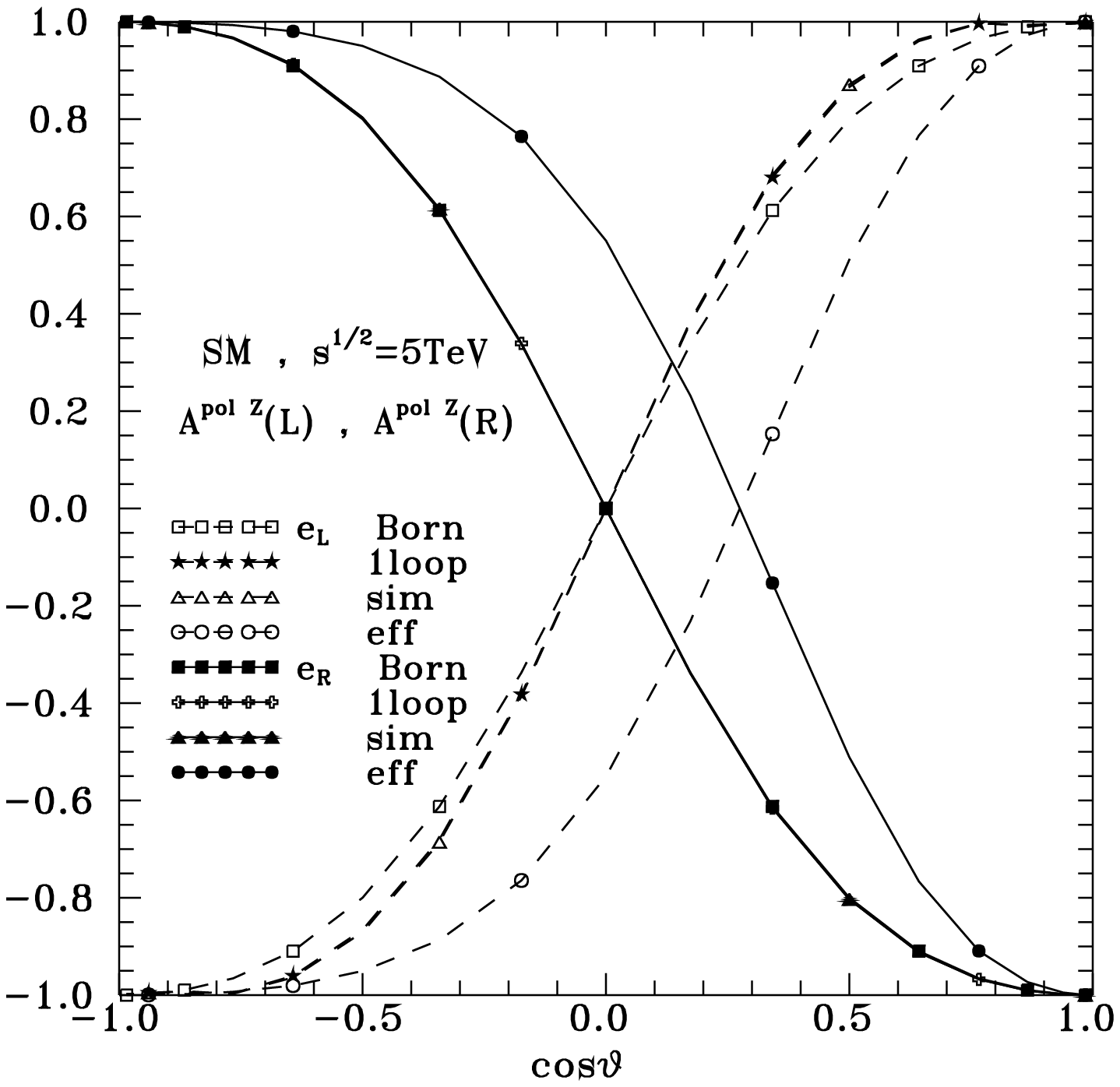, height=6.cm}\hspace{1.cm}
\epsfig{file=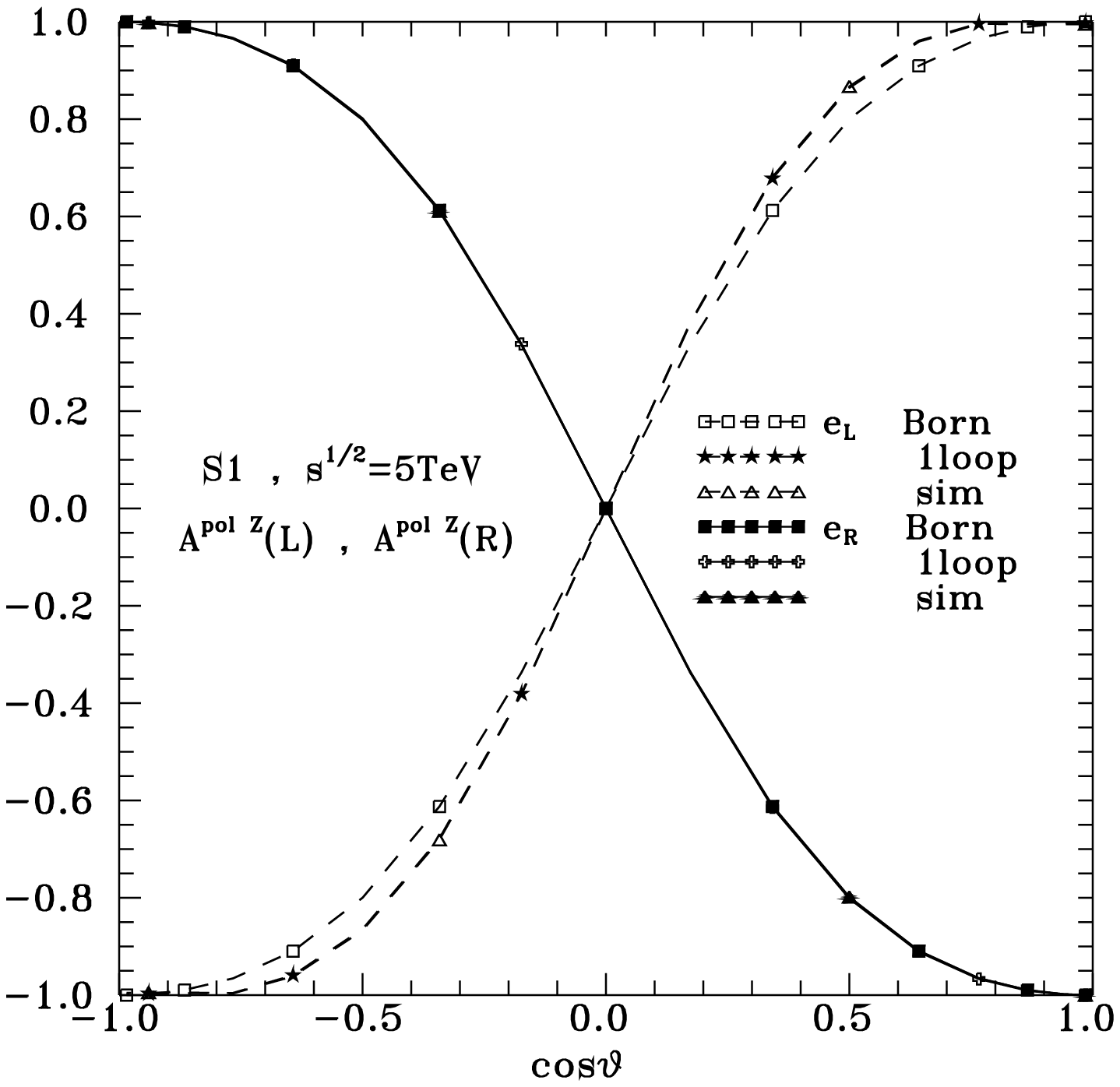,height=6.cm}
\]
\caption[1]{$A^{{\rm pol}~ Z}(\lambda)$ asymmetries defined in (\ref{Alambdatau-polZ})  for  $e^-$-beams with definite helicity $\lambda=L,R$. See caption  in Fig.\ref{sigmas-fig}.}
\label{C4-fig}
\end{figure}

 Fig.\ref{C3-fig} shows $A^{{\rm pol}~Z}$ defined in (\ref{A-polZ})  for unpolarized $e^{\mp}$-beams,
 while Fig.\ref{C4-fig} shows $A^{pol~Z}(\lambda)$ defined in (\ref{Alambdatau-polZ}) for polarized $e^{\mp}$-beams with electron helicity $\lambda=L,R$. Left panels correspond to the SM prediction for
 $e^-e^+\to  Z H$, and right panels to the MSSM result for  $e^-e^+\to  Z h^0$. Upper panels give the energy dependence at $60^\circ$, while middle and lower panels  present the angular distributions at 1 and 5 TeV respectively.

It may be interesting to note that  at Born level  one has
\bq
A^{{\rm pol}~Z ~{\rm Born}}(\lambda)=-A^{{\rm pol}~Z ~{\rm Born}}(-\lambda) ~~,
\label{A-polZ-Born}
\eq
because of the simple Z exchange with a $\eesl$ coupling.
At one loop level though, this is no more  the case,
because of various diagrams differing  for $e^-_L$ and for $e^-_R$.
If CP is conserved, as is the case for SM and the S1 benchmark \cite{bench},
$A^{{\rm pol}~Z}(L)$ and $A^{{\rm pol}~Z}(R)$ are both forward-backward antisymmetric; see (\ref{AF-polZ}). But as their Born parts are opposite and cancel  each-other,
$A^{{\rm pol}~Z}$ is finally smaller
than $A^{pol~Z}(R)$ or $A^{pol~Z}(L)$ separately. Consequently $A^{{\rm pol}~Z}$
it is very sensitive to one loop contributions, as shown in Fig.\ref{C3-fig}.
For these reasons, the sim accuracy, although already good at 1 TeV
for $A^{{\rm pol}~Z}(L,R)$, needs higher energies for $A^{{\rm pol}~Z}$; compare  Fig.\ref{C4-fig}
with  Fig.\ref{C3-fig}.

So finally the comparison of the various polarized asymmetries in the spirit of the
relations written in Sect.6 should produce fruitful tests of the natures
of the corrections to the Born terms and in particular of their CP conservation
property.

\subsection{Integrated Asymmetries}

The above illustrations show the angular dependencies at given
energy. In order to accumulate more statistics we can make
angular integrations, over all forward or backward angles,
or only appropriate  domain of  angles.

As a non trivial example we give in Table 1 the asymmetries $A^{{\rm pol}~Z}_{FB}(L)$,
 $A^{{\rm pol}~Z}_{FB}(R)$  and $A^{{\rm pol}~Z}_{FB}$,
in  SM and S1  \cite{bench}, integrated in the forward region at 5 TeV; compare
(\ref{AFB-polZ-tau}, \ref{AFB-polZ-lambdatau}).
\begin{table}[hbt]
\begin{center}
{Table 1: The asymmetries $A^{{\rm pol}~Z}_{FB}(L)$,  $A^{{\rm pol}~Z}_{FB}(R)$
 and $A^{{\rm pol}~Z}_{FB}$, \\
in  SM and S1  \cite{bench}, integrated in the forward region at 5 TeV.  }\\
  \vspace*{0.3cm}
\begin{small}
\begin{tabular}{||c|| c| c| c||c|c|c||}
\hline \hline
& \multicolumn{3}{|c||}{ SM:  $H$ }  &
\multicolumn{3}{c||}{ MSSM S1:  $h^0$} \\
 \hline
 & Born & one loop & sim & Born  & one loop & sim    \\
  \hline
 $A^{{\rm pol}~Z}_{FB}(L)$ & 0.74  & 0.72 & 0.73 & 0.74  & 0.71 & 0.72 \\
 $A^{{\rm pol}~Z}_{FB}(R)$ &-0.74 & -0.73  & -0.73  & -0.74 & -0.73 & -0.73 \\
 $A^{{\rm pol}~Z}_{FB}$ &  0.11 & -0.22 & -0.21 & 0.11 & -0.20  &-0.21 \\
 \hline
\end{tabular}
 \end{small}
\end{center}
\end{table}
As seen there,  the individual $L,R$ contributions
are similar but of opposite sign. When they are combined though, in the unpolarized $e^\mp$-beam case, a larger sensitivity
to the one loop corrections appears. Table 1 confirms what has been seen at angular level in
Figs.\ref{C3-fig},\ref{C4-fig}. Note that the supersimple expressions reproduce correctly the exact one loop
contributions.

Turning to CP conservation tests,  one may also experimentally  check
the relations for $A_{FB}(\lambda, \tau)$, $A^F_{LR}(\tau)$
and $A^{{\rm pol}~Z}(\lambda)$, written respectively in (\ref{CP-conservation-AFBlambdatau},
 \ref{ALRFB-tau1}, \ref{AF-polZ}).  In particular
the vanishing of $A_{FB}$ when the $e^\mp$-beams are not polarized
and the $Z$ polarization is not measured, and the vanishing  of  $A^{{\rm pol}~Z}(\lambda)$ when
integrated   over all angles, are rather striking.\\

Summarizing, we have shown in this Section 7, that the polarized asymmetries are very
powerful for identifying small corrections to Born contributions, both in SM
and in MSSM cases. In addition the comparison of these various asymmetries
in the spirit of the relations written in Sect.6, should produce fruitful
tests of CP conservation for these corrections.\\

\section{Conclusions and possible developments}

We have analyzed in detail the   Born and one loop corrections to
the amplitudes for the process $e^-e^+\to  ZH$ in SM and $e^-e^+\to  Zh^0$ in MSSM,  at high energies.
We have separately discussed the behavior of the helicity
conserving and of the helicity violating amplitudes, and we have shown
that they can be reasonably approximated by simple expressions, called supersimple (sim),
both in SM and in MSSM cases.

In the case of the HC amplitudes,
the simplicity arises from gauge cancelations of
contributions coming from various diagrams, and in addition,
for MSSM, from  typical cancelations between standard and partner
contributions, as we have already observed in previous  "supersimplicity" studies \cite{super, ttbar, WW}.
The sim expressions thus obtained, allow to immediately see the dynamical contents
and to make a quick estimate of the size of the amplitudes.
We have illustrated how much and in
which high energy domain these expressions are useful.

While doing this study we have noticed several interesting properties
of the process $e^-e^+\to  ZH$, for $H=H_{SM}, h^0, H^0$, and we have emphasized the importance
of doing a complete amplitude analysis. It can be obtained by
using the various observables i.e. the cross section measurements
for different $e^{\mp}$ and $Z$ polarizations.
We have illustrated the large sensitivity of the asymmetries
$A_{LR}$, $A^{{\rm pol}~Z}$, $A^{{\rm pol}~Z}_{FB}$ to the one loop contributions. These
asymmetries should allow to singularize and identify the nature
(SM or MSSM or other BSM) of the corrections to Born predictions.

This work  should be useful to the working groups
on the various projects of high energy $e^-e^+$ colliders,
for studying  the identification of the nature of the Higgs boson
and of its interactions.

\newpage

\renewcommand{\thesection}{A}
\renewcommand{\theequation}{A.\arabic{equation}}
\setcounter{equation}{0}

\noindent
{\Large \bf Appendix A: The one-loop  sim expressions \\
 for the HC amplitudes   in SM and  MSSM}\\

As has been observed in \cite{super, WW},  there exist only
four different forms that appear in the sim expressions. These are the two augmented
Sudakov forms\footnote{
The augmented Sudakov definitions used in this paper are more precise than  those used in our previous work, but fully consistent with them.}
\bqa
\overline{\ln s_{ij}(a)} & \equiv & \ln{-s-i\epsilon \over m_im_j}+b_0^{ij}(m^2_a)-2 ~~
\label{Sud-ln-form} \\
\overline{\ln^2s_{Via}}& \equiv  & \ln^2{-s-i\epsilon \over m^2_V}+4L_{aVi} ~~,  \label{Sud-ln2-form}
\eqa
 where explicit expressions for $b_0^{ij}(m^2_a)$ and $L_{aVi}$ are given in e.g.  Eqs.(A.6, A.5) of \cite{WW}.
 In them, quantity $a$  describes  an  on-shell particle,
  $(i,j)$ denote internal exchanges, and    $V$  an internal Vector (gauge) exchange;
the existence of non-vanishing tree-order vertices for $aij$ and $aVi$ is assumed.

The other two forms entering the sim expressions, involve ratios
   of the Mandelstam-variables $s,t,u$. Denoting any of them by
   $x,y$, these ratios are given by
 \bq
   r_{xy} \equiv \frac{-x-i\epsilon}{-y-i\epsilon} ~~, \label{rxy-form}
 \eq
  and the two relevant forms by
 \bq
\overline{\ln^2r_{xy}}=\ln^2r_{xy}+ \pi^2 ~~~,~~~ \ln r_{xy} ~~~.~~ \label{lnr-form}
\eq\\

 Using the definitions introduced in Sect.2, the complete sim expressions, to the one-loop order, for the various HC amplitudes in SM are
\bqa
F_{-0}&= &F^{\rm Born}_{-0} \Bigg \{ 1+ {\alpha\over4\pi}  \Big \{ {1\over4s^2_Wc^2_W}
 \Big [-\overline{\ln^2s_{Zee}}+3\overline{\ln s_{Ze}(e)}-1 \Big ]\nonumber\\
&& +{1\over2s^2_W(2s^2_W-1)}
\Big [-\overline{\ln^2s_{W\nu e}}+3\overline{\ln s_{W\nu}(e)}-1 \Big ]\nonumber\\
&& - {c^2_W\over s^2_W(2s^2_W-1)}
\Big [ \overline{\ln s_{W\nu}(e)}+1+4\overline{\ln s_{WW}(Z)} \Big ]\nonumber\\
&& - {c^2_W\over 2s^2_W(2s^2_W-1)}
\Big [{1\over2}\overline{\ln^2s_{WZH}}+~\overline{\ln s_{WW}(H)}+~\overline{\ln s_{WW}(Z)} \Big]
\nonumber\\
&&-3 \Big [{m^2_t \ln s_{tt}(Z)\over2s^2_Wm^2_W} +{m^2_b\ln s_{bb}(Z)\over2s^2_Wm^2_W} \Big ]\nonumber\\
&&+ {1\over 2s^2_W(2s^2_W-1)}
\Big [-~\overline{\ln^2s_{WZH}}+2\overline{\ln s_{WW}(H_{SM})}+2\overline{\ln s_{WW}(Z)} \Big ]\nonumber\\
&&+ {1\over 4s^2_Wc^2_W}
\Big [-~\overline{\ln^2s_{ZWH}}+~2\overline{\ln s_{ZZ}(H)}
+~2\overline{\ln s_{HZ}(Z)}\Big ]\nonumber\\
&& + {c^2_W\over8s^2_W(2s^2_W-1)}
\Big [2\overline{\ln^2s_{WZH}}+8\overline{\ln^2t_{WZH}}+8\overline{\ln^2u_{WZH}}
-2\overline{\ln t_{WW}(H)}\nonumber\\
&&-2\overline{\ln u_{WW}(H)}-2\overline{\ln t_{WW}(Z)}-2\overline{\ln u_{WW}(Z)}
\nonumber\\
&&
-~{4(u-t)\over u}\overline{\ln^2r_{ts}}-~{4(t-u)\over t}\overline{\ln^2r_{us}}
+4\overline{\ln r_{ts}}+4\overline{\ln r_{us}} \Big ] \Big \}
\nonumber\\
&&+{2s_Wc_W\over(2s^2_W-1)}{\hat{\Sigma}_{\gamma Z}(s)\over s}-{\hat{\Sigma}_{Z Z}(s)\over s}+C_P
 \Bigg \} ~, \label{SM-FSIMm0}
\eqa
\bqa
F_{+0} &= & F^{\rm Born}_{+0} \Bigg \{ 1+ {\alpha\over4\pi} \Big \{ {1\over c^2_W}
\Big [-\overline{\ln^2s_{Zee}}+3\overline{\ln s_{Ze}(e)}-1 \Big ]\nonumber\\
&&+ {1\over 2s^2_W}
\Big [-~\overline{\ln^2s_{WZH}}+2\overline{\ln s_{WW}(H)}+2\overline{\ln s_{WW}(Z)} \Big ]
\nonumber\\
&&+ {1\over 4s^2_Wc^2_W}
\Big [-~\overline{\ln^2s_{ZWH}}+~2\overline{\ln s_{ZZ}(H)}
+~2\overline{\ln s_{H_{SM}Z}(Z)} \Big ]
\nonumber \\
&& -3 \Big [{m^2_t \overline{\ln s_{tt}(Z)}\over2s^2_Wm^2_W}
 +{m^2_b \overline{\ln s_{bb}(Z)}\over2s^2_Wm^2_W} \Big ] \Big \}
 +{c_W\over s_W}{\hat{\Sigma}_{\gamma Z}(s)\over s}-{\hat{\Sigma}_{Z Z}(s)\over s} \Bigg \} ~.
 \label{SM-FSIMp0}
\eqa
where    $\hat{\Sigma}$  denotes  renormalized self-energies\footnote{See e.g. \cite{ttbar, WW}},
and  $C_P$ is the "pinch term" \cite{pinch}
\bq
C_P={\alpha\over\pi} {c^2_W\over s^2_W(2s^2_W-1)} \overline{\ln s_{WW}(Z)} ~.  \label{pinch-term}
\eq \\

The corresponding complete sim expressions for the MSSM HC amplitudes with $H=h^0,H^0$ are
\bqa
&& F_{-0} =  F^{\rm Born}_{-0} \Bigg \{ 1 + {\alpha\over4\pi} \Big \{ {1\over4s^2_Wc^2_W}
\Big [-\overline{\ln^2s_{Zee}}+3\overline{\ln s_{Ze}(e)}-1 \Big ]\nonumber\\
&& +{1\over2s^2_W(2s^2_W-1)}
\Big [-\overline{\ln^2s_{W\nu e}}+3\overline{\ln s_{W\nu}(e)}
-\sum_{i} |Z^+_{1i}|^2 \overline{\ln s_{\chi_i\tilde{\nu}_{eL}}(e)} \Big ]
\nonumber\\
&& - {c^2_W\over s^2_W(2s^2_W-1)}
\Big [\overline{\ln s_{W\nu}(e)} +4\overline{\ln s_{WW}(Z)} -\sum_{i} |Z^+_{1i}|^2
\overline{\ln s_{\chi_i\tilde{\nu}_{eL}}(e)} \Big ]\nonumber\\
&& - \sum_{i} {|Z^N_{1i}s_W+Z^N_{2i}c_W|^2\over4s^2_Wc^2_W}
\Big [\overline{\ln s_{\chi_i\tilde{e}_{L}}(e)}-1 \Big]
\nonumber\\
&& -{c^2_W\over 4s^2_W(2s^2_W-1)}
\Big [\overline{\ln^2s_{WZH}}+2\overline{\ln s_{WW}(H)}+2\overline{\ln s_{WW}(Z)} \Big ]
\nonumber\\
&&+{1\over 2s^2_W(2s^2_W-1)}
\Big [-~\overline{\ln^2s_{WZH}}+2\overline{\ln s_{WW}(H)}+2\overline{\ln s_{WW}(Z)} \Big]\nonumber\\
&&-3 \Big [{m^2_t \overline{\ln s_{tt}(Z)}\over2s^2_Wm^2_W}
\Big (h^0\to {\cos\alpha\over\sin\beta\sin(\beta-\alpha)} ~,~
H^0\to {\sin\alpha\over\sin\beta\cos(\beta-\alpha)} \Big )
\nonumber\\
&&+ {m^2_b \overline{\ln s_{bb}(Z)}\over2s^2_Wm^2_W}
\Big (h^0 \to -~{\sin\alpha\over\cos\beta\sin(\beta-\alpha)} ~,~
H^0\to {\cos\alpha\over\cos\beta\cos(\beta-\alpha)} \Big )
\Big ]
\nonumber\\
&&+{1\over s^2_W} \sum_{ijk} X_{ijk}(H) \overline{\ln s_{jk}(Z)}
\Big (h^0\to  {1\over\sin(\beta-\alpha)} ~,~
H^0\to {1\over\cos(\beta-\alpha)} \Big ) \nonumber\\
&&+{1\over2s^2_Wc^2_W} \sum_{ijk} X^0_{ijk}(H)\overline{\ln s_{jk}(Z)}
\Big (h^0\to  {1\over\sin(\beta-\alpha)} ~,~
H^0\to {1\over\cos(\beta-\alpha)} \Big ) \nonumber\\
&& +{c^2_W\over8s^2_W(2s^2_W-1)}
\Big [\overline{\ln^2s_{WZH}}+8\overline{\ln^2t_{WZH}}
-2\overline{\ln t_{WW}(H)}-2\overline{\ln t_{WW}(Z)}\nonumber\\
&&+\overline{\ln^2s_{WZH}}+8\overline{\ln^2u_{WZH}}
-2\overline{\ln u_{WW}(H)}-2\overline{\ln u_{WW}(Z)}\nonumber\\
&&-~{4(u-t)\over u}\overline{\ln^2r_{ts}}-~{4(t-u)\over t}\overline{\ln^2r_{us}}
+4\overline{\ln r_{ts}}+4\overline{\ln r_{us}} \Big ]
\nonumber\\
&& +{c^2_W\over2s^2_W(2s^2_W-1)} \Big [{s\over 2u}\overline{\ln^2r_{ts}}+
{s\over 2t}\overline{\ln^2r_{us}} \Big ]
\Big (h^0 \to {-\cos\beta\sin\alpha\over\sin(\beta-\alpha)} ~,~
H^0\to {\cos\beta\cos\alpha\over\cos(\beta-\alpha)} \Big )
\nonumber\\
&& - {(2s^2_W-1)\over4s^2_Wc^2_W} \Big [{s\over 2u}\overline{\ln^2r_{ts}}+
{s\over 2t}\overline{\ln^2r_{us}} \Big ]
\Big (h^0\to {\sin(\beta+\alpha)\over\sin(\beta-\alpha)}~,~
H^0\to {-\cos(\beta+\alpha)\over\cos(\beta-\alpha)} \Big )
\nonumber\\
&&+ {1\over 4s^2_Wc^2_W} AZ(H) \Big \}
+{2s_Wc_W\over(2s^2_W-1)}{\hat{\Sigma}_{\gamma Z}(s)\over s}-{\hat{\Sigma}_{Z Z}(s)\over s}
+C_P \Bigg \} ~~, \label{MSSM-FSIMm0}
\eqa
\bqa
&&F_{+0}=F^{\rm Born}_{+0} \Bigg \{ 1+ {\alpha\over4\pi} \Big \{  {1\over c^2_W}
\Big [-\overline{\ln^2s_{Zee}}+3\overline{\ln s_{Ze}(e)}-1 \Big ]\nonumber\\
&&+{1\over 2s^2_W}
\Big [-~\overline{\ln^2s_{WZH}}+2\overline{\ln s_{WW}(H)}+2\overline{\ln s_{WW}(Z)}\Big ]\nonumber\\
&& -\sum_{i} {|Z^N_{1i}|^2\over c^2_W}
\Big [\overline{\ln s_{\chi_i\tilde{e}_{R}}(e)}-1 \Big ]
+{1\over 4s^2_Wc^2_W} AZ(H) \nonumber\\
&&-3 \Big [{m^2_t \overline{\ln s_{tt}(Z)}\over2s^2_Wm^2_W}
\Big ( h^0 \to {\cos\alpha\over\sin\beta\sin(\beta-\alpha)}~,~
H^0\to {\sin\alpha\over\sin\beta\cos(\beta-\alpha)} \Big )
\nonumber\\
&&+ {m^2_b \overline{\ln s_{bb}(Z)}\over2s^2_Wm^2_W}
\Big (h^0 \to -{\sin\alpha\over\cos\beta\sin(\beta-\alpha)}~,~
H^0\to {\cos\alpha\over\cos\beta\cos(\beta-\alpha)} \Big ) \Big ] \nonumber\\
&&+ {1\over s^2_W} \sum_{ijk} X_{ijk}(H)\overline{\ln s_{jk}(Z)}
\Big (h^0 \to {1\over\sin(\beta-\alpha)}~,~
H^0 \to {1\over\cos(\beta-\alpha)} \Big ) \nonumber\\
&&+{1\over2s^2_Wc^2_W} \sum_{ijk}
X^0_{ijk}(H)\overline{\ln s_{jk}(Z)} \Big  ( h^0 \to {1\over\sin(\beta-\alpha)}~,~
H^0\to {1\over\cos(\beta-\alpha)}\Big )
\nonumber\\
&& - {1\over2c^2_W} \Big  [{s\over 2u}\overline{\ln^2r_{ts}}+
{s\over 2t}\overline{\ln^2r_{us}}\Big  ]
\Big  ( h^0 \to {-\sin(\beta+\alpha)\over\sin(\beta-\alpha)} ~,~
{H^0 \to \cos(\beta+\alpha)\over\cos(\beta-\alpha)} \Big ) \Big  \}
\nonumber\\
&&+{c_W\over s_W}{\hat{\Sigma}_{\gamma Z}(s)\over s}-{\hat{\Sigma}_{Z Z}(s)\over s}
\Bigg \} ~~, \label{MSSM-FSIMp0}
\eqa
where the renormalized gauge self energies  $\hat{\Sigma}$
and the pinch term (\ref{pinch-term}) have been used.

The exact expressions for $AZ(H)$ for $H=h^0, ~H^0$ used in
(\ref{MSSM-FSIMm0},\ref{MSSM-FSIMp0}) are
\bqa
 AZ(h^0) &= & \sin^2(\beta-\alpha)A_{h^0h^0}+\cos^2(\beta-\alpha)A_{H^0h^0} \nonumber \\
          && + \cos^2(\beta-\alpha)(A_{Ah^0h^0}-A_{AH^0h^0}) \nonumber \\
AZ(H0) &= & \sin^2(\beta-\alpha)A_{h^0H^0}+\cos^2(\beta-\alpha)A_{H^0H^0} \nonumber \\
&& +\sin^2(\beta-\alpha)(A_{AH^0H^0}-A_{Ah^0H^0}) ~~, \label{AZH-exact}
\eqa
with
\bqa
A_{h^0h^0} &= &-\overline{\ln^2s_{Zh^0h^0}}+2\overline{\ln s_{ZZ}(h^0)}
+2\overline{\ln s_{Zh^0}(Z)} ~~, \nonumber \\
A_{H^0h^0} &= & -\overline{\ln^2s_{ZH^0h^0}}+2\overline{\ln s_{ZZ}(h^0)}
+2\overline{\ln s_{ZH^0}(Z)} ~~, \nonumber \\
A_{Ah^0h^0} &= & -\overline{\ln^2s_{ZAh^0}}+2\overline{\ln s_{AZ}(h^0)}
+2\overline{\ln s_{Zh^0}(Z)} ~~, \nonumber \\
A_{Ah^0H^0} &=& -\overline{\ln^2s_{ZAH^0}}+2\overline{\ln s_{AZ}(h^0)}
+2\overline{ln s_{ZH^0}(Z)} ~~, \nonumber \\
A_{h^0H^0} &= & -\overline{\ln^2s_{Zh^0H^0}}+2\overline{\ln s_{ZZ}(H^0)}
+2\overline{ln s_{Zh^0}(Z)} ~~, \nonumber \\
A_{H^0H^0} &=& -\overline{\ln^2s_{ZH^0H^0}}+2\overline{\ln s_{ZZ}(H^0)}
+2\overline{\ln s_{ZH^0}(Z)} ~~, \nonumber \\
A_{Ah^0H^0} &=& -\overline{\ln^2s_{ZAh^0}}+2\overline{\ln s_{AZ}(H^0)}
+2\overline{\ln s_{Zh^0}(Z)} ~~, \nonumber \\
A_{AH^0H^0} &=& -\overline{\ln^2s_{ZAH^0}}+2\overline{\ln s_{AZ}(H^0)}
+2\overline{\ln s_{ZH^0}(Z)} ~~. \label{A-auxiliary-terms}
\eqa
When the mass differences among $(h^0, H^0, A^0, Z)$, which  determine  the scales of  the various
augmented Sudakov terms in (\ref{AZH-exact}, \ref{A-auxiliary-terms}), are neglected we obtain
\bq
AZ(h^0)\simeq AZ(H^0) \simeq -\overline{\ln^2s}+4\overline{\ln s} ~~. \label{AZH-approx}
\eq
with a common mass scale M.\\

Similarly,  the $X^0_{ijk}(H), X_{ijk}(H)$ expressions in
(\ref{MSSM-FSIMm0},\ref{MSSM-FSIMp0}) are given by  the following equations,
where the last lines give the approximate  results in case the  $(h^0, H^0, A^0, Z)$-mass
differences are negligible:
\bqa
&&X^0_{ijk}(h^0) =-{1\over2} \Bigg [(Z^{N*}_{4i}Z^{N}_{4k}-Z^{N*}_{3i}Z^{N}_{3k})
\Big \{ \Big [(-\cos\beta Z^{N*}_{3k}-\sin\beta Z^{N*}_{4k})(Z^{N*}_{1j}s_W
- Z^{N*}_{2j}c_W)\nonumber \\
&& +(j\to k) \Big ](-\sin\alpha Z^{N}_{3i}-\cos\alpha Z^{N}_{4i})(Z^{N}_{1j}s_W-Z^{N}_{2j}c_W)+(i\to j)
\Big \} \nonumber\\
&&-(Z^{N}_{3i}Z^{N*}_{3k}-Z^{N}_{4i}Z^{N*}_{4k}) \Big \{
\Big [(-\cos\beta Z^{N}_{3j}-\sin\beta Z^{N}_{4j})(Z^{N}_{1k}s_W
- Z^{N}_{2k}c_W) \nonumber\\
&& + (j\to k) \Big ](-\sin\alpha Z^{N*}_{3j}-\cos\alpha Z^{N*}_{4j})(Z^{N*}_{1i}s_W-Z^{N*}_{2i}c_W)
+(i\to j) \Big \} \Bigg ] \nonumber \\
&& \simeq  -\sin(\beta-\alpha) ~~, \label{X0ijk-h0}  \\[0.2cm]
&&X^0_{ijk}(H^0) =-{1\over2} \Bigg [(Z^{N*}_{4i}Z^{N}_{4k}-Z^{N*}_{3i}Z^{N}_{3k})\Big \{
\Big [(-\cos\beta Z^{N*}_{3k}-\sin\beta Z^{N*}_{4k})(Z^{N*}_{1j}s_W
- Z^{N*}_{2j}c_W)\nonumber\\
&& +(j\to k) \Big ](\cos\alpha Z^{N}_{3i}-\sin\alpha Z^{N}_{4i})(Z^{N}_{1j}s_W-Z^{N}_{2j}c_W)
 +(i\to j) \Big \}\nonumber\\
&& -(Z^{N}_{3i}Z^{N*}_{3k}-Z^{N}_{4i}Z^{N*}_{4k})
\Big \{ \Big [(-\cos\beta Z^{N}_{3j}-\sin\beta Z^{N}_{4j})(Z^{N}_{1k}s_W
- Z^{N}_{2k}c_W)\nonumber\\
&& +(j\to k) \Big ](\cos\alpha Z^{N*}_{3j}-\sin\alpha Z^{N*}_{4j})(Z^{N*}_{1i}s_W-Z^{N*}_{2i}c_W)+(i\to j)
\Big \} \Bigg ]  \nonumber \\
&& \simeq -\cos(\beta-\alpha)  ~~, \label{X0ijk-H0}  \\[0.2cm]
&&X_{ijk}(h^0) =[Z^{-}_{1i}Z^{-*}_{1k}+\delta_{ik}(c^2_W-s^2_W)](\cos\beta Z^{-}_{2k}Z^{+}_{1j}
-\sin\beta Z^{-}_{1k}Z^{+}_{2j})(-\sin\alpha Z^{-*}_{2i}Z^{+*}_{1j}
\nonumber \\
&& +\cos\alpha Z^{-*}_{1i}Z^{+*}_{2j})
-[Z^{+*}_{1i}Z^{+}_{1k}+\delta_{ik}(c^2_W-s^2_W)](\cos\beta Z^{-*}_{2j}Z^{+*}_{1k}
\nonumber \\
&& -\sin\beta Z^{-*}_{1j}Z^{+*}_{2k})(-\sin\alpha Z^{-}_{2j}Z^{+}_{1i}
 +\cos\alpha Z^{-}_{1j}Z^{+}_{2i}) \nonumber \\
&& \simeq -\sin(\beta-\alpha)  ~~, \label{Xijk-h0}  \\[0.2cm]
&&X_{ijk}(H^0) =[Z^{-}_{1i}Z^{-*}_{1k}+\delta_{ik}(c^2_W-s^2_W)](\cos\beta Z^{-}_{2k}Z^{+}_{1j}
-\sin\beta Z^{-}_{1k}Z^{+}_{2j})(\cos\alpha Z^{-*}_{2i}Z^{+*}_{1j}
\nonumber\\
&& +\sin\alpha Z^{-*}_{1i}Z^{+*}_{2j})
-[Z^{+*}_{1i}Z^{+}_{1k}+\delta_{ik}(c^2_W-s^2_W)](\cos\beta Z^{-*}_{2j}Z^{+*}_{1k}
\nonumber\\
&& -\sin\beta Z^{-*}_{1j}Z^{+*}_{2k})(\cos\alpha Z^{-}_{2j}Z^{+}_{1i}
+\sin\alpha Z^{-}_{1j}Z^{+}_{2i}) \nonumber \\
&& \simeq -\cos(\beta-\alpha) ~~. \label{Xijk-H0}
\eqa\\

\vspace*{1cm}

\renewcommand{\thesection}{B}
\renewcommand{\theequation}{B.\arabic{equation}}
\setcounter{equation}{0}

\noindent
{\Large \bf Appendix B: The one-loop  sim expressions \\
for the HV amplitudes in SM and  MSSM}\\

 Assuming CP conservation implying (\ref{CP-conservation}), which for the Helicity Violating (HV)
  amplitudes enforces the relation
 \bq
F_{-+}(\theta)=F_{--}(\pi-\theta)~~, ~~F_{++}(\theta)=F_{+-}(\pi-\theta) ~~, \label{CP-HV-constraint}
\eq
we end up with the two independent HV amplitudes
\bqa
F_{--} &=& F^{\rm Born}_{--}
+ \alpha^2 m_W \left [{u\sqrt{2}\over\sqrt{s}}N^L_1+{ut\over\sqrt{2s}}(N^L_3-N^L_4) \right ]  ~~, \label{FHVmm} \\
F_{+-} &=& F^{\rm Born}_{+-} +
  \alpha^2m_W \left [{t\sqrt{2}\over\sqrt{s}}N^R_1-{ut\over\sqrt{2s}}(N^R_3-N^R_4) \right ] ~~, \label{FHVpm}
\eqa
 expressed in terms of the Born amplitudes in (\ref{FBorn-T}) and the
  quantities $N^L_1, N^L_3-N^L_4$ and  $N^R_1, N^R_3-N^R_4$ given below.

 Since the HV amplitudes are  vanishing  at high energies though, (albeit rather slowly, like $M/\sqrt{s}$, as we have seen in the main text) the mass depended corrections
 in the augmented Sudakov terms   (\ref{Sud-ln-form}, \ref{Sud-ln2-form})
 can often be suppressed, and a common mass scale $M$ may be used in the logarithms,
 leading to the simplification
 \bqa
 \overline{\ln s}  & = & \ln {-s -i\epsilon \over M^2} ~~, \nonumber \\
 \overline{\ln^2 s } & = & \ln^2 {-s-i\epsilon \over M^2} ~~. \label{Sud-HV}
 \eqa
In cases  this is not sufficient though, the complete augmented Sudakov forms of Appendix A, as well as
(\ref{rxy-form}, \ref{lnr-form}) are used. We thus  obtain  \\

\underline{in SM}
\bqa
  sN^L_1&=& {(2s^2_W-1)\over2s^2_Wc^3_W} \Big \{{1\over4s^2_Wc^2_W}
[-\overline{\ln^2s}+3\overline{\ln s}-1]
+{1\over2s^2_W(2s^2_W-1)} [-\overline{\ln^2s}+3\overline{\ln s}-1]\nonumber\\
&& - {c^2_W\over s^2_W(2s^2_W-1)}
[\overline{\ln s}+1+4\overline{\ln s}]\Big \}
+{(1-2s^2_W)m^2_{H}\over8c^3_Ws^4_Wm^2_W}(1-\overline{\ln s})
\nonumber\\
&&+{1\over2s^4_Wc_W}\overline{\ln^2s}
-{(1+9c^4_W+13c^6_W)\over8s^4_Wc^5_W}\overline{\ln s}-{(1-2c^4_W-2c^6_W) \over8s^4_Wc^6_W}
\nonumber\\
&&+{6\over s_Wc_W}\Big [{m^2_t\over m^2_W}C^R_t+{m^2_b\over m^2_W}C^R_b \Big ]
+\Big [{2s^2_W-1\over 2s^4_Wc_W}+{(2s^2_W-1)^3\over 4s^4_Wc^5_W} \Big ]
\Big [{s\overline{\ln t}\over t}+{s\overline{\ln u}\over u} \Big ]
\nonumber\\
&&-{c_W\over 2s^4_W}\Big [{(t-2s)\over2t}\overline{\ln^2t}+{(u-2s)\over2u}\overline{\ln^2u}
+{(2s-u)\over2u}\overline{\ln^2r_{ts}}+{(2s-t)\over2t}\overline{\ln^2r_{us}} \Big ]
\nonumber\\
&&+{1\over 4c_Ws^2_W}\Big [-\overline{\ln^2t}-\overline{\ln^2u}
+\overline{\ln^2r_{ts}}+\overline{\ln^2r_{us}} \Big ]
\nonumber\\
&&-\Big ( {1\over 4c_Ws^4_W}+{(2s_W^2-1)^3\over 8c^5_Ws^4_W}\Big )
\Big [{s\over t}\overline{\ln^2t}+{s\over u}\overline{\ln^2u}+2\overline{\ln^2r_{tu}} \Big]
~~, \label{NL1SM}
\eqa
with
\bqa
C^L_t &= & {(-9+18s^2_W+8s^4_W)\over72s^3_Wc^3_W}\Big [{\overline{\ln^2s}\over4}-{\overline{\ln s}\over2}
+1 \Big ]-{s_W\over18c^3_W}\overline{\ln s} ~~, \nonumber \\
C^L_b &= &{(-9+18s^2_W-4s^4_W) \over72s^3_Wc^3_W}\Big [{\overline{\ln^2s}\over4}-{\overline{\ln s}\over2}
+ 1 \Big ]+{s_W\over36c^3_W}\overline{\ln s} ~~, \label{CLtb}
\eqa
and
\bqa
s(N^L_3-N^L_4)&=&-{2c_W\over s^4_W} \Big [{(t-u)\over2tu}\overline{\ln^2s}
+{\overline{\ln^2t}\over 4t}-{\overline{\ln^2u}\over 4u}\nonumber\\
&&
+{s\overline{\ln r_{tu}}\over ut}+{(2u^2-2t^2+ut)\over4tu^2}\overline{\ln^2r_{ts}}
+{(2u^2-2t^2-ut)\over4ut^2}\overline{\ln^2r_{us}} \Big ]\nonumber\\
&&
+{1\over 2c_Ws^2_W}\Big [{\overline{\ln^2u}\over u}-{\overline{\ln^2t}\over t}
+{\overline{\ln^2r_{us}}\over t}-{\overline{\ln^2r_{ts}}\over u} \Big ]\nonumber\\
&&-\Big [{1\over 2c_Ws^4_W}+{(2s_W^2-1)^3\over 4c^5_Ws^4_W} \Big ]
\Big [{s\over ut}(\overline{\ln^2t}-\overline{\ln^2u})\nonumber\\
&&+{2s\over ut}(\overline{\ln u}
-\overline{\ln t})+{\overline{\ln^2r_{tu}}\over u}-{\overline{\ln^2r_{tu}}\over t} \Big ]
~~, \label{NL34SM} \\[0.2cm]
sN^R_1&=&{-1\over c^5_W} \Big [-\overline{\ln^2s}+3\overline{\ln s}-1 \Big ]
+{\overline{\ln^2s}\over c_Ws^2_W}
-{\overline{\ln s}\over 4c^5_Ws^2_W} (1+2c^2_W+4c^2_W) \nonumber\\
&&+{1\over 4c^5_Ws^2_W}[1+2c^2_W+4c^2_W]
+{(1-2s^2_W)m^2_{H}\over4c^3_Ws^2_Wm^2_W}(1-\overline{\ln s})
\nonumber\\
&&+{6s_W\over c_W} \Big [{m^2_t\over s^2_Wm^2_W}C^R_t+{m^2_b\over s^2_Wm^2_W}C^R_b \Big ]
\nonumber\\
&&+{2s^2_W\over c^5_W}\Big [{s\overline{\ln t}\over t}+{s\overline{\ln u}\over u} \Big ]
-{s^2_W\over c^5_W} \Big [{s\overline{\ln^2t}\over t}+{s\overline{\ln^2u}\over u}
+2\overline{\ln^2r_{tu}} \Big ]~~, \label{NR1SM}
\eqa
with
\bqa
C^R_t &= &{(-3+20s^2_W) \over36s_Wc^3_W}\Big [{\overline{\ln^2s}\over4}-{\overline{\ln s}\over2}
+1\Big ]+{(6-10s^2_W)\over36s_Wc^3_W}\overline{\ln s} ~~, \nonumber \\
C^R_b &= & {(3+2s^2_W)\over36s_Wc^3_W}\Big [{\overline{\ln^2s}\over4}-{\overline{\ln s}\over2}
+1\Big ]+{(3-s^2_W) \over36s_Wc^3_W}\overline{\ln s} ~~, \label{CRtb}
\eqa
and
\bqa
s(N^R_3-N^R_4) = -{2s^2_W\over c^5_W} \left [{s(\overline{\ln^2t}-\overline{\ln^2u})\over ut}
+2s\left ({\overline{\ln u}-\overline{\ln t}\over ut} \right )
+{\overline{\ln^2r_{tu}}\over u}-{\overline{\ln^2r_{tu}}\over t} \right ] ~~. \label{NR34SM}
\eqa\\

\vspace*{1cm}
 \underline{Correspondingly in MSSM},  using the parameters in Table B.1 for  $H=h^0, H^0$,

\begin{table}[hbt]
\begin{center}
{ Table B.1: Parameters for the HV amplitudes in MSSM  }\\
  \vspace*{0.3cm}
\begin{small}
\begin{tabular}{||c|c|c||}
\hline \hline
  & $h^0$ & $H^0$  \\
 \hline
 $C^-_H$ & $\sin(\beta-\alpha) $  &   $\cos(\beta-\alpha)$     \\
  $C^+_H$ & $\sin(\beta+\alpha)$  & $-\cos(\beta+\alpha)$        \\
   $f_{G^+G^-H}$ & ${m_W\over2s_Wc^2_W} \cos(2\beta)\sin(\beta+\alpha)$   &
   $ -{m_W\over2s_Wc^2_W} \cos(2\beta)\cos(\beta+\alpha)$       \\
    $f_{H^+H^-H}$ & $-~{m_W\over s_W}\left [{\cos(2\beta)\sin(\beta+\alpha)\over2c^2_W}
+\sin(\beta-\alpha) \right ] $ &   ${m_W\over s_W} \left [{\cos(2\beta)\cos(\beta+\alpha)\over2c^2_W}
-\cos(\beta-\alpha) \right ]$       \\
   $f_{tH}$ & ${\cos\alpha\over\sin\beta}$ &  ${\sin\alpha\over\sin\beta}$             \\
     $f_{bH}$ & $-{\sin\alpha\over\cos\beta}$ & ${\cos\alpha\over\cos\beta}$            \\
 \hline  \hline
\end{tabular}
 \end{small}
\end{center}
\end{table}

\noindent
and the sfermion couplings\footnote{In (\ref{Hsfermion}), the  Higgs-fermion coupling
in the last two lines of Table B.1  are used. }
\bqa
f_{Z\tilde{f}} &= & -~{1\over s_Wc_W}(I^3_{\tilde{f}}-s^2_WQ_{\tilde{f}}) ~~, \label{Zsfermion}  \\
f_{H\tilde{f}}&=&{m_W\over s_Wc^2_W}(I^3_{\tilde{f}}-s^2_WQ_{\tilde{f}})C^+_H
-{m^2_f\over s_Wm_W}f_{fH} ~~, \label{Hsfermion}
\eqa
we get
\bqa
 && sN^L_1={(2s^2_W-1)\over2s^2_Wc^3_W} \Bigg \{
 {[-\overline{\ln^2s}+3\overline{\ln s}-1]\over4s^2_Wc^2_W}
  +{[-\overline{\ln^2s} +3\overline{\ln s}-1] \over2s^2_W(2s^2_W-1)}
\nonumber\\
&& -  {c^2_W [\overline{\ln s}+1+4\overline{\ln s}]\over s^2_W(2s^2_W-1)}
- \sum_{i} Z^+_{1i}Z^{+*}_{1i} {[\overline{\ln s_{\chi_i\tilde{\nu}_{eL}}(e)}-1] \over 2s^2_W(2s^2_W-1)}
\nonumber\\
&& - \sum_{i} {|Z^N_{1i}s_W+Z^N_{2i}c_W|^2\over4s^2_Wc^2_W}
[\overline{\ln s_{\chi_i\tilde{e}_{L}}(e)}-1]
+ \sum_{i} {c^2_W |Z^+_{1i}|^2\over s^2_W(2s^2_W-1)}
[\overline{\ln s_{\chi_i\tilde{\nu}_{eL}}(e)}+1] \Bigg \}
\nonumber\\
&&  +C^-_H \Bigg \{{1\over2s^4_Wc_W}\overline{\ln^2s}
-{1+9c^4_W+13c^6_W\over8s^4_Wc^5_W}\overline{\ln s}-{1-2c^4_W-2c^6_W\over8s^4_Wc^6_W}
\Bigg \}\nonumber\\
&&
-{(f_{G^+G^-H}+f_{H^+H^-H})(1-2s^2_W)\over4c^3_Ws^3_Wm_W}(1-\overline{\ln s})
+{6s_W\over c_W}\Big [{m^2_t\over s^2_Wm^2_W}C^R_tf_{tH}+{m^2_b\over s^2_Wm^2_W}C^R_bf_{bH} \Big ]
\nonumber\\
&&+C^-_H\Bigg\{
-{c_W\over 2s^4_W}\Big [{(t-2s)\over2t}\overline{\ln^2t}+{(u-2s)\over2u}\overline{\ln^2u}
+{(2s-u)\over2u}\overline{\ln^2r_{ts}}+{(2s-t)\over2t}\overline{\ln^2r_{us}} \Big ]
\nonumber\\
&&+{1\over 4c_Ws^2_W}\Big [-\overline{\ln^2t}-\overline{\ln^2u}
+\overline{\ln^2r_{ts}}+\overline{\ln^2r_{us}} \Big ]
\nonumber\\
&&-\Big ( {1\over 4c_Ws^4_W}+{(2s_W^2-1)^3\over 8c^5_Ws^4_W}\Big )
\Big [{s\over t}\overline{\ln^2t}+{s\over u}\overline{\ln^2u}+2\overline{\ln^2r_{tu}} \Big]
\Bigg \}\nonumber\\
&&+\Big( {(2s^2_W-1)\sin\alpha<M^+_{12}>\over2\sqrt{2}c_Ws^4_Wm_W}
+{(2s^2_W-1)^2<M^{0}_{L+}>\over8c^4_Ws^4_Wm_W}\Big)
\Big [{s\overline{\ln^2t}\over 2t}-2{s\overline{\ln t}\over t}
+{s\overline{\ln^2u}\over 2u}-2{s\overline{\ln u}\over u} \Big ]\nonumber\\
&&
+\Big ( {1-2s^2_W\over 4c^3_Ws_W^4}-{(1-2s^2_W)^2\over 8c^5_Ws_W^4}\Big )
C^+_{H}\Big [{s\overline{\ln t}\over t}+{s\overline{\ln u}\over u}\Big ]\nonumber\\
&& +{2\over m_W}\Sigma_{\tilde{f}} f_{Z\tilde{f}}f_{H\tilde{f}} \Big [Q_{\tilde{f}}
+{2s^2_W-1\over 2s_Wc_W}f_{Z\tilde{f}} \Big ](1-\overline{\ln s})\nonumber\\
&&-{\sqrt{2}\over m_W}\Big \{{(1-4c^2_W)\over s^2_Wc_W}(<M^+_{21}>\cos\alpha-<M^+_{12}>\sin\alpha)
\Big [ {1\over4}\overline{\ln^2s}-\overline{\ln s}+1\Big ]\nonumber\\
&&-{(1-2s^2_W)\over4s^4_Wc^3_W} \Big [2c^2_W(1-2s^2_W)(<M^+_{21}>
\cos\alpha-<M^+_{12}>\sin\alpha)\overline{\ln s}
\nonumber\\
&&-(8c^4_W-4c^2_W-1)(<M^+_{21}>\cos\alpha-<M^+_{12}>\sin\alpha)
\Big ({1\over4}\overline{\ln^2s}-{1\over2}\overline{\ln s}+1 \Big ) \Big ]
\Big \}\nonumber\\
&&-{(1-2s^2_W)\over4s^4_Wc^4_Wm_W}<M^{'0}_{L-}>
\Big [{1\over4}\overline{\ln^2s}-{1\over2}\overline{\ln s}+1 \Big ]
\nonumber\\
&&-\Big ({1\over 8c^3_Ws^4_W}+{(1-2s^2_W)^2\over16s^4_Wc^5_W}\Big )
C^+_{H}\Big [{s\over u}\overline{\ln^2r_{ts}}+{s\over t}\overline{\ln^2r_{us}} \Big ]
\nonumber\\
&&-{<M^+_{12}> c_W \sin\alpha \over2\sqrt{2}s^4_Wm_W} \Big [\overline{\ln^2t}
+\overline{\ln^2u}+{t\over u}\overline{\ln^2r_{ts}}+{u\over t}\overline{\ln^2r_{us}}\nonumber\\
&&
-2\overline{\ln^2s}-{s\over u}\overline{\ln^2r_{ts}}
-{s\over t}\overline{\ln^2r_{us}}+{(1-2s^2_W)\over2c^2_W}(-\overline{\ln^2t}
-\overline{\ln^2u}+\overline{\ln^2r_{ts}}+\overline{\ln^2r_{us}}) \Big ]
\nonumber\\
&& +{(1-2s^2_W)\over16s^4_Wc^4_W}{<M^0_{L-}>\over m_W} [-\overline{\ln^2t}+\overline{\ln^2r_{ts}}
-\overline{\ln^2u}+\overline{\ln^2r_{us}}]\nonumber\\
&&+{\sin\alpha\over2\sqrt{2}s^4_Wc_W}{<M^+_{12}>\over m_W}\overline{\ln^2r_{tu}}
+{(1-2s^2_W)^2\over8s^4_Wc^4_W}{<M^0_{L+}>\over m_W}\overline{\ln^2r_{tu}} ~~, \label{NL1MSSM}
\eqa
\bqa
&& s(N^L_3-N^L_4)=C^-_{H}\Bigg \{-{2c_W\over s^4_W} \Big [{(t-u)\over2tu}\overline{\ln^2s}
+{\overline{\ln^2t}\over 4t}-{\overline{\ln^2u}\over 4u}\nonumber\\
&&
+{s\overline{\ln r_{tu}}\over ut}+{(2u^2-2t^2+ut)\over4tu^2}\overline{\ln^2r_{ts}}
+{(2u^2-2t^2-ut)\over4ut^2}\overline{\ln^2r_{us}} \Big ]\nonumber\\
&&
+{1\over 2c_Ws^2_W}\Big [{\overline{\ln^2u}\over u}-{\overline{\ln^2t}\over t}
+{\overline{\ln^2r_{us}}\over t}-{\overline{\ln^2r_{ts}}\over u} \Big ]\nonumber\\
&&-\Big ({1\over 2c_Ws^4_W}+{(2s_W^2-1)^3\over 4c^5_Ws^4_W} \Big )
\Big [{s\over ut}(\overline{\ln^2t}-\overline{\ln^2u})+{2s\over ut}(\overline{\ln u}
-\overline{\ln t})+{\overline{\ln^2r_{tu}}\over u}-{\overline{\ln^2r_{tu}}\over u}\Big ] \Bigg \}
\nonumber\\
&&-\Big ({{1\over 4c^3_Ws^4_W}+{(2s^2_W-1)^2\over8s^4_Wc^5_W}} \Big ) C^+_{H}
\Bigg [{s\overline{\ln^2r_{ts}}\over u^2}
-{s\overline{\ln^2r_{us}}\over t^2}
+2s{\overline{\ln r_{ts}}-\overline{\ln r_{us}}\over ut} \Bigg ]
\nonumber\\
&&-{<M^+_{12}>c_W \sin\alpha \over\sqrt{2}s^4_Wm_W}\Bigg [{\overline{\ln^2s}\over t}
-{\overline{\ln^2s}\over u}+{\overline{\ln^2u}\over u}
-{\overline{\ln^2t}\over t}-{2s\overline{\ln r_{ts}}\over ut}\nonumber\\
&&
+{2s\overline{\ln r_{us}}\over ut}+{t\over u^2}\overline{\ln^2r_{us}}
-{u\over t^2}\overline{\ln^2r_{us}}+{s(u-t)\over tu^2}\overline{\ln^2r_{ts}}
+{s(u-t)\over ut^2}\overline{\ln^2r_{us}}
\nonumber\\
&&+{(1-2s^2_W)\over2c^2_W}\Bigg ({\overline{\ln^2t}\over t}-{\overline{\ln^2u}\over u}
+{\overline{\ln^2r_{ts}}\over u}-{\overline{\ln^2r_{us}}\over t} \Bigg ) \Bigg ]
+{C^+_H\over s^4_W}({s\over tu})\ln r_{tu}\nonumber\\
&&
+{(1-2s^2_W) \over8s^4_Wc^4_W}{<M^0_{L-}>\over m_W} \Bigg ({\overline{\ln^2t}\over t}
+{\overline{\ln^2r_{ts}}\over u}
-{\overline{\ln^2u}\over u}-{\overline{\ln^2r_{us}}\over t} \Bigg )\nonumber\\
&&+\Bigg [{<M^+_{12}>\sin\alpha \over\sqrt{2}s^4_Wc_Wm_W}
+{(1-2s^2_W)^2\over4s^4_Wc^4_W}{<M^0_{L+}>\over m_W} \Bigg ]
\Bigg [{s(\overline{\ln^2t} -\overline{\ln^2u})\over 2tu}
\nonumber\\
&& -{2s\overline{\ln^2r_{tu}}\over tu}+{\overline{\ln^2r_{tu}}\over 2u}
-{\overline{\ln^2r_{tu}}\over 2t} \Bigg ] ~~, \label{NL34MSSM}
\eqa

\bqa
&& sN^R_1= {-1\over c^3_W}\Big \{ {[-\overline{\ln^2s}+3\overline{\ln s}-1]\over c^2_W}
-\sum_{i} {|Z^N_{1i}|^2\over c^2_W}
[\overline{\ln s_{\chi_i\tilde{e}_{R}}(e)}-1] \Big \}\nonumber\\
&&
+C^-_H\Big \{{\overline{\ln^2s}\over c_Ws^2_W}
-{\overline{\ln s}\over 4c^5_Ws^2_W}[1+2c^2_W+4c^2_W]+{1\over 4c^5_Ws^2_W}[1+2c^2_W+4c^2_W] \Big \}
\nonumber\\
&& -{f_{G^+G^-H}(1-2s^2_W)\over2c^3_Ws_Wm_W}(1-\overline{\ln s})
+{6s_W\over c_W}\Big [{m^2_t\over s^2_Wm^2_W}C^R_tf_{tH}+{m^2_b\over s^2_Wm^2_W}C^R_bf_{bH} \Big ]
\nonumber\\
&&+C^-_H \Big \{{2s^2_W\over c^5_W} \Big [{s\overline{\ln t}\over t}+{s\overline{\ln u}\over u} \Big ]
-{s^2_W\over c^5_W}\Big [{s\overline{\ln^2t}\over t}+{s\overline{\ln^2u}\over u}
+2\overline{\ln^2r_{tu}} \Big ] \Big\}\nonumber\\
&&+{s_W\over c^4_W}{<M^0_{R+}>\over m_W} \Big [{s\overline{\ln^2t}\over 2t}-2{s\overline{\ln t}\over t}
+{s\overline{\ln^2u}\over 2u}-2{s\overline{\ln u}\over u} \Big ]
+{2s^2_W\over c^5_W}C^+_{H} \Big [{s\overline{\ln t}\over t}+{s\overline{\ln u}\over u} \Big ]
\nonumber\\
&&+{2\over m_w}\Sigma_{\tilde{f}} f_{Z\tilde{f}}f_{H\tilde{f}} \Big [Q_{\tilde{f}}
+{s_W\over c_W}f_{Z\tilde{f}} \Big ](1-\overline{\ln s})\nonumber\\
&&-{\sqrt{2}\over m_W} \Big \{{(1-4c^2_W)\over s^2_Wc_W}(<M^+_{21}>\cos\alpha-<M^+_{12}>\sin\alpha) \Big [
{1\over4}\overline{\ln^2s}-\overline{\ln s}+1 \Big ]\nonumber\\
&&+{1\over2s^2_Wc^3_W} \Big [2c^2_W(1-2s^2_W)(<M^+_{21}>\cos\alpha-<M^+_{12}>\sin\alpha)\overline{\ln s}
\nonumber\\
&&-(8c^4_W-4c^2_W-1)(<M^+_{21}>\cos\alpha-<M^+_{12}>\sin\alpha)
\Big ({1\over4}\overline{\ln^2s}-{1\over2}\overline{\ln s}+1 \Big ) \Big ] \Big \}
\nonumber\\
&&+{1\over2s^2_Wc^4_Wm_W}<M^{'0}_{L-}> \Big [{1\over4}\overline{\ln^2s}-{1\over2}\overline{\ln s}+1 \Big ]\nonumber\\
&& -~{<M^0_{R-}>\over4s_Wc^4_Wm_W} [-\overline{\ln^2t}+\overline{\ln^2r_{ts}}
-\overline{\ln^2u}+\overline{\ln^2r_{us}}]
\nonumber\\
&&+{s^2_W\over c^5_W}C^+_{H}\Big [{s\over u}\overline{\ln^2r_{ts}}+{s\over t}\overline{\ln^2r_{us}} \Big ]
+{s_W<M^0_{R+}>\over c^4_Wm_W} \overline{\ln^2r_{us}} ~~, \label{NR1MSSM}
\eqa
\bqa
&& s(N^R_3-N^R_4) = -{2s^2_W\over c^5_W}C^-_{H}\Big  [{\overline{\ln^2t}-\overline{\ln^2u}\over ut}
+2({\overline{\ln u}-\overline{\ln t}\over ut})
+{\overline{\ln^2r_{tu}}\over su}-{\overline{\ln^2r_{tu}}\over st} \Big  ]
\nonumber\\
&&+{2s^2_W\over c^5_W}C^+_{H}\Big [{\overline{\ln^2r_{ts}}\over u^2}
-{\overline{\ln^2r_{us}}\over t^2}
+2{\overline{\ln r_{ts}}-\overline{\ln r_{us}}\over ut} \Big ]
\nonumber\\
&&-{<M^0_{R-}>\over 2s_Wc^4_Wm_W}\Big [{\overline{\ln^2t}\over t}
-{\overline{\ln^2u}\over u}+{\overline{\ln^2r_{ts}}\over u}
-{\overline{\ln^2r_{us}}\over t} \Big ]\nonumber\\
&&
+{s_W<M^0_{R+}>\over c^4_Wm_W} \Big [{s\overline{\ln^2t}\over ut}-{s\overline{\ln^2u}\over ut}
-4s{\overline{\ln r_{tu}}\over ut}+{\overline{\ln^2r_{tu}}\over u}
-{\overline{\ln^2r_{tu}}\over t} \Big ] ~~. \label{NR34MSSM}
\eqa

In (\ref{NL1MSSM}-\ref{NR34MSSM}) some mass averages appear, which for $h^0$ are
\bqa
&& <M^+_{12}>={m_W\sqrt{2}\over\sqrt{1+\tan^2\beta}}
~~, ~~<M^+_{21}>={m_W\sqrt{2}\tan\beta\over\sqrt{1+\tan^2\beta}}~~, \nonumber \\
&& <M^0_{L+}>=\sin\alpha(s_W<M^N_{13}>+c_W<M^N_{23}>)+\cos\alpha(s_W<M^N_{14}>+c_W<M^N_{24}>)~~, \nonumber \\
&& <M^0_{L-}>=-\sin\alpha(s_W<M^N_{13}>+c_W<M^N_{23}>)+\cos\alpha(s_W<M^N_{14}>+c_W<M^N_{24}>)~~, \nonumber \\
&& <M^0_{R-}>=\sin\alpha <M^N_{13}>-\cos\alpha <M^N_{14}> ~~, \nonumber \\
&& <M^0_{R+}>=\sin\alpha <M^N_{13}>+\cos\alpha <M^N_{14}>~~, \nonumber \\
&& <M^{'0+}_{L-}>=\sin\alpha(s_W<M^N_{13}>-c_W<M^N_{23}>)+\cos\alpha(s_W<M^N_{14}>-c_W<M^N_{24}>)~~, \nonumber \\
&& <M^N_{13}>={-m_Ws_W\over c_W\sqrt{1+\tan^2\beta}}
~~,~~<M^N_{23}>={m_W\over \sqrt{1+\tan^2\beta}}~~, \nonumber \\
&& <M^N_{14}>={m_Ws_W\tan\beta\over c_W\sqrt{1+\tan^2\beta}}
~~,~~<M^N_{24}>={-m_W\tan\beta\over \sqrt{1+\tan^2\beta}}~~, \label{chargino-neutralino-param}
\eqa
where   $M^{+\top}$ denotes the  $\tilde \chi^+$ mass matrix and  $M^N$
the neutralino one.  The corresponding expressions for $H^0$ may be obtained from them by replacing
\bq
\sin\alpha \to -\cos\alpha ~~~,~~~  \cos\alpha \to \sin\alpha ~~. \label{H0-mass-averages}
\eq\\

\subsection{ A simpler approximation for the HV amplitudes }

As the HV amplitudes are less important than the HC ones,
we can try an even simpler approximation at intermediate energies,
by  neglecting mass differences and using a common mass-scale $m_W$.
Thus at energies like e.g. in the range 0.6-5 TeV,  we try fitting
the complete Born+1loop  contributions by a Sudakov-type expression\footnote{Although there is
no rigorous Sudakov rule for such suppressed amplitudes.} of the form
\bq
F^{\rm HV}=F^{\rm Born} \left \{1+a \ln^2{s\over m^2_W} +b \ln{s\over m^2_W} +c \right \}~~. \label{simple-HV-fit}
\eq
Using this, the fitted constants are given in Table B.2 for the benchmarks S1(S2) of
ref.\cite{bench}.

\begin{table}[hbt]
\begin{center}
{ Table B.2: Fits of the approximate expression (\ref{simple-HV-fit}) to the    HV amplitudes \\
for the S1 (S2) benchmarks of \cite{bench},  in the energy range   0.6-5 TeV.}\\
  \vspace*{0.3cm}
\begin{small}
\begin{tabular}{||c| c| c| c||c|c|c||}
\hline \hline
& \multicolumn{3}{|c||}{ SM:  $H$ }  &
\multicolumn{3}{c||}{ MSSM S1(S2):  $h^0$} \\
 \hline
 & a & b & c & a & b & c    \\
  \hline
 $F_{--}, ~ F_{-+}$ & -0.0075  & 0.024 & -0.058 &-0.015(-0.013)  &  0.12(0.10)  & -0.28(-0.22) \\
 $F_{+-} , ~F_{++}$ & -0.0046  & 0.017  & 0.050  & -0.0027(-0.0026)  & -0.005(-0.008) & 0.14(0.16) \\
 \hline
\end{tabular}
 \end{small}
\end{center}
\end{table}

\clearpage

\clearpage

\clearpage

\clearpage

\clearpage

\clearpage

\clearpage

\clearpage

\clearpage

\newpage

\begin{thebibliography}{99}

%
\bibitem{Higgs} P. Higgs, Phys. Lett. 12 (1964) 132; \prl{13}{508}{1964}; \pr{145}{1156}{1966)};
F. Englert and R. Brout, \prl{13}{321}{1964}; G. Guralnik, C. Hagen and
T. Kibble, \prl{13}{585}{1964}.
%
\bibitem{Higgsdiscov} G. Aad et al. [ATLAS Collaboration], \pl{B716}{1}{2012} [arXiv:1207.7214 [hep-ex]].
S. Chatrchyan et al. [CMS Collaboration], \pl{B716}{30}{2012}  [arXiv:1207.7235
[hep-ex]]. Gavin J. Davies for the  CDF and D0 Collaborations, published in Front.Phys.China {\bf 8}, 270 (2013), [arXiv:1207.0449 [hep-ex]].
ATLAS Collaboration, see:
https://twiki.cern.ch/twiki/bin/view/AtlasPublic/HiggsPublicResults.
CMS Collaboration, see:
 https://twiki.cern.ch/twiki/bin/view/CMSPublic/PhysicsResultsHIG.
%
\bibitem{Higgsearch}John Ellis, 1312.5672;
S. Dawson et al (Higgs Working Group)  1310.8361;
M Klute et al 1301.1322;
A. Djouadi, \prep{459}{1}{2008} 1 , arXiv:hep-ph/0503173;
J. Gunion, H. Haber, G. Kane and S. Dawson, The Higgs Hunter's Guide, Addison-Wesley,
1990;
S. Heinemeyer, \ijmp{A 21}{2659}{2006} , arXiv:hep-ph/0407244.
%
\bibitem{ILC} Higgs Working Group Report, A. Ajaib et al, arXiv:(hep-ex)1310.8361;
T.Behnke et al, The Int.Lin.Coll.Tech.Design Report, arXiv:1306.6327.
%
\bibitem{heli1}  G.J. Gounaris and F.M. Renard,
\prl{94}{131601}{2005},  hep-ph/0501046.
%
\bibitem{heli2} G.J. Gounaris and F.M. Renard,  \pr{D73}{097301}{2006},
hep-ph/0604041, (an Addendum).
%
\bibitem{super} G.J. Gounaris and F.M. Renard,
 \polon{42}{2107}{2011}, arXiv:1106.2707[hep-ph].
%
\bibitem{ttbar} G.J. Gounaris and F.M. Renard, \pr{D86}{013003}{2012},
arXiv:1205.4547 [hep-ph].
%
\bibitem{WW} G.J. Gounaris and F.M. Renard, \pr{D88}{113003}{2013},
arXiv:1309.3177 [hep-ph].
%
\bibitem{Denner} A. Denner, J. Kubleck, R. Mertig, M. Bohm, \zp{C56}{261}{1992}.
B.A. Kniehl, \zp{C55}{605}{1992}. A. Denner , B.A. Kniehl, \npps{29A}{263}{1992}.
R. Hempfling, B.A. Kniehl, \zp{C59}{263}{1933}.
%
\bibitem{MSSMrules} M. Beccaria, M. Melles, F. M. Renard,
S. Trimarchi, C. Verzegnassi, \ijmp{A18}{5069}{2003}, hep-ph/0304110.
M. Beccaria, F.M. Renard and C. Verzegnassi, Nucl.Phys. B663 (2003) 394, hep-ph/0304175.
M. Beccaria, E. Mirabella, \pr{D71}{115016}{2005},
  [hep-ph/0505172].
%
\bibitem{equivalence1} J.M. Cornwall, D.N. Levin, and G. Tiktopoulos,
\pr{D10}{1145}{1974}; C.E.Vayonakis, \lnc{17}{383}{1976}.
%
\bibitem{equivalence2} M.S. Chanowitz, and M.K. Gaillard, \np{B261}{379}{1985};
G.J. Gounaris, R. K\"ogerler and H. Neufeld, \pr{D34}{3257}{1986}.
%
\bibitem{anom} G.J. Gounaris, F.M. Renard and N.D. Vlachos,
\np{B459}{51}{1996}; G.J. Gounaris, D.T. Papadamou and F.M. Renard,
\zp{C76}{333}{1997}.
%
\bibitem{JW} M. Jacob and G.C. Wick, \aop{7}{404}{1959}, \aop{281}{774}{2000}.
%
\bibitem{PV} G. Passarino and M. Veltman \np{B160}{151}{1979}.
%
\bibitem{OS} W. Hollik, \fortp{38}{165}{1990}.
%
\bibitem{Freitas} A. Freitas and D. Stockinger, \pr{D66}{095014}{2002};
see also arXiv: hep-ph/0210372.
%
\bibitem{HH} M. Beccaria, A. Ferrari, F.M. Renard, C. Verzegnassi,
LC-TH-2005-005, arXiv:(hep-ph) 0506274.
%
\bibitem{asPV} M. Beccaria, G.J. Gounaris, J. Layssac and F.M. Renard,
\ijmp{A23}{1839}{2008}.
%
\bibitem{bench} M Arana-Catania, S. Heinemeyer, M.J. Herrero,  \pr{D88}{015026}{2013} arXiv:1304.2783[hep-ph]. See also arXiv:1405.6960[hep-ph].
%
\bibitem{pinch} G. Degrassi and A. Sirlin, \np{B383}{73}{1992}, \pr{D46}{3104}{1992}.
%


\end{thebibliography}
\end{document}